\documentclass[12pt]{article}
\pdfoutput=1

\usepackage[pdftex]{graphicx}
\usepackage{amssymb}
\usepackage{amsmath,ulem}
\usepackage{cite}
\usepackage{slashed}
\usepackage{lscape} 
\usepackage{longtable}
\usepackage{booktabs}
\usepackage{array}
\usepackage{multirow}
\usepackage{rotating}

\usepackage{xcolor}
\definecolor{tit}{rgb}{0.1,0.2,0.4}

\newcommand{\eq}[1]{\begin{equation} #1 \end{equation}}

\newcommand{\eqa}[1]{\begin{eqnarray} #1 \end{eqnarray}}

\newcommand{\av}[1]{\langle #1 \rangle}

\newcommand{\GeV}{\,{\rm GeV}}
\newcommand{\MeV}{\,{\rm MeV}}

\newcommand{\op}{\mathcal{O}}

\newcommand{\Eq}[1]{Eq.~(\ref{#1})}
\newcommand{\Eqs}[1]{Eqs.~(\ref{#1})}
\newcommand{\Ref}[1]{Ref.~\cite{#1}}
\newcommand{\Refs}[1]{Refs.~\cite{#1}}
\newcommand{\App}[1]{Appendix~\ref{#1}}
\newcommand{\Sec}[1]{Section~\ref{#1}}
\newcommand{\Tab}[1]{Table~\ref{#1}}
\newcommand{\Fig}[1]{Figure~\ref{#1}}
\newcommand{\im}{{\rm Im}}
\newcommand{\F}{{\cal F}}

\newcommand{\W}{{\cal W}}
\newcommand{\I}{{\cal I}}
\renewcommand{\P}{{\cal P}}
\newcommand{\A}{{\cal A}}
\newcommand{\cL}{{\cal L}}
\newcommand{\B}{{\cal B}}
\newcommand{\D}{{\cal D}}
\newcommand{\N}{{\cal N}}
\renewcommand{\H}{{\cal H}}

\DeclareMathOperator{\Tr}{Tr}

\begin{document}


\begin{flushright}
LPT-ORSAY/19-31, SI-HEP-2019-11, QFET-2019-08 \\
TUM-HEP-1161/18, MIT-CTP/5058, NIOBE-2019-01 \\[1cm]
\end{flushright}

$\ $
\vspace{-2mm}
\begin{center}
\fontsize{16.8}{20}\selectfont
\bf 
Light-Cone Sum Rules for $B\to K\pi$ Form Factors\\[2mm]
and Applications to Rare Decays
\end{center}

\vspace{2mm}

\begin{center}
{\rm S\'ebastien Descotes-Genon{$^{\, a}$}, Alexander Khodjamirian{$^{\, b}$} and Javier Virto{$^{\, c,\, d}$}} \\[5mm]
{\it\small
{$^{\, a}$} Laboratoire de Physique Th\'eorique (UMR 8627), CNRS, Univ. Paris-Sud,\\
Universit\'e Paris-Saclay,
91405 Orsay, France\\[2mm]
{$^{\, b}$} Theoretische Physik 1, Naturwissenschaftlich-Technische Fakult\"at,\\
Universit\"at Siegen, Walter-Flex-Stra\ss{}e 3, D-57068 Siegen, Germany\\[2mm]
{$^{\, c}$} Center for Theoretical Physics, Massachusetts Institute of Technology,\\
77 Mass. Ave., Cambridge, MA 02139, USA\\[2mm]
{$^{\, d}$} Physics Department T31, Technische Universit\"at M\"unchen,\\ James Frank-Stra{\ss}e 1, D-85748 Garching, Germany
}
\end{center}

\vspace{1mm}
\begin{abstract}\noindent
\vspace{-5mm}

We derive a set of light-cone sum rules relating the hadronic form factors 
relevant for $B\to K\pi\ell^+\ell^-$ decays to the $B$-meson light-cone distribution amplitudes (LCDAs). We obtain the sum rule relations for all 
$B\to K\pi$ form factors of (axial)vector and (pseudo)tensor $b\to s$ currents with a $P$-wave $K\pi$ 
system. Our results reduce  to the known 
light-cone sum rules for $B\to K^*$ form factors 
in the limit of a single narrow-width resonance.
We update the operator-product expansion for the underlying correlation function
by including a more complete set of $B$-meson LCDAs with higher twists, and produce numerical results for all $B\to K^*$ form factors in the narrow-width limit. 
We then use the new sum rules to estimate the effect of a non-vanishing $K^*$ width
in $B\to K^*$ transitions, and find that this effect is universal and increases the factorizable part of the rate of $B\to K^*X$ decays by a factor of $20\%$. This effect, by itself, goes in the direction of increasing the current tension in the differential $B\to K^*\mu\mu$ branching fractions.
We also discuss $B\to K\pi$ transitions outside the $K^*$ window, and explain how measurements of $B\to K\pi\ell\ell$ observables above the $K^*$ region can be used to further constrain the $B\to K^*$ form factors.

\end{abstract}

\newpage

\setcounter{tocdepth}{2}
\tableofcontents

\section{Introduction}
\label{sec:intro}

Exclusive $b\to s$ decay modes have been under intense experimental and theoretical scrutiny since the era of the $B$-factories and the Tevatron,
and have had a recent revival with the new data from the LHC.
Compared to the corresponding inclusive decay modes they have smaller branching ratios, but they are easier to measure, and with the much larger numbers
of $B$ mesons produced at the LHC experiments and the larger number of observables they provide, the exclusive modes lead the quest for indirect searches
for New Physics~\cite{Descotes-Genon:2015uva,Bifani:2018zmi}.

On the theory side, predictions for the exclusive $b\to s$ observables
-- and thus our ability to interpret the experimental data --
require the knowledge of hadronic form factors, defined as the matrix elements of flavour-changing quark currents between an initial $B$ meson and a final exclusive mesonic state:
$$F_i^{B\to M}(q^2) = \av{M(k)| \bar s \Gamma_i b |B(q+k)}\,.$$ 
A very substantial effort has been devoted in 
the past to the study of form factors where $M$ is a single pseudoscalar ($P=K,\pi,\eta$) or vector
($V=K^*\!,\rho,\phi,\omega$) meson.
In the attempt to calculate these form factors directly from QCD, two largely complementary approaches stand out.
Lattice QCD~(see~\Ref{Aoki:2016frl} and references therein) applies to the region of small momenta of the final meson $M$, where the momentum transfer $q^2$ is such that $0\ll q^2 \lesssim (m_B-m_M)^2$, typically $q^2 \gtrsim 20\ \rm{GeV}^2$. 
The method of QCD Light-Cone Sum Rules (LCSRs)~\cite{Balitsky:1986st, Balitsky:1989ry, Chernyak:1990ag} applied to the form factors is on the other hand limited to the region of 
large and intermediate recoil of~$M$, or $0< q^2\ll (m_B-m_M)^2$,
usually $q^2\lesssim 10\ \rm{GeV}^2$.
In addition, two versions of the LCSRs exist, depending on whether they relate the form factors to the light-cone distribution amplitudes (LCDAs) of the light meson  (see e.g.  \Refs{Belyaev:1993wp,Duplancic:2008ix}
and \cite{Ball:1998kk,Ball:2004rg} for the LCSRs with $P$-meson and $V$-meson LCDAs, respectively), or 
to the $B$-meson LCDAs~\cite{Khodjamirian:2005ea,Khodjamirian:2006st} (for more recent applications see e.g. \Refs{Wang:2015vgv,Lu:2018cfc,Gubernari:2018wyi,Gao:2019lta}).
We will use the latter approach in this article.

Our main focus here is on the $B\to K^*$ form factors.
Current state-of-the-art calculations of $B\to K^*$ form factors from lattice QCD~\cite{Horgan:2013hoa}
and LCSRs~\cite{Straub:2015ica} are compatible with each other 
and have estimated uncertainties at the level of $5\%-20\%$. While systematic improvement
in the lattice calculations will eventually render the LCSRs not competitive, 
the latter will remain an important alternative and a complementary method.

As we shall discuss below, the sum rules with $B$-meson LCDAs help to assess
the following limitation of the current calculations of $B\to V$ form factors. 
Both approaches, lattice QCD and LCSRs work under the assumption that the 
vector mesons are stable. This is certainly a reasonable approximation if one assumes that corrections to the narrow-width limit are suppressed by the width-to-mass ratio,
which is $\sim 5\%$ for the~$K^*$ and below the percent for the~$\phi$.
However, once the uncertainties of the calculated form factors get below the ten-percent level 
(specially for the $B\to K^*$ form factors), a proper estimation of the finite-width effects becomes mandatory.
In addition, other non-resonant backgrounds  such as the interference among $P$ and $S$ waves in the $K\pi$ system are known to be important. 
While these are disentangled at the level of the experimental analysis, a fully consistent match between experimental  measurements and theoretical predictions
requires addressing all effects beyond the narrow-width approximation from the theory point of view, too.

Thus, ideally one needs to consider the form factors with two stable mesons in the final state, e.g. $\av{K\pi | \bar s \Gamma b | B}$ instead of
$\av{K^* | \bar s \Gamma b | B}$. As discussed originally in \Ref{Cheng:2017smj}, the LCSRs with $B$-meson LCDAs provide
a natural framework to perform this generalization, making also the relationship with the narrow-width limit completely transparent.
This method was used in \Ref{Cheng:2017smj} to study the $B\to \pi\pi$ form factors and their relation to $B\to \rho$ transitions.
The purpose of the present article is to derive the
corresponding LCSRs for $B\to K\pi$ form factors, and to study some of the phenomenological implications, with a focus on 
the semileptonic decay  $B\to K\pi \ell\ell$ both on and off resonance.

The LCSRs derived here are valid for small and intermediate momentum transfer squared $0<q^2\lesssim 10$ GeV$^2$ 
and, simultaneously, small invariant mass of the $K\pi$ system, typically up to the mass of the 
second radial excitation of $K^*$,  $k^2 \equiv m_{K\pi}^2\lesssim m(K^*(1680))^2$.
In other regions of phase space different calculational tools apply~\cite{Faller:2013dwa}.
For example, for large invariant masses of the $K\pi$ system the $B\to K\pi$ form factors can be factorized into $B\to K$ form factors and convolutions of perturbative kernels 
with pion light-cone distribution amplitudes~\cite{Boer:2016iez}. In the same region of small $q^2$ and $k^2$, 
one may use LCSRs with $K\pi$ distribution amplitudes
analogous to the ones derived in \Refs{Hambrock:2015aor,Cheng:2017sfk} for the $B\to\pi\pi$ case. The drawback of these sum rules is  our currently
limited knowledge of the generalized dimeson LCDAs.

A summary of the main points and novelties of this article is the following:

\begin{itemize}

\item We derive the LCSRs with $B$-meson LCDAs for the $P$-wave $B\to M_1 M_2$ form factors, with $M_{1,2}$ pseudoscalar mesons. This generalizes the results of \Ref{Cheng:2017smj} to the case where the two mesons have different masses. We also include the tensor form factors that were not considered in that reference. The focus is put on $B\to K\pi$ form factors but the analytic results are general to other two-body final states. The LCSRs 
are summarized in \Eq{eq:AllLCSRs}.

\item We recalculate the Operator Product Expansion (OPE) side of the sum rules including two- and three-particle contributions up to twist four, within a new parametrization of $B$-meson LCDAs with definite conformal
twist, proposed recently in \Ref{Braun:2017liq}. This in turn updates the results for $B\to V$ form factors in \Refs{Khodjamirian:2006st,Khodjamirian:2010vf}.
These results are collected in Appendices~\ref{app:BtoVLCSRs}~and~\ref{OPEexpressions}.
A brief review of the new parametrization and of the OPE calculation is given in Appendix~\ref{BLCDAs}. This update has also been done recently in~Refs.\cite{Gubernari:2018wyi,Gao:2019lta}.

\item We use our sum rules to constrain a resonance model for the form factors, and demonstrate explicitly how the narrow-width limit of the
LCSRs leads analytically to the known $B\to K^*$ sum rules. The models for all form factors are summarized in~Eqs.~(\ref{eq:FFmodels})
and~(\ref{eq:FFmodelsFinal}).
We use these models with the correct narrow-width limit to estimate the finite-width effects. We find that this correction is universal and encoded in a multiplicative factor $\W_{K^*}\simeq 1.1$. Thus the finite width effects are a $20\%$ correction at the level of the decay rate.

\item We use the data from the Belle collaboration on the $\tau \to K\pi \nu$ spectrum to determine the decay constants $f_{K^*}$ and $f_{K^*(1410)}$, and update the hadronic part of the two-point QCD sum rule to determine the threshold parameter $s_0$.
We  find that the threshold parameter is lower than the one used in the literature for $B\to K^*$ form factors in the narrow-width limit.
 
\item We apply our results to the rare decay $B\to K\pi \ell\ell$ on and off resonance.
We rederive the angular distribution and we demonstrate that in the narrow-width limit we recover the angular coefficients of the $B\to K^* \ell\ell$ decay. With this generalization at hand, we show how to use the LHCb data on the angular moments around the $K^*(1410)$ region to improve our control on the $B\to K^*$ form factors.

\end{itemize}

The $B\to K\pi \ell\ell$ decay off resonance has also been discussed in~\Refs{Das:2014sra,Das:2015pna}, focusing on the other end of the physical kinematic region, that is, at large dilepton masses. In this case the form factors can be parametrized using Heavy Hadron Chiral Perturbation Theory. In addition, the $B\to K\pi \ell\ell$ decay at large recoil with a soft pion has been discussed in~\Ref{Grinstein:2005ud}.
These analyses are complementary to the one presented here. 

While we will focus on the application to the semileptonic flavour-changing neutral-current decays, this is not the only case to which the results of this article apply. Analogously to non-leptonic decays of the type $B\to K^* M$
where factorization in the heavy-quark limit reduces the amplitudes to simpler objects including $B\to K^*$ form factors~\cite{Beneke:1999br,Bauer:2004tj},
the three body non-leptonic decays such as $B\to K \pi\pi$ are also reducible to $B\to K\pi$ form factors in certain regions of phase space~\cite{Krankl:2015fha,Virto:2016fbw,Klein:2017xti}. Thus the results presented here are also needed to compute predictions for non-leptonic three-body decays. Our results can be generalized to form factors of charged currents, such as those appearing in $B_s \to K \pi \ell \nu$. In this case naive factorization is exact and the form factors are the only non-perturbative input, allowing for a simultaneous study of $V_{ub}$ and right-handed currents~\cite{Feldmann:2015xsa}.

\bigskip
The plan of this article is the following.
We begin in \Sec{sec:FormFactors} reviewing the basic definitions, kinematics and partial-wave expansions of the form factors.
In \Sec{sec:LCSRs} we derive the light-cone sum rules for vector, timelike-helicity and tensor form factors.
In \Sec{sec:LCSRs@Work} we construct a phenomenological model for the $B\to K\pi$ form factors and rewrite the LCSRs in the context of this model. Using this model we demonstrate that the narrow-width limit leads to the well-known LCSRs for $B\to K^*$ form factors.
\Sec{sec:numerical} contains a comprehensive numerical analysis.
Applications of the results derived in this  are discussed in~\Sec{sec:applications}, with a focus on the rare $B\to K\pi \ell\ell$ decay. \Sec{sec:conclusions} contains a summary and a brief discussion on future directions. The various appendices include further material that complements the main part of the article: \App{app:BtoVLCSRs} contains a collection of the LCSRs with $B$-meson LCDAs for the $B\to V$ form factors in the narrow-width limit. In \App{BLCDAs} we give the definitions for the $B$-meson LCDAs and a discussion of the models used in the article. In \App{app:OPEcalc} we review our calculation of the correlation functions in the OPE, and the results are collected in \App{OPEexpressions}. \App{sec:OPEnumerics} contains some numerical results that complement the numerical analysis of \Sec{sec:numerical}.
Finally, in \App{app:beyondNWL} we review the formalism for corrections to the narrow-width limit in Breit-Wigner models. \App{app:kinematics} contains some details on the kinematics of the $B\to K\pi\ell\ell$ decay.


\section{$B\to K\pi$ Form Factors: Definitions, Kinematics and Partial Waves}
\label{sec:FormFactors}

A general Lorentz decomposition of the  $\bar B^0\to K^-\pi^+$ hadronic matrix elements of the $b\to s$ currents 
consistent with parity invariance is given by\,\footnote{
Our conventions are $\epsilon_{0123}=-\epsilon^{0123}=+1$ and
$\gamma_5\equiv (i/4!)\, \epsilon^{\mu\nu\rho\sigma} \gamma_\mu \gamma_\nu \gamma_\rho \gamma_\sigma$.
}:
\eqa{
i\langle K^-(k_1) \pi^+(k_2)|\bar{s}\gamma^\mu b|\bar{B}^0(q+k)\rangle
&=& F_\perp \, k^\mu_\perp \,,
\label{FV}\\
-i\langle K^-(k_1) \pi^+(k_2)|\bar{s}\gamma^\mu \gamma_5 b|\bar{B}^0(q+k)\rangle
&=&  F_t \, k^\mu_t + F_0 \, k^\mu_0 + F_\parallel \, k^\mu_\|\,, 
\label{FA}\\
\langle K^-(k_1) \pi^+(k_2)|\bar{s}\sigma^{\mu\nu} q_\nu b|\bar{B}^0(q+k)\rangle
&=&  F_\perp^T \, k^\mu_\perp \,, \\
\langle K^-(k_1) \pi^+(k_2)|\bar{s} \sigma^{\mu\nu}q_\nu \gamma_5  b|\bar{B}^0(q+k)\rangle
&=& F_0^T \, k^\mu_0 + F_\parallel^T \, k^\mu_\|\,,
\label{FTA}
}
where $k\equiv k_1+k_2$ is the total dimeson momentum and $q$ is the momentum transfer. 
The $B\to K\pi$ form factors $F_i^{(T)}$ $(i=\perp,t,0,\parallel)$ are Lorentz-invariant scalar functions of three kinematic invariants, $F^{(T)}_i=F^{(T)}_i(k^2,q^2,q\cdot \overline{k})$, with
\eq{
\overline k^\mu =
\bigg(1- \frac{\Delta m^2}{k^2}\bigg)\ k_1^\mu
- \bigg(1+ \frac{\Delta m^2}{k^2}\bigg)\ k_2^\mu\ ,
\label{eq:kbar}
} 
and $\Delta m^2 \equiv k_1^2 - k_2^2 = m_K^2 - m_\pi^2$, such that $k\cdot \overline k=0$.
In the Lorentz decomposition (\ref{FV}$-$\ref{FTA}) defining the form factors $F_i^{(T)}$ we use the following set of orthogonal Lorentz vectors:
\begin{align}
k^\mu_\perp &= \frac{2}{\sqrt{k^2} \sqrt{ \lambda}} \,
i\epsilon^{\mu\alpha\beta\gamma} \, q_\alpha \, k_{\beta} \, \bar{k}_{\gamma}\ ,   &
k^\mu_t &= \frac{q^\mu}{\sqrt{q^2}}\ , \nonumber \\
k^\mu_0 &= \frac{2\sqrt{q^2}}{\sqrt{\lambda}} \, \Big(k^\mu - \frac{k \cdot q}{q^2} q^\mu\Big)\ ,  &
k^\mu_\| &= \frac{1}{\sqrt{k^2}} \,
\Big(\overline{k}^\mu - \frac{4 (q\cdot k) (q \cdot \overline{k})}{\lambda} \, k_\mu
+ \frac{4 k^2 (q\cdot \overline{k})}{\lambda} \, q_\mu \Big)\ ,
\label{eq:kpar}
\end{align}
where $\lambda \equiv\lambda(m_B^2,q^2,k^2) = m_B^4 + q^4 + k^4 - 2(m_B^2 q^2 + m_B^2 k^2 + q^2 k^2 )$ is the kinematic K\"all\'en function.
Some useful relations are:
\begin{align}
q\cdot k &=\frac12(m_B^2-q^2-k^2)\ ,   &
q\cdot \overline k &=
\frac{\sqrt{\lambda\,\lambda_{K\pi}}\ \cos \theta_K}{2k^2}\ ,  \nonumber \\
\lambda &= 4(q\cdot k)^2 - 4q^2 k^2\ ,   &
k^2 \overline k^2  &=  -\lambda_{K\pi}  \ ,
\label{eq:kinematicrels}
\end{align}
where $\lambda_{K\pi} \equiv \lambda(k^2,m_K^2,m_\pi^2)$, and $\theta_K$ is the angle between the 3-momenta of the  pion and the
$B$-meson in the $(K\pi)$ rest frame. Occasionally, for example in the sum rules, the functions $\lambda$ and $\lambda_{K\pi}$ will depend
on a variable $s$ or $m_R^2$ instead of $k^2$. In these cases we will use the notation
$\lambda(x)\equiv\lambda(m_B^2,q^2,x)$ and $\lambda_{K\pi}(x)\equiv \lambda(x,m_K^2,m_\pi^2)$.

\bigskip

Other $B\to K\pi$ form factors with different flavour quantum numbers such as $\langle \overline K^0 \pi^0 |\bar{s} \Gamma b|\bar{B}^0\rangle$
or $\langle K^- \pi^0 |\bar{s} \Gamma b|B^-\rangle$ are related to $\langle K^- \pi^+ |\bar{s} \Gamma b|\bar{B}^0\rangle$ by isospin.
Since the $b\to s$ current is an isosinglet, in the isospin symmetry limit the $I=3/2$ component of the final $K\pi$ state does not contribute
to the form factor.
In the case of the neutral $K\pi$ state, the isospin decomposition is given by
\eq{
|K^-\pi^+\rangle = \sqrt{1/3} | K\pi \rangle_{3/2} + \sqrt{2/3} | K\pi \rangle_{1/2} \ ,\quad
|\overline K^0\pi^0\rangle = \sqrt{2/3} | K\pi \rangle_{3/2} - \sqrt{1/3} | K\pi \rangle_{1/2} \ ,
\label{eq:isospin}
}
which implies the isospin relation
\eq{
\langle \overline K^0(k_1) \pi^0(k_2)|\bar{s} \Gamma b|\bar{B}^0(p)\rangle = -\frac1{\sqrt{2}} \langle K^-(k_1) \pi^+(k_2)|\bar{s} \Gamma b|\bar{B}^0(p)\rangle\ .
}
From~\Eq{FA} also follows that
\eq{
-i\langle K^-(k_1) \pi^+(k_2)|\bar{s}\slashed{q} \gamma_5 b|\bar{B}^0(p)\rangle
= \langle K^-(k_1) \pi^+(k_2)|i (m_b+m_s) \bar{s}\gamma_5 b|\bar{B}^0(p)\rangle
= \sqrt{q^2}\ F_t\ ,
}
so that the timelike-helicity form factor $F_t$ is simply related to the matrix element of the pseudoscalar current, which is not independent.

\bigskip

The form factors $F_i^{(T)}(k^2,q^2, q\cdot \bar k)$ can be expanded in partial waves, by expressing the invariant $q\cdot \bar k$
in terms of the polar angle $\theta_K$ by virtue of \Eq{eq:kinematicrels} :
\eqa{
F_{0,t}(k^2,q^2, q\cdot \bar k) &=&
\sum_{\ell=0}^{\infty} \sqrt{2 \ell+1}\ F_{0,t}^{(\ell)}(k^2,q^2)\ 
P_{\ell}^{(0)}(\cos\theta_K)\, , 
\label{eq:PWE0t} \\[2mm]
F_{\bot,\|}(k^2,q^2, q\cdot \bar k) &=&
\sum_{\ell=1}^{\infty} \sqrt{2 \ell+1}\ F_{\bot,\|}^{(\ell)}(k^2,q^2)\ 
\frac{P_{\ell}^{(1)}(\cos\theta_K)}{\sin \theta_K}\, ,
\label{eq:PWEperppar} \
}
and similarly for the tensor form factors $F^T_{\perp,0,\|}$. Our conventions for the Legendre polynomials are such that
$P_0^{(0)}(\cos\theta)=1$, $P_1^{(0)}(\cos\theta)=\cos\theta$, $P_1^{(1)}(\cos\theta)=-\sin\theta$, etc. The sum rules derived below will project out
the $P$-wave components $F_i^{(\ell=1)}$. Starting from different correlation
functions one may derive analogous sum rules for other partial waves. These other sum rules will be discussed in a forthcoming publication~\cite{WIPSwave}.

We will also need the definition of the $K\pi$ form factors:
\eq{
\langle K^-(k_1)\pi^+(k_2) |\bar{s}\gamma_\mu d | 0 \rangle =
f_+(k^2) \ \overline k_\mu
+\frac{m_K^2 - m_\pi^2}{k^2}  f_0(k^2) \ k_\mu\ .
\label{eq:piFF}
}
In the isospin limit, $f_{+,0}(k^2)$ can be related to the vector and scalar form factors 
$F_{V,S}$ accessible from $\tau$ decays, which have been measured e.g. by Belle~\cite{Epifanov:2007rf} (see~\Sec{sec:taudata} for details).

We end this section by noting that the definitions for the form factors $F_i^{(\ell)}$ depend on the choice of the polar angle $\theta_K$.
In \Ref{Cheng:2017smj} (which focused on $B\to \pi\pi$ form factors) the partial-wave expansion was performed with respect to the angle $\theta_\pi$ between the $B$-meson and the particle with momentum $k_2$, and corresponds to $\theta_K$ here. Thus the definitions for $F_{0,t}^{(\ell=1)}$ in~\Ref{Cheng:2017smj} agree with the ones used here. One can derive the results in~\Ref{Cheng:2017smj} from the results in this article by taking $m_K\to m_\pi$, and including an isospin factor as discussed below.

\section{LCSRs with $B$-meson Distribution Amplitudes}
\label{sec:LCSRs}

\subsection{General framework}
\label{sec:general}

We follow the method used in \Refs{Khodjamirian:2006st,Cheng:2017smj}, which we review here very briefly.
We consider correlation functions of the type:
\eq{
\P_{ab}(k,q) = i\int d^4x\, e^{i k \cdot x} \langle 0 | {\rm T} \{ j_a(x), j_b(0) \} | \bar{B}^0(q+k) \rangle \ ,
}
where $j_{a}$ is the current interpolating the final state and $j_b$ is the quark transition current.
Within a kinematic region where both invariant variables $k^2$ and $q^2$ are far below the hadronic thresholds
in the channels of interpolating and transition currents, respectively, these correlation functions can be calculated by means of a light-cone OPE in terms of $B$-meson LCDAs:
$$\P_{ab}(k, q) = \P_{ab}^\text{OPE}(k, q).$$ A dispersion relation in the variable $k^2$ relates this object to the spectral density of the correlation function:
\eq{
\P_{ab}^\text{OPE}(k^2, q^2) = \frac{1}{\pi} \int_{s_{\rm th}}^\infty ds \,\frac{\im \P_{ab}(s,q^2)}{s-k^2}\ .
}
The imaginary part of the correlation function can be obtained from unitarity by inserting a full set of states between the two currents in the T-product:

\eq{
2\, \im \P_{ab}(k,q)= \sum_{h} \int d\tau_{h}\ 
\langle 0 | j_a \,|h(k) \rangle
\langle h(k) | j_b | \bar{B}^0(q+k)\rangle.
\label{eq:unitarity}
}
The interpolating current $j_a$ is chosen such that the desired quantum numbers of the states~$h$ are projected out. In our case this will be
a vector current with the flavour of a $K\pi$ state. 
The standard assumption when deriving $B\to K^*$ form factors is that the $K^*$ state is stable and thus contributes
as a one-particle state in~\Eq{eq:unitarity}, and as a simple pole at $k^2=m_{K^*}^2$ in the correlation function
$\P_{ab}(k^2,q^2)$. 
The hadronic representation with $h=K\pi,\dots$ goes beyond the single-pole approximation.
This is the generalization proposed in~\Ref{Cheng:2017smj}, and  the representation that we will use in this article. 

One then performs a Borel transformation in the variable $k^2$ and uses the quark-hadron duality approximation to equate the hadronic integral above the effective threshold $s_0$ with its OPE expression. The sum rule becomes:
\eq{
\frac{1}{\pi} \int_{s_{\rm th}}^{s_0} ds \,e^{-s/M^2}\, \im \P_{ab}(s,q^2)
= \P_{ab}^\text{OPE}(q^2,\sigma_0,M^2)\,,
\label{eq:GenericLCSR}
}
where $\P_{ab}^\text{OPE}(q^2,\sigma_0,M^2)$ is the Borel-transformed OPE expression after subtracting the contribution from the dispersion integral above the effective threshold, with $\sigma_0$ depending on $s_0$. The choice of the effective threshold $s_0$ is made such that the main contribution to $\im \P_{ab}(s,q^2)$ in the integral
comes from $h=K\pi$, and other higher states (e.g. $h = K\pi\pi\pi$) are suppressed.
The Borel transformation makes the sum rule less sensitive to the duality approximation, and improves the convergence of the OPE.
One typically checks that the integral from $s_0$ to
infinity in the OPE side is a small fraction of the total integral, thus minimizing the dependence on the duality ansatz. The specific values of the effective threshold $s_0$ in the numerical analysis will be taken by fitting the corresponding QCD (SVZ) sum rule for the vacuum-to-vacuum correlation function 
of two interpolating currents (see \Sec{sec:2pt-s0}).

\subsection{Sum Rules for $P$-wave $B\to K\pi$ Form Factors}

In order to project out the specific $P$-wave $K^-\pi^+$ state in the sum rules,
we need to take as an interpolating current  the vector 
current with strangeness, $j_a= \bar{d}\gamma_\mu s$.
The choice for the transition current $j_b$ depends on the type of the form factor.
We consider first the form factors of the V$-$A current $\bar{s}\gamma^\nu (1-\gamma_5) b$, and start from the correlation function:
\eqa{
\label{eq:corrV}
\P_{\mu\nu}(k,q)&=&i\int d^4x\, e^{i k \cdot x} \langle 0 | {\rm T} \{\bar{d}(x) \gamma_\mu s(x),
\bar{s}(0)\gamma_\nu (1-\gamma_5) b(0) \} | \bar{B}^0(q+k) \rangle \\[1mm]
&=&\varepsilon_{\mu\nu\rho\sigma}q^{\rho}k^{\sigma}\P_\perp(k^2,q^2)
+i g_{\mu\nu} \P_\|(k^2,q^2)
+iq_\mu k_\nu \P_{-}(k^2,q^2)\nonumber\\[1mm]
&&
+ik_\mu k_\nu \P_{(kk)}(k^2,q^2)+iq_\mu q_\nu \P_{(qq)}(k^2,q^2) +ik_\mu q_\nu \P_{(kq)}(k^2,q^2)\,, \nonumber}
where the notation used for the invariant amplitudes $\P_{\perp,\|,-}$ indicates the form factors that will be extracted from them.
The hadronic spectral function of the correlation function is obtained from unitarity, i.e. inserting the complete set of states with the quantum numbers of the interpolating current between the two currents in \Eq{eq:corrV}:
\eq{
2\, \im \P_{\mu\nu}(k,q)= \sum_{ \{K\pi\} } \int d\tau_{K\pi}\ 
\langle 0 |\bar{d} \gamma_\mu s \,| K\pi \rangle
\langle K\pi | \bar{s} \gamma_\nu (1-\gamma_5) b | \bar{B}^0(q+k)\rangle
+ \cdots\,.
\label{eq:unit}
}
Here the lightest intermediate states  $  \{K\pi\} = \{K^-\pi^+, \overline K^0\pi^0\}$
are included explicitly and the ellipsis denotes the contributions from other intermediate states with higher thresholds, such as
$K\pi\pi,K \pi\pi\pi,\ K\pi\bar K K$, etc. 

As mentioned above, the interpolating $\bar d \gamma_\mu s$ current in~\Eq{eq:unit} projects out the $I=1/2$ components of the $K\pi$ states
-- given by \Eq{eq:isospin} -- and this allows us to relate the two contributions:
\eq{
 \sum_{ \{K\pi\} } \langle 0 |\bar{d} \gamma_\mu s \,| K\pi \rangle\langle K\pi |
 = \langle 0 |\bar{d} \gamma_\mu s \,| K\pi \rangle_{1/2} \langle K\pi |_{1/2}
 = \frac32 \,\langle 0 |\bar{d} \gamma_\mu s \,| K^-\pi^+ \rangle\langle K^-\pi^+ |\ ,
 }
such that
\eq{
2\, \im \P_{\mu\nu}(k,q)= \frac32 \int d\tau_{K\pi} \  
\langle 0 |\bar{d} \gamma_\mu s \,| K^-\pi^+ \rangle
\langle K^-\pi^+ | \bar{s} \gamma_\nu (1-\gamma_5) b | \bar{B}^0(q+k)\rangle
+ \cdots\,.
\label{eq:unit2}
}
This factor $3/2$ must be taken into account when comparing the sum rules in this article with the ones derived in~\Ref{Cheng:2017smj} for $\bar B\to \pi^+\pi^0$ form factors.

We consider now the invariant functions $\P_\perp$, $\P_\|$ and $\P_{-}$ in~\Eq{eq:corrV}.
Using the definition for the $K\pi$ form factors in~\Eq{eq:piFF} and the ones for the $B\to K\pi$ form factors in~Eqs.(\ref{FV}) and~(\ref{FA}),
and following closely the derivation in \Ref{Cheng:2017smj} (recalled in \Sec{sec:general}), we find the sum rules:
\eq{
\label{eq:LCSRV}
\int_{s_{\rm th}}^{s_0} ds~e^{-s/M^2}
\omega_{i}(s,q^2) \,f^\star_+(s) \,F_{i}^{(\ell=1)}(s,q^2)
= \P_i^\text{OPE}(q^2,\sigma_0,M^2)\,, \\[2mm]
}
for $i=\{\perp,\|,-\}$, where we have defined the combination of form factors:
\eq{
F_{-}^{(\ell =1)}(s,q^2) \equiv (m_B^2-q^2-s)  F^{(\ell=1)}_{\parallel}(s,q^2)
+ \frac{2 s^{3/2} \sqrt{q^2}}{\sqrt{\lambda_{K\pi}(s)}} F^{(\ell=1)}_{0}(s,q^2)
\ ,
\label{eq:FminFparF0}
}
and the functions:
\eq{
2\, \omega_{\|}(s,q^2)
= \sqrt{\lambda(s)}\, \omega_{\perp}(s,q^2)
= -\lambda(s)\, \omega_{-}(s,q^2)
= -\frac{\sqrt{3}\,\lambda^{3/2}_{K\pi}(s)}{16\pi^2 s^{5/2}}\ .
\label{eq:omegai}
}
These sum rules contain implicitly the factor 3/2 that accounts for the two intermediate $K\pi$ states in the isospin limit.
The functions $\P_{i}^\text{OPE}(q^2,\sigma_0,M^2)$ are given explicitly in Appendix~\ref{OPEexpressions} and can be compared to~\Ref{Khodjamirian:2006st}.

In the limit $m_K\to m_\pi$ the expressions in \Ref{Cheng:2017smj} are  recovered~\footnote{In this limit, $\lambda_{K\pi}(s) \to  s^2 [\beta_\pi(s)]^2 = s(s - 4m_\pi^2)$, and $f_+^\star(s) F^{(\ell=1)}_i(s,q^2) \to -\frac{2\sqrt2}3 F_\pi^\star(s) F^{(\ell=1)}_i(s,q^2)$, which takes into account the various isospin factors.\label{foot2}}. The relative sign difference in the $F_0$ term with respect to \Ref{Cheng:2017smj} has been discussed at the end
of \Sec{sec:FormFactors}. We also note that due to the chosen Lorentz structures, the scalar $K\pi$ form factor $f_0(s)$ does not contribute.

\bigskip

For the timelike-helicity form factor we start from the correlation function
with a pseudoscalar transition current $j_b=i(m_b+m_s)\,\bar{s}\gamma_5 b$:
\eqa{
\P_{\mu}(k,q)&=& i\int d^4x\, e^{i k \cdot x} \langle 0 | {\rm T} \{\bar{d}(x) \gamma_\mu s(x),
i(m_b+m_s)\bar{s}(0)\gamma_5 b(0) \} | \bar{B}^0(q+k) \rangle\nonumber\\[2mm]
&=&  q_\mu \P_t(k^2,q^2) + k_\mu \P_{(k)}(k^2,q^2)\ .
\label{eq:corrt}
}
Focusing on the invariant amplitude $\P_t$, and following the same procedure as before we find the sum rule
\eq{
\int_{s_{\rm th}}^{s_0} ds~e^{-s/M^2}~
\omega_t(s,q^2)
\,f^\star_+(s)\,F_{t}^{(\ell=1)}(s,q^2)
=\P_t^{\rm OPE}(q^2,\sigma_0,M^2) \,,
\label{eq:LCSRFt}
}
where
\eq{
\omega_t(s,q^2) =  -\frac{\sqrt{3q^2}\ \lambda_{K\pi}(s)}{16 \pi^2\, s\, \sqrt{\lambda(s)}} \ .
\label{eq:omegat}
}
The function $\P_t^{\rm OPE}$, also given in Appendix~\ref{OPEexpressions},
was not derived in ~\Ref{Khodjamirian:2006st}. 
It was, however, considered in~\Ref{Cheng:2017smj} for the $B\to\pi\pi$ case.
As before, the result in \Ref{Cheng:2017smj} is recovered when $\lambda_{K\pi}(s) \to s^2 [\beta_\pi(s)]^2$ and the proper isospin factors are included.
Also in this case the scalar $K\pi$ form factor does not contribute.

\bigskip

For the tensor form factors we start from the correlation function
with a tensor transition current $j_b=\bar{s}\sigma_{\nu\rho} q^\rho (1+\gamma_5) b$:
\eqa{
\P^T_{\mu\nu}(k,q)&=&
i\int d^4x\, e^{i k \cdot x} \langle 0 | {\rm T} \{\bar{d}(x) \gamma_\mu s(x),
\bar{s}(0)\sigma_{\nu\rho} q^\rho (1+\gamma_5) b(0) \} | \bar{B}^0(q+k) \rangle
\label{eq:corrT}
\\[1mm]
&=& i\, \varepsilon_{\mu\nu\rho\sigma}q^{\rho}k^{\sigma}\P^T_\perp(k^2,q^2)
+ \big[ q_\mu q_\nu - q^2 g_{\mu\nu} \big] \P^T_{(qq)}(k^2,q^2)
\nonumber\\[1mm]
&&
+ \big[ q^2 k_\mu k_\nu - (k\cdot q) k_\mu q_\nu \big] \P^T_{(kk)}(k^2,q^2)
+ \big[ q_\mu k_\nu - (k\cdot q) g_{\mu\nu} \big] \P^T_{(qk)}(k^2,q^2)
\,. \nonumber
}
Proceeding analogously to the vector form factors, but focusing on the three invariant functions 
$\P^T_\perp$,
$\P^T_\| \equiv q^2 \P^T_{(qq)} + (k\cdot q) \P^T_{(qk)}$
and $\P^T_{-} \equiv 2 \P^T_{(qq)} - \P^T_{(qk)}$
we find
\eq{
\label{eq:LCSRT}
\int_{s_{\rm th}}^{s_0} ds~e^{-s/M^2}
\omega_i(s,q^2) \,f^\star_+(s) \,F_{i}^{T(\ell=1)}(s,q^2)
= \P_i^{T,\text{OPE}}(q^2,\sigma_0,M^2)\,,
}
where the functions $\omega_i(s,q^2)$ are given in~\Eq{eq:omegai}, and we have defined
\eq{
F^{T(\ell=1)}_{-}(s,q^2) \equiv
\frac{2s^{3/2} (m_B^2 - s)}{\sqrt{q^2}\, \sqrt{\lambda_{K\pi}(s)}}  F^{T(\ell=1)}_{0}(s,q^2)
+ (m_B^2-q^2+3s)   F^{T(\ell=1)}_{\parallel}(s,q^2)\ .
\label{eq:FminFparF0T}
}
As before, the scalar $K\pi$ form factor does not contribute.
While the tensor $B\to\pi\pi$ form factors were not considered in \Ref{Cheng:2017smj}, these can be derived by
performing the replacement indicated in footnote~\ref{foot2}
in the results given here.
The functions $\P_{\perp,\|,-}^{T,\text{OPE}}$ can be found in Appendix~\ref{OPEexpressions}
($\P_{\|,-}^{T,\text{OPE}}$ are new with respect to \Ref{Khodjamirian:2006st}).

\bigskip

The sum rules in~Eqs.~(\ref{eq:LCSRV}),~(\ref{eq:LCSRFt}) and~(\ref{eq:LCSRT}) can be written compactly as:

\eq{
\int_{s_{\rm th}}^{s_0} ds \ e^{-s/M^2}
\omega_i(s,q^2) \,f^\star_+(s)\, F_i^{(T)(\ell=1)}(s,q^2)
= \P_i^{(T),\text{OPE}}(q^2,\sigma_0,M^2)
\label{eq:AllLCSRs}
}
for\,\footnote{For the tensor form factors $F_i^T$ we only have $i = \{\perp,\|,-\}$, since there is no timelike-helicity form factor for the tensor current $F_t^T$. 
For simplicity, we will not remind this difference between the vector-axial and tensor cases whenever we quote the form factors $F_i^{(T)}$ generically.}
$i = \{\perp,\|,-,t\}$, with the functions $\omega_i(s,q^2)$ given in~Eqs.~(\ref{eq:omegai}) and~(\ref{eq:omegat}).

The functions $\P^{(T),\text{OPE}}_i(q^2,\sigma_0,M^2)$ are all collected in \App{OPEexpressions}.
The set of sum rules given by  \Eq{eq:AllLCSRs} together with the OPE functions collected in \App{OPEexpressions} are the main results of this article.
They generalize and update the results in \Refs{Khodjamirian:2006st,Cheng:2017smj} in four directions:
1) going beyond the single-pole (narrow-width) approximation considered in \Ref{Khodjamirian:2006st},
2) allowing the two mesons in the final state to have different masses (thus generalizing the $B\to\pi\pi$ case of \Ref{Cheng:2017smj}),
3) including the tensor form factors, and
4) using the new parametrization of the $B$-meson LCDAs from~\Ref{Braun:2017liq}.
In the rest of the article we will discuss how to exploit these sum rules,
how to interpret the narrow-width limit and how to go beyond it, and we will present a few applications of these results.


\section{Parametrization of $B\to K\pi$ Form Factors}
\label{sec:LCSRs@Work}

The sum rules derived in the previous section and summarized in compact form in \Eq{eq:AllLCSRs} relate the OPE functions $\P_i^{(T), \rm OPE}$ depending on the $B$-meson LCDAs to the $P$-wave form factors $F_i^{(T)(\ell=1)}(s,q^2)$ weighted by the timelike $K\pi$ form factor $f_+(s)$ and integrated over the duality interval $s_{\rm th}<s<s_0$.
As such, these sum rules do not provide closed-form expressions for the $B\to K\pi$ form factors.
The dominance of the $K^*$ resonance in $f_+(s)$ and the choice of the Borel exponent makes the integral over the hadronic spectral density in the sum rules dominated by the region $s\sim m_{K^*}^2$. 
As will be shown below, in the narrow-width limit the product $F_i^{(T)(\ell=1)}(s,q^2) f_+(s) \sim \delta(s- m_{K^*}^2)$ removes the integral on the l.h.s. of the sum rules and reduces them to a set of equations for the $B\to K^*$ form factors, which coincide with the $B\to K^*$ sum rules in~\Ref{Khodjamirian:2006st}.

The fact that the sum rules  (\ref{eq:AllLCSRs}) only provide weighted integrals of the form factors means that no ``local" information on the $s$ dependence can be obtained.
One needs to start from an ansatz or model for the form factors, using the sum rules to constrain its parameters~\cite{Cheng:2017smj}.
In this section we consider one particular set of models for the form factors, and show how their free parameters can be constrained using the sum rules.

\subsection{Resonance models for $B\to K\pi$ form factors}

For all $P$-wave form factors we will consider a resonance model similar to the one considered in~\Ref{Cheng:2017smj}.
The starting point is to assume that the $P$-wave $K\pi$ state couples to its interpolating current $\bar s \Gamma d$ resonantly,
through a set of Breit-Wigner-type vector resonances.

Consider first the $K\pi$ vector form factor. The resonance ansatz then implies that
\eq{
\langle K(k_1) \pi(k_2) | \bar s \gamma^\mu d | 0 \rangle = 
\sum_{R,\eta}  BW_R(k^2) \langle K(k_1) \pi(k_2) | R(k,\eta) \rangle \langle R(k,\eta) | \bar s \gamma^\mu d | 0 \rangle
\label{eq:ResSumKpi}
}
where the sum runs over $R = \{K^\star(892),K^\star(1410)\}$
and the vector-meson polarization states~$\eta$.
The third factor in the right-hand side is related to the $R$ decay constants $f_R$:
\eq{
\langle R(k,\eta) | \bar s \gamma^\mu d | 0 \rangle = \epsilon_\eta^{*\mu} \,m_R\,f_R
\label{eq:decayconstant}
}
which are defined to be real and positive by definition of the phases of the states $\langle R(k,\eta) |$.
Here $\epsilon_\eta^{*\mu}$ is the polarization vector of $R$ with polarization $\eta$.
The second factor in (\ref{eq:ResSumKpi}) is related to the strong coupling of the resonances to the $K^- \pi^+$  state:
\eq{
g_{RK\pi} \,e^{i\varphi_R} \ \bar k \cdot \epsilon_\eta = \ \langle K^- \pi^+ | R(k,\eta) \rangle = -\sqrt2\ \langle \bar K^0 \pi^0 | R(k,\eta) \rangle\ ,
\label{eq:gRKpi}
}
where  we include a phase $\varphi_R$ related to the normalization of the hadronic state.
This phase will be merged later on with the relative phases between the separate resonance contributions to the form factors.
The first factor in (\ref{eq:ResSumKpi}) is a Breit-Wigner function with an energy-dependent width, which is assumed in our simple model to describe well the line shapes of the $K^*$ resonances:
\eq{
BW_R(s) = \frac1{m_R^2 - s -i\sqrt{s}\, \Gamma_R(s)}\ ,
}
with
\eq{
\Gamma_{R}(s)
=\Gamma^{\rm tot}_{R} \left[\frac{\lambda_{K\pi}(s)}{\lambda_{K\pi}(m_{R}^2)}
\right]^{3/2}\frac{m_{R}^5}{s^{5/2}} \,\theta\big(s- s_{\rm th} \big)\,.
\label{eq:Gamma(s)}
}
The strong coupling $g_{RK\pi}$ is determined by the total width of the resonance $R$ by
\eq{
\Gamma^{\rm tot}_{R}  = \frac{g_{RK\pi}^2}{48\pi} \frac{\lambda_{K\pi}^{3/2}(m_{R}^2)}{m_{R}^5} \, \frac{1}{\B(R\to K^- \pi^+)}\ ,
\label{eq:Gammatot}
}
where $\B(K^*(892) \to K^- \pi^+)= 2/3$ is the isospin-limit prediction (assuming a $100\%$ branching ratio to $K\pi$~\cite{PDG}), and $\B(K^*(1410) \to K^- \pi^+)\simeq 2/3 \times 0.06 = 0.04$~\cite{PDG}. In our computations, it proves useful to write the strong coupling in terms of the energy-dependent width:
\eq{
g_{RK\pi}^2 = \frac{48\pi\,s^{5/2}\,\Gamma_R(s)}{\lambda_{K\pi}^{3/2}(s)} \B(R\to K^- \pi^+)\ ,
\label{eq:gKpi}
}
which follows from \Eqs{eq:Gamma(s)} and~(\ref{eq:Gammatot}), and where the $s$ dependence in the r.h.s. cancels out. 

Plugging Eqs.~(\ref{eq:gRKpi})~and~(\ref{eq:decayconstant}) into \Eq{eq:ResSumKpi}, and summing over the three polarizations of the vector resonance:
$\sum_\eta \epsilon^{*\mu}_\eta \epsilon_\eta^\nu = -g^{\mu\nu} + k^\mu k^\nu/k^2$ , we obtain
\eq{
\langle K^-(k_1) \pi^+(k_2) | \bar s \gamma^\mu d | 0 \rangle = 
- \bar k^\mu \sum_R \ g_{RK\pi} \, m_R\,f_R\, BW_R(k^2)\,e^{i\phi_R(k^2)}
}
and comparing to~\Eq{eq:piFF} we get
\eq{
f_+(s) =  - \sum_R \frac{m_{R} \,f_R\, g_{RK\pi} \ e^{i\phi_R(s)}}{m_{R}^2 - s -i \sqrt{s}\, \Gamma_{R}(s)}\ .
\label{eq:f+model}
}
The relative $s$-dependent phases between the $R$-resonance terms emerge due to the admixture with other resonances via $R\to K\pi\to R'$ strong transitions. This effect complements the diagonal $R\to K\pi \to R$ transitions which, after resummation, generate the energy-dependent width of the resonance $R$~\footnote{A more detailed analysis of the mixing 
between resonances and meson loops demands a coupled channel approach, which is beyond our scope here.}.

The  model in~\Eq{eq:f+model} for the vector $K\pi$ form factor is equivalent to the one used by Belle in~\Ref{Epifanov:2007rf},
which fits well the data for the $\tau\to K\pi \nu$ decay for the relevant energy range. This model is thus
phenomenologically justified (see \Sec{sec:taudata}).

In the case of the $B\to K\pi$ form factors and along the same lines we have:
\eq{
\langle K(k_1) \pi(k_2) | \bar s \Gamma b |B(q+k) \rangle = 
\sum_{R,\eta}  BW_R(k^2) \langle K(k_1) \pi(k_2) | R(k,\eta) \rangle \langle R(k,\eta) | \bar s \Gamma b |B(q+k) \rangle\ ,
\label{eq:ResSum}
}
for a generic Dirac structure $\Gamma$.
The third factor in the right-hand side is related to $B\to R$ form factors $\F_{R,i}^{(T)}(q^2)$ (as defined in \Tab{tab:DictionaryFFs} and \App{app:BtoVLCSRs}).
Plugging \Eq{eq:gRKpi} and the definition of the various form factors from \App{app:BtoVLCSRs} into \Eq{eq:ResSum}, and summing over the three polarizations we obtain the following compact expression for all $P$-wave $B\to K\pi$ form factors:
\eq{
F_i^{(T),(\ell=1)}(s,q^2) =
\sum_R  \frac{Y_{R,i}^{(T)}(s,q^2)\, g_{RK\pi} \, \F_{R,i}^{(T)}(q^2) \,e^{i\phi_R(s)}}{m_R^2 - s - i\sqrt{s}\, \Gamma_R(s)}
\label{eq:FFmodels}
}
with $i=\{ \perp,\|,-,t \}$ and
\begin{align}
Y_{R,\perp}^{T} &= (m_B+m_R) Y_{R,\perp} = \frac{m_R\sqrt{\lambda}}{\sqrt3}  \ ,   &   Y_{R,\|}^{T} &= (m_B-m_R) Y_{R,\|} = \frac{(m_B^2-m_R^2)\,m_R}{\sqrt3}  \ , \nonumber \\
Y_{R,-}^{T} &= (m_B+m_R) Y_{R,-} = \frac{m_R\lambda}{\sqrt3}  \ ,   & Y_{R,t} &= -\frac{\sqrt{\lambda\, \lambda_{K\pi}}}{m_R\sqrt{3q^2}}\ ,
\label{eq:Ys}
\end{align}
where $Y_{R,i}^{(T)}(s,q^2)$ depend on $s$ and $q^2$ also implicitly through the functions $\lambda\equiv \lambda(m_B,q^2,s)$
and $\lambda_{K\pi}\equiv \lambda(s,m_K^2,m_\pi^2)$. In \Eq{eq:FFmodels} we are assuming that the relative phase $\phi_R(s)$ is a genuine characteristic of the resonance $R$ -- such as the width -- and thus independent of the process where $R$ is produced (interpolating current or $B$-meson decay).

\begin{table}
\centering
\setlength{\tabcolsep}{10pt}
\begin{tabular}{@{}lccccccc@{}}
\toprule[0.7mm]
Traditional Notation~\cite{Khodjamirian:2006st}\  &
$V^{BR}$ & $A_1^{BR}$ & $A_2^{BR}$ & $A_0^{BR}$ & $T_1^{BR}$ & $T_2^{BR}$ & $T_3^{BR}$\\
\midrule[0.2mm]
This work  &
$\F_{R,\perp}$ & $\F_{R,\|}$ & $\F_{R,-}$ & $\F_{R,t}$ & $\F^T_{R,\perp}$ & $\F^T_{R,\|}$ & $\F^T_{R,-}$\\
\bottomrule[0.7mm]
\end{tabular}
\caption{\it Notation for the various $B\to R$ form factors used in this article (for $R$ a vector resonance), as compared to the ``traditional" notation (see e.g.~\Ref{Khodjamirian:2006st}). The notation used in this article has a closer correspondence to the notation for the ``parent" $B\to P_1P_2$ form factors.}
\label{tab:DictionaryFFs}
\end{table}

It is now useful to rewrite the LCSRs of~\Eq{eq:AllLCSRs} in the framework of this resonance model.
Putting~\Eq{eq:FFmodels} into~(\ref{eq:AllLCSRs}), we find:
\eq{
\sum_R \F_{R,i}^{(T)}(q^2)\,d_{R,i}^{(T)}\,I_R(s_0,M^2) = \P_i^{(T),\rm OPE}(q^2,\sigma_0,M^2)\ ,
\label{eq:LCSRsmodel}
}
with
\eq{
I_R(s_0,M^2) = \frac{m_R}{16\, \pi^2} \int_{s_{\rm th}}^{s_0} ds \ e^{-s/M^2}\  \frac{g_{RK\pi}\,\lambda_{K\pi}^{3/2}(s) \, |f_+(s)|}{s^{5/2} \sqrt{(m_R^2-s)^2 + s\,\Gamma_R^2(s)}}\ ,
\label{eq:IR}
}
and
\eqa{
&&d_{R,\perp}= -d_{R,-} = (m_B+m_R)^{-1}\ ,\quad
d_{R,\|} =  \frac{(m_B+m_R)}2\ ,\quad
d_{R,t} = -m_R\ , \nonumber \\
&& d_{R,\perp}^T= - d_{R,-}^T = 1\ ,\quad
d_{R,\|}^T = \frac{(m_B^2 - m_R^2)}2\ .
\label{eq:di}
}
We refrain from replacing $f_+(s)$ by its model expression~(\ref{eq:f+model}), since one may choose to use another model or a direct experimental determination of $f_+(s)$ inside the integral in~\Eq{eq:IR}. In deriving \Eq{eq:LCSRsmodel}
we have also adopted the ansatz that the phase cancellation between $f_+$ and the form factors $\F_{R,i}^{(T)}$ that follows from unitarity happens at the level of the individual resonances~\cite{Cheng:2017smj}.
This is enforced by imposing that the phases $\phi_R(s)$ in
\Eq{eq:FFmodels} are such that
\eq{
\tan \big[ \delta_{K\pi}(s) - \phi_R(s) \big] = \frac{\sqrt{s}\, \Gamma_R(s)}{m_R^2 - s}
\ ,
\label{eq:phasecondition}
}
which follows from the more general unitarity condition $\im[F_i^{(\ell=1)}(s,q^2) f_+^*(s)] = 0$~\cite{Cheng:2017smj}. Here we have defined $\delta_{K\pi}(s)$ as the phase of the $K\pi$ form factor:
\eq{
f_+(s) = |f_+(s)| e^{i\delta_{K\pi}(s)}\ .
}
Note that this assumption also implies that the phases $\phi_R(s)$ are
$q^2$-independent. One can see that the model for $f_+(s)$ in~\Eq{eq:f+model} satisfies the condition~(\ref{eq:phasecondition}) trivially.
This is also true for the models used by Belle~\cite{Epifanov:2007rf} (see \Sec{sec:taudata}), which are equivalent to that in~\Eq{eq:f+model}.

\subsection{Narrow-width limit}
\label{sec:NWL}

The narrow-width limit is model-independent as long as the model used leads to a resonance pole at the right position (mass).
Inserting the model in \Eq{eq:f+model} for $f_+(s)$ into the integrand in~\Eq{eq:IR}
and considering for the moment the case of a single resonance, we have
\eq{ \label{eq:defIR}
I_R(s_0,M^2) =
3\,m_R f_R\,\B(R\to K^+\pi^-) \int_{s_{\rm th}}^{s_0} ds \ e^{-s/M^2}\,\frac{m_R}{\sqrt{s}}
\Bigg[\frac1\pi \frac{\sqrt{s}\,\Gamma_R(s) }{(m_R^2-s)^2 + s\,\Gamma_R^2(s)}\Bigg]\ .
}
Here we have used~\Eq{eq:gKpi} in order to write $g_{RK\pi}$ in terms of $\Gamma(s)$.
The expression inside square brackets becomes a delta function $\delta(s-m_R^2)$ in the narrow-width limit, $\Gamma_R^{\rm tot}\to 0$. Thus, the narrow-width limit is simply recovered by the substitution
\eq{
I_R(s_0,M^2) \xrightarrow{\Gamma_R^{\rm tot}\to 0} 3\,m_R f_R\,\B(R\to K^+\pi^-)\,e^{-m_R^2/M^2}\ .
}
Implementing this limit in~\Eq{eq:LCSRsmodel} leads to the LCSRs in the narrow-width limit:
\eq{
3\,m_R f_R\,d_{R,i}^{(T)}\, \F_{R,i}^{(T)}(q^2)\,e^{-m_R^2/M^2}\,\B(R\to K^+\pi^-) = \P_i^{(T),\rm OPE}(q^2,\sigma_0,M^2)\ .
}
This agrees with the LCSRs for $B\to V$ form factors collected in \App{app:BtoVLCSRs}. 
For example, in the case of $\F_{R,\perp}\equiv V^{BR}$, we have $d_\perp=(m_B+m_R)^{-1}$ and 
\eq{
e^{-m_R^2/M^2}\, \frac{3f_{R} m_R \F_{R,\perp}(q^2)}{ (m_B+m_R)}
\ \B(R\to K^+\pi^-)
 = \P_\perp^{\rm OPE}(q^2,\sigma_0,M^2)\ ,
}
which agrees exactly with \Eq{eq:LCSRBVV} when $\B(R\to K^+\pi^-)=2/3$.
In the narrow-width limit all sum rules for $B\to V$ form factors in \App{app:BtoVLCSRs} are reproduced in the same way.

The same exercise goes through if we keep the various resonances and send all widths to zero (all of them scaling with the same factor $\Gamma\to 0$).
In this case the cross-terms in the sum have a vanishing contribution in the narrow-width limit:
\eq{\label{eq:NWL3resonances}
\frac{g_{RK\pi} \, g_{R'K\pi}}{[m_{R}^2 - s +i \sqrt{s}\, \Gamma_{R}(s)][m_{R'}^2 - s -i \sqrt{s}\, \Gamma_{R'}(s)]} + c.c.
\ \xrightarrow{\Gamma_{R,R'}^{\rm tot}\to 0} \ 0\ ,
}
so that the limit produces one simple pole for each resonance, {\it e.g.}
\eq{\label{eq:NWL3resonancessumrules}
\sum_R 3\,m_R f_R\,d_{R,i}^{(T)}\, \F_{R,i}^{(T)}(q^2)\,e^{-m_R^2/M^2}\,\B(R\to K^+\pi^-) = \P_i^{(T),\rm OPE}(q^2,\sigma_0,M^2)\ .
}
This is exactly what is expected from the LCSRs with several stable vector resonances.

The $\op(\Gamma)$ corrections to \Eq{eq:NWL3resonancessumrules} can also be calculated. The arguments given in \Eq{eq:NWL3resonances} indicate that the overlap of two different resonances from the $K\pi$ form factor and the $B\to K\pi$ form factor yield a contribution of order $\op(\Gamma^2)$, so we can focus to the contributions coming from the same resonance in both cases, i.e. the $\op(\Gamma)$ correction to the integral defined in \Eq{eq:defIR}.
Using the formalism reviewed in \App{app:beyondNWL} and denoting $\tilde{I}_R(s_0,M^2)$ 
as the value of $I_R$ in the limit $\Gamma_R^{\rm tot}\to 0$, we have
\eq{
\frac{I_R(s_0,M^2)-\tilde{I}_R(s_0,M^2)}{\tilde{I}_R(s_0,M^2)}
= \frac{\Gamma_R^{\rm tot}}{m_R}\,\Delta_R(s_0,M^2) + \op(\Gamma^2)\ ,
}
with
\eqa{
\Delta_R(s_0,M^2)
&=&\frac{1}{\pi} \Bigg[-\frac{m_R^2(s_0-s_{\rm th})}{(s_0-m_R^2)(m_R^2-s_{\rm th})}
+m_R^2\big[\phi'(m_R^2)+\rho'(m_R^2)\big] \log\frac{s_0-m_R^2}{m_R^2-s_{\rm th}}
\nonumber\\
&&\hspace{2cm}
+\tilde{F}(s_0,m_R)-\tilde{F}(s_{\rm th},m_R)\Bigg]\ ,
\\
\tilde{F}(s,m_R)&=&\int_{1}^{s/m_R^2}   \frac{d\tau}{(\tau-1)^2}\Big[\phi(m_R^2\tau)\rho(m_R^2\tau)-1-(\tau-1)m_R^2[\phi'(m_R^2)+\rho'(m_R^2)]\Big]\ ,\quad
}
and with 
\eqa{
\rho(m_R^2\tau)&=&\gamma(m_R^2\tau)\sqrt{\tau}\ , \qquad
\gamma(m_R^2\tau)=\left[\frac{\lambda_{K\pi}(m_R^2\tau)}{\lambda_{K\pi}(m_R^2)}\right]^{3/2}\frac{1}{\tau^{5/2}}\theta(m_R^2\tau-s_{\rm th})\ ,\\[2mm]
\rho'(m_R^2)&=&\frac{-2(m_K^2-m_\pi^2)^2+m_R^2(m_R^2+m_\pi^2+m_K^2)}
{m_R^2\lambda_{K\pi}(m_R^2)}\ ,\\
\phi(m_R^2\tau )&=&\frac{1}{\sqrt{\tau}}e^{-\frac{m_R^2}{M^2}(\tau-1)}\ ,\qquad
\phi'(m_R^2)=-\frac{1}{M^2}-\frac{1}{2m_R^2}\ .
}
Up to order $\op(\Gamma)$, the l.h.s. of \Eq{eq:NWL3resonancessumrules} then reads
\eq{\label{eq:beyondNWL3resonancessumrules}
\sum_R 3\,m_R f_R\,d_{R,i}^{(T)}\, \F_{R,i}^{(T)}(q^2)\,e^{-m_R^2/M^2}\,
\bigg[1+\frac{\Gamma_R^{\rm tot}}{m_R}\, \Delta_R(s_0,M^2)+\cdots \bigg]\,\B(R\to K^+\pi^-)\ .
}
We see, in particular, that the generalization of the LCSRs formulae for the $B\to K^*$ form factors (in the one-resonance approximation) to include $\op(\Gamma_{K^*})$ corrections is rather simple: one must multiply all form factors in \Eqs{eq:LCSRBVV}-(\ref{eq:LCSRBVT3}) by $(1-\Delta_{K^*} \Gamma_{K^*}/m_{K^*})$.
The numerical impact of this correction will be discussed in~\Sec{sec:finitewidtheffects}.

\subsection{$z$-parametrization and $q^2$ dependence}

Following \Ref{Cheng:2017smj}, we parametrize the $q^2$-dependence of the $B\to R$ form factors $\F_{R,i}^{(T)}(q^2)$ entering~\Eq{eq:FFmodels} with the standard $z$-series expansion~\cite{Bourrely:2008za}, as adopted in \Ref{Khodjamirian:2010vf}. 
One defines the following function in the complex plane:
\eq{
z(q^2)=\frac{\sqrt{t_+-q^2}-\sqrt{t_+-t_0}}{\sqrt{t_+-q^2}+\sqrt{t_+-t_0}}\,,
\label{eq:zdef}
}
with $t_{\pm} \equiv (m_B \pm m_{K^*})^2$ and $t_0=t_+(1-\sqrt{1-t_-/t_+})$.
This function maps the segment $t_+\le q^2 < \infty$ onto the unit circle $|z|=1$, and the rest of the complex $q^2$ plane onto the interior of the disc $|z|<1$. The form factors $\F_{K^*,i}^{(T)}(q^2)$ are meromorphic in the first Riemann sheet and have a branch cut for $q^2\ge t_+$, so they admit a Pad\'e representation as a function of $z$ for $|z|<1$. In practice one only needs to subtract the subthreshold $B_s^{(*)}$ poles, and the remainder can be Taylor expanded. In the case of $K^*(1410)$ form factors, one would normally choose a different form for $z(q^2)$ where $m_{K^*}$ is replaced by $m_{K^*(1410)} $, since in this case the branch cut starts at the higher threshold $(m_B \pm m_{K^*(1410)})^2$.
However, keeping the same form of $z(q^2)$ for both sets of form factors will prove advantageous, and this choice does not invalidate any of the properties of the $z$-expansion for the $B\to K^*(1410)$ form factors.

Thus, for a generic form factor $\F_{R,i}^{(T)}(q^2)$,
with $i = \{\perp,\|,-,t\}$ and $R=\{ K^\star(892), K^\star(1410) \}$, 
we write\,\footnote{We use a slightly different notation compared to~\Ref{Khodjamirian:2010vf} as we do not normalize the slope parameters $b_{R,i}^{(T)}$ to the form factors at $q^2=0$. In this way the slope parameters are well behaved when any of the $\F_{R,i}^{(T)}(0)$ assume very small values in the scans, with the errors not blowing up. In addition, the coefficients $b_{R,i}$ coincide then with the parameters $\alpha_1^{(i)}$ in~Refs.\cite{Straub:2015ica,Gubernari:2018wyi}.}:
\eq{\F_{R,i}^{(T)}(q^2)= \frac{1}{1-q^2/m^2_{i}}
\Big\{ \F_{R,i}^{(T)}(0) + b_{R,i}^{(T)}\ \zeta(q^2) + \cdots \Big\}\,,
\label{eq:zpar}
}
where
\eq{
\zeta(q^2)=z(q^2) - z(0) + \frac12 [z(q^2)^2-z(0)^2] \,,
\label{eq:zeta}
}
and 
$m_{i}$ is the lowest heavy-light pole mass in the $q^2$ channel
with a spin-parity depending on the type of the form factor~\cite{PDG}:
\eqa{
m_{\perp} =m_{B_s^*} &=&5.415~\GeV\ \ (J^P=1^-)\ ,
\nonumber\\
m_{\|,-}= m_{B_{s1}}&=&5.829~\GeV\ \ (J^P=1^+)\ ,
\label{eq:mF}\\
m_{t} =m_{B_s}&=&5.366 \GeV\ \ (J^P=0^-)\ .\nonumber
}
The sum rules in~\Eq{eq:LCSRsmodel} can then be written in the form of the $z$-parametrization:
\eq{
\sum_R \frac{\F_{R,i}^{(T)}(0)+b_{R,i}^{(T)} \ \zeta(q^2)}{1-q^2/m^2_{i}}\,d_{R,i}^{(T)}\,I_R(s_0,M^2) = \P_i^{(T),\rm OPE}(q^2,\sigma_0,M^2)\ .
\label{eq:LCSRszpar}
}

It is now useful (and advantageous from the point of view of the $z$-expansion) to write the r.h.s. of the sum rules in the same $z$-expanded form:
\eq{
\P_i^{(T),\rm OPE}(q^2,\sigma_0,M^2) =
\frac{\kappa_{i}^{(T),\rm OPE}+\eta_{i}^{(T),\rm OPE} \ \zeta(q^2)}{1-q^2/m^2_{i}}\ ,
\label{eq:OPEz}
}
where $\kappa_{i}^{(T),\rm OPE}(\sigma_0,M^2)$ and $\eta_{i}^{(T),\rm OPE}(\sigma_0,M^2)$ both depend on the effective threshold and the Borel parameter. Note that this is merely a convenient reparametrization, and does not imply that the OPE functions depend on the resonance masses $m_i^2$ in any way.
Thus, one arrives to the final form of the sum rules in the resonance model with $z$-expansion:
\eqa{
\sum_R \F_{R,i}^{(T)}(0)\ d_{R,i}^{(T)}\ I_R(s_0,M^2)
&=& \kappa_i^{(T),\rm OPE}(\sigma_0,M^2)\ ,
\nonumber\\
\sum_R b_{R,i}^{(T)}\ d_{R,i}^{(T)}\ I_R(s_0,M^2)
&=& \eta_i^{(T),\rm OPE}(\sigma_0,M^2)\ ,
\label{eq:fitrel}
}
with $I_R$ and $d_{R,i}^{(T)}$ given in~Eqs.~(\ref{eq:IR}) and~(\ref{eq:di}).

This form of the sum rules can be used to fit the parameters $\F_{R,i}^{(T)}(0)$ and  $b_{R,i}^{(T)}$  which determine the $P$-wave $B\to K\pi$ form factors through \Eqs{eq:FFmodels} and~(\ref{eq:zpar}). 
These equations can also be compared to Eq.~(36) of~\Ref{Cheng:2017smj},
to which they reduce in the limit $m_K \to m_\pi$, up to the appropriate isospin factors.
In this form, it becomes clear that the sum rules can only be used to determine the combination of resonant contributions in~\Eq{eq:fitrel}. 
This illustrates explicitly the point made at the beginning of this section, i.e. the sum rules only give information on a weighted integral over the $K\pi$ invariant mass. Thus, in the context of the resonance model, some information on the relative importance of each resonance is necessary.

\subsection{Phenomenological formula for $B\to K\pi$ form factors}
\label{sec:recap}

We summarize here the main points in this section.
We use a two-resonance model for the $P$-wave $B\to K\pi$ form factors which matches the
model for the form factor $f_+(s)$ used by Belle and which fits well the $\tau$-decay data (see~\Sec{sec:taudata}). This model reproduces in the narrow-width limit the
$B\to K^*$ form factors derived in~\Ref{Khodjamirian:2006st}, calculated in \Ref{Khodjamirian:2010vf}, and rederived in \App{app:BtoVLCSRs}.
Putting together~Eqs.~(\ref{eq:FFmodels}),~(\ref{eq:zpar}) and~(\ref{eq:phasecondition}), the $B\to K\pi$ form factors take the following form:
\eq{
F_{i}^{(T),(\ell=1)}(s,q^2) 
= \sum_R  \frac{g_{RK\pi}\,Y_{R,i}^{(T)}(s,q^2)\, \big[\F_{R,i}^{(T)}(0)+b_{R,i}^{(T)} \ \zeta(q^2)\big] \,e^{i\delta_{K\pi}(s)}}
{(1-q^2/m^2_{{i}})\sqrt{(m_R^2-s)^2 + s\,\Gamma_R^2(s)}}
\label{eq:FFmodelsFinal}
}
for $i=\{\perp,\|,-,t\}$, where the sum runs over  $R = \{K^\star(892),K^\star(1410)\}$, the functions $Y_{R,i}^{(T)}(s,q^2)$ are given in~\Eq{eq:Ys},
the function $\zeta(q^2)$ is given in~\Eq{eq:zeta}, $\delta_{K\pi}(s)$ is the phase of the $K\pi$ form factor $f_+(s)$~(see~\Eq{eq:phasecondition}), and all the numerical parameters are collected in \Tab{tab:parameters}.
The form factors $F_0^{T,(\ell=1)}$ can in turn be obtained from $F_{\|,-}^{(T),(\ell=1)}$ by means of Eqs.~(\ref{eq:FminFparF0}) and~(\ref{eq:FminFparF0T}). Note that the phase $\delta_{K\pi}(s)$ could in principle be extracted directly from data.

The sum rules in the form of \Eq{eq:fitrel} together with the OPE expressions in~\App{OPEexpressions}
(or the numerical values collected in \Tab{tab:OPEpars}) and the Belle determination of $|f_+(s)|$
from~\Ref{Epifanov:2007rf} (see~\Sec{sec:taudata}) can be used to fit for the parameters $\F_{R,i}^{(T)}(0),\, b_{R,i}^{(T)}$. This is discussed further in \Sec{sec:numerical} below.

For phenomenological applications, one can take the form factor models in~\Eq{eq:FFmodelsFinal} together with the parameters in Table~\ref{tab:parameters}.
Alternatively, one may use the sum rules \Eq{eq:fitrel} to refit the parameters of the models under different assumptions, or for more specific analyses one can write new motivated models for the form factors and use the sum rules in~\Eq{eq:AllLCSRs} to fit their free parameters.

\section{Numerical Analysis}
\label{sec:numerical}


\subsection{Numerical input and strategy}

In this section we perform a numerical study of the $B\to K\pi$ form factors employing the LCSRs that we have derived. For this purpose, we start discussing the numerical input in the sum rules, both in the hadronic and OPE sides.
All input parameters are collected in~\Tab{tab:parameters}.
Meson and quark masses are taken from the PDG review~\cite{PDG}. 
For $m_b$ and $m_s$ we use the running $\overline{\rm MS}$ masses. We neglect uncertainties on the masses since they are negligible with respect to other sources of uncertainty. For the mass and width of the $K^*\equiv K^*(892)$ we take the central values and uncertainties obtained in the fit of~\Ref{Epifanov:2007rf}. Since this reference does not provide fit results for the $K^*(1410)$, we take its mass and width (with uncertainties) from the PDG review.

The $B$-meson decay constant is taken from the 2-point QCD sum rule determination in~\Ref{Gelhausen:2013wia}. This determination is consistent with lattice QCD computations,
e.g.  $f_B^{2+1+1}=190.0(1.3)\MeV$~\cite{FLAG,Dowdall:2013tga,Bussone:2016iua,Bazavov:2017lyh,Hughes:2017spc}. Probably the most important parameter regarding the OPE determination of the correlation functions in this article is $\lambda_B$, the inverse moment  of the 2-particle $B$-meson LCDA. Unfortunately this quantity is still known rather poorly. As in~\Ref{Cheng:2017smj}, we use the following interval:
\eq{
\lambda_B\equiv \lambda_B(1 \GeV)= 460\pm 110 ~\mbox{MeV}\,,
\label{eq:lambdaB}
}
derived from QCD sum rules~\cite{Braun:2003wx}. This determination satisfies the lower limit $\lambda_B> 238$~MeV (at $90\%$~C.L.) obtained by the Belle collaboration~\cite{Heller:2015vvm}, which combines the search for $B\to \gamma\ell \nu_\ell$ with the theory prediction for its branching ratio~\cite{Beneke:2011nf,Braun:2012kp}. It is worth recalling that this limit is a challenge for the lower values ($\lambda_B \sim 200-250 $~MeV) preferred by the QCD factorization analysis of $B\to \pi\pi$ decays  (see e.g. \Refs{Beneke:2015wfa,Beneke:2009ek}).
The estimate $\lambda_B= 358^{+38}_{-30}$~MeV~\cite{Wang:2015VGV} has been obtained by comparing the LCSRs with pion~\cite{Khodjamirian:2011ub} and $B$-meson LCDAs for the \mbox{$B\to\pi$} form factor and using a similar model for the $B$-meson DA as the Model I used here (see~\App{sec:modelsLCDAs}).
Since we are not including next-to-leading-order (NLO) corrections in the correlation functions,
we do not take into account the renormalization of $\lambda_B$.
Concerning the 3-particle $B$-meson LCDAs, the models that we will consider depend on one single parameter $R$ corresponding to a ratio of LCDA moments. As discussed in more detail at the end of~\App{sec:modelsLCDAs}, the value adopted for this parameter is $R=0.4^{+0.5}_{-0.3}$.

\begin{table}
\centering
\setlength{\tabcolsep}{10pt}
\begin{tabular}{@{}cc c @{\hspace{1cm}} cc c@{}}
\toprule[0.7mm]
Parameter &Value & Ref & Parameter & Value & Ref \\
\midrule[0.7mm]
$m_{\pi^\pm}$                &  $140\MeV$  		&  \cite{PDG}			& $m_{K^\pm}$ 			&  $494 \MeV$ 				&\cite{PDG}\\
\midrule
$m_{\perp}$                  &  $5.415\GeV$  	&  \cite{PDG}			& $m_{B^0}$ 				&  $5.28 \GeV$  			& \cite{PDG}\\
$m_{\|,-}$   &  $5.829\GeV$  	&  \cite{PDG}			&  $f_B$              			&  $207^{+17}_{-9} \MeV$  	&  \cite{Gelhausen:2013wia} \\
$m_{t}$                     &  $5.366\GeV$  	&  \cite{PDG}			&  $\overline m_b (m_b)$ 		&  $4.2\GeV$   				&  \cite{PDG}\\
\midrule
$m_{K^*(892)}$             &  $895.4(2)\MeV$ 	&  \cite{Epifanov:2007rf}	&  $\Gamma_{K^*(892)}$ 		&  $46.1(6)\MeV$  		& \cite{Epifanov:2007rf}\\
$m_{K^*(1410)}$   	     &  $1421(9)\MeV$  	&  \cite{PDG}			&  $\Gamma_{K^*(1410)}$	&  $236(18) \MeV$  		&  \cite{PDG} \\
\midrule
$\lambda_B$  	     	     &    $460\pm 110 \MeV$	&   	 \cite{Braun:2003wx}		&  $R$ 	&     	$0.4^{+0.5}_{-0.3}$	&  \cite{Braun:2017liq} \\
\midrule
\multirow{3}{*}{$\{M^2,s_0\}$}  	    &  $\{1.00,1.26(18)\}\GeV^2$ 		&   	\multirow{3}{*}{Sec.\ref{sec:2pt-s0}}  	& \multirow{3}{*}{$m_s(1\GeV)$}  & \multirow{3}{*}{$123(14) \MeV$}  & \multirow{3}{*}{\cite{PDG}} \\
					 	    &  $\{1.25,1.31(12)\}\GeV^2$ 		&   								&   &   & \\
					 	    &  $\{1.50,1.35(09)\}\GeV^2$ 			&   							 	&   &   & \\
\bottomrule[0.7mm]
\end{tabular}
\caption{\it Compendium of numerical inputs used in the analysis.}
\label{tab:parameters}
\end{table}

The $K\pi$ form factor $f_+(s)$ is a key input to the sum rule, which can be extracted from data. This is done in~\Sec{sec:taudata} employing two models used by Belle which fit well the $\tau\to K\pi \bar\nu_\tau$ data:
the first one (Model 1) contains only a $K^*(892)$ vector resonance plus two scalar ones, and the second one (Model 2) contains the $K^*(892)$ and the $K^*(1410)$, plus one scalar resonance. We will consider the impact of both models in the LCSRs for the $B\to K\pi$ form factors.

Another key input to the sum rules is the effective threshold $s_0$. In order to fix this input we follow~\Ref{Cheng:2017smj} and use the SVZ QCD sum rules to relate an integral of the $K\pi$ form factor
to the vacuum correlation function of two $K^*$ interpolating currents. In this way, the duality interval that satisfies the SVZ sum rule correlates the effective threshold $s_0$ to the Borel parameter $M^2$. This analysis is carried out in~\Sec{sec:2pt-s0}.

For the Borel parameter $M^2$ we take values inside the interval $M^2 = 1.0 - 1.5 \GeV^2$, in all the sum rules, following~\Ref{Cheng:2017smj}. This is slightly narrower than the one used in~\Ref{Khodjamirian:2005ea}.
Within this interval, the convergence of the OPE is manifested by relatively small three-particle DA contributions:
\eq{
\frac{\P_{i,\text{[3-particle]}}^{(T),\rm OPE}}{\P_{i,\text{[2-particle]}}^{(T),\rm OPE}} \lesssim 15\%
\qquad \text{for}\quad q^2 \in [0, 5] \GeV^2\ ,
}
for all form factors and all three models for the LCDAs considered.
Simultaneously, the duality-subtracted part of the integral over the spectral density of the correlation function (l.h.s. of~\Eq{eq:GenericLCSR} with the integral above $s=s_0$) does not exceed $\sim 40\%$ of the total integral, making the result weakly sensitive to
the quark-hadron duality approximation.

Once the numerical input has been fixed, we produce numerical results for the OPE side of the sum rules, in the context of the $z$-expansion. This is done in~\Sec{sec:OPEz}. With these results at hand, we put the LCSRs to work and study the $B\to K\pi$ form factors in three steps: the $B\to K^*$ form factors in the narrow-width limit (\Sec{sec:BtoK*NWL}), the finite-width corrections to $B\to K^*$ form factors (\Sec{sec:finitewidtheffects}), and discuss the $B\to K\pi$ form factors beyond the $K^*$ window
(\Sec{sec:K*1410effects}).


\subsection{The vector $K\pi$ form factor from $\tau$ data}
\label{sec:taudata}

The $K\pi$ form factors in the time-like region can be extracted from the measurement of the $\tau \to K_S\, \pi^- \nu_\tau$ spectrum by Belle~\cite{Epifanov:2007rf}.
This spectrum provides the scalar and vector form factors $f_{+,0}^{K_S\pi^-}(s)$ -- actually, one particular combination -- which are related to the ones in the $K^-\pi^+$ channel by isospin symmetry:
\eq{
f_{+,0}(s) \equiv f_{+,0}^{K^-\pi^+}(s) = - f_{+,0}^{\bar K^0\pi^-}(s) = - \sqrt2 \ f_{+,0}^{K_S\pi^-}(s) \ .
\label{eq:isospinf+0}
}
$CP$ invariance is assumed and the $K_S$, which is the mass eigenstate of the neutral kaon system with shorter lifetime, is identified through its decay into two pions.

Belle measures the binned spectrum of events in $\tau \to K_S\, \pi^- \nu_\tau$ as a function of the invariant mass of the $K_S\, \pi^-$ pair
$\sqrt{s} = \sqrt{ (p_K + p_\pi)^2}$, which is related to the differential decay rate by:
\eq{
\frac{N^{\rm bin}_{\rm events}}{N^{\rm total}_{\rm events}}
= 
\frac1{\Gamma} \int_{\rm bin} d\sqrt{s} \, \frac{d \Gamma}{d\sqrt{s}}
\simeq
\frac{\Delta_{\rm bin}}{\Gamma} \frac{d \Gamma}{d\sqrt{s}} \bigg|_{\rm bin}
=
\frac{\Delta_{\rm bin}}{\Gamma}\ 2\sqrt{s}\ \frac{d \Gamma}{ds} \bigg|_{\rm bin}
\ ,
\label{eq:spectrum}
}
where $\Gamma \equiv \Gamma(\tau \to K_S\, \pi^- \nu_\tau)$, $\Delta_{\rm bin}$ is the size of the $\sqrt{s}$-bin, 
and in the second step we have assumed that the bin
is small enough such that the differential rate does not change sensibly within it. In the Belle analysis the bin size is fixed to $\Delta_{\rm bin}=11.5\,\MeV$,
and the total number of events is $N^{\rm total}_{\rm events} = 53\,113.2\,$.

On the other hand, the differential decay rate is given by (see \Refs{Pich:2013lsa,Escribano:2014joa}):
\eq{
\frac{d \Gamma}{ds} =
\frac{N_\tau}{s^3} \Big( 1- \frac{s}{m_\tau^2} \Big)^2 
\Big( 1+ 2 \frac{s}{m_\tau^2} \Big)\, \lambda_{K\pi}^{3/2} \, |\widetilde f_+(s)|^2
\bigg\{  1
+ \frac{3 (\Delta m^2)^2}{(1+2s/m_\tau^2) \,\lambda_{K\pi}} \, |\widetilde f_0(s)|^2 \bigg\} \ ,
\label{eq:dGammads}
}
with the normalization
\eq{
N_\tau = \frac{G_F^2\, |V_{us}|^2\, |f_+(0)|^2\, m_\tau^3}{1536 \pi^3} S_{EW}^{\rm had} \ .
}
Here $S^{\rm had}_{EW} = 1.0201\pm 0.0003$ accounts for the short-distance electroweak corrections~\cite{Erler:2002mv}, and we have included in the normalization the vector form factor at zero momentum transfer to
define the ``normalized" form factors $\widetilde f_{+,0}(s)$, such that $\widetilde f_+(0)=1$
and $\widetilde f_0(s) = f_0(s)/f_+(s)$.
While the normalization $f_+(0)$ is well known from the lattice~\cite{FLAG}, we will not assume that the model used here for $f_+(s)$
is very precise at $s=0$. We will rather fix this normalization from the total $\tau\to K_S\pi^- \nu_\tau$ decay rate as measured by Belle~\cite{Epifanov:2007rf}. 
The CKM element $|V_{us}|$ is extracted from a
global fit to $K_{\ell3}$ observables~\cite{PDG,FLAG}:
$|V_{us}| = 0.2243 \pm 0.0005$~\footnote{We are assuming that there is no physics beyond the Standard Model (BSM) affecting the $\tau$ decay, directly or indirectly. This includes any BSM contributions affecting the extraction of $G_F$ or $V_{us}$, see e.g.~\Ref{Descotes-Genon:2018foz}.}.
We will neglect all uncertainties in $N_\tau$.

\bigskip

Belle~\cite{Epifanov:2007rf} uses the following model for the vector and scalar form factors in their fits (in our notation)\,\footnote{This model does not fulfill the theoretical constraint $f_+(0)=f_0(0)$, which is however not a concern for us: we are mostly focused on the behaviour of the $f_+$ form factor at $s>(m_K+m_\pi)^2$, whose fit to the data is unlikely to be altered significantly if we changed the behaviour of the $f_0$ form factor at zero.}:
\eq{
\widetilde f_+(s) = \sum_R \frac{\xi_R \,m_R^2}{m_R^2 - s - i\sqrt{s}\, \Gamma_R(s)}\ , \quad
f_0(s) = f_+(0) \cdot \sum_{R_0} \frac{\xi_{R_0} \,s}{m_{R_0}^2 - s - i\sqrt{s}\, \Gamma_{R_0}(s)}\ ,
}
with $R=\{ K^*(892),K^*(1410),K^*(1680) \}$ and $R_0=\{ K^*(800),K^*(1430) \}$.
For the case of $S$-wave resonances, the energy-dependent width is modified with respect to~\Eq{eq:Gamma(s)}:
\eq{
\Gamma_{R_0}(s) 
=\Gamma^{\rm tot}_{R_0} \left[\frac{\lambda_{K\pi}(s)}{\lambda_{K\pi}(m_{R_0}^2)}
\right]^{1/2}\frac{m_{R_0}^3}{s^{3/2}} \ \theta\big(s- s_{\rm th} \big)\,.
}
In the notation of Belle, $\xi_{K^*(892)} = 1/(1+\beta+\chi)$, $\xi_{K^*(1410)} = \beta/(1+\beta+\chi)$, $\xi_{K^*(1680)} = \chi/(1+\beta+\chi)$,
$\xi_{K_0^*(800)} = \varkappa$ and $\xi_{K_0^*(1430)} = \gamma$.

Belle finds two models that fit well the data, the first one (hereon Model 1) with the $K^*(892)$ plus the two scalar resonances $K_0^*(800)$ and $K_0^*(1430)$,
and the second one (hereon Model 2) where the scalar $K_0^*(1430)$ is replaced by the vector $K^*(1410)$. In both fits the mass and width of the $K^*(892)$
are left as free parameters, but in both cases the fits give results which are essentially equal to the numbers given in \Tab{tab:parameters}.
For the other parameters Belle finds: $\varkappa = 1.27$ and $\gamma = 0.954\,e^{i\,0.62}$ (Model 1),
and $\varkappa = 1.57$ and $\beta = 0.075\,e^{i\,1.44}$ (Model 2), where we have ignored the uncertainties. In our notation, this implies:
\eqa{
\text{\bf Model 1}: && \xi_{K^*(892)} = 1\,,\ \xi_{K_0^*(800)} = 1.27\,,\ \xi_{K_0^*(1430)}= 0.954\,e^{i\,0.62}\ ,\\
\text{\bf Model 2}: &&  \xi_{K^*(892)} = 0.988\,e^{-i\,0.07}\,,\ \xi_{K^*(1410)}=0.074\,e^{i\,1.37}\,, \  \xi_{K_0^*(800)} = 1.57\ ,
}
with the omitted parameters set to zero.

\begin{figure}
\includegraphics[height=8.26cm,width=8.5cm]{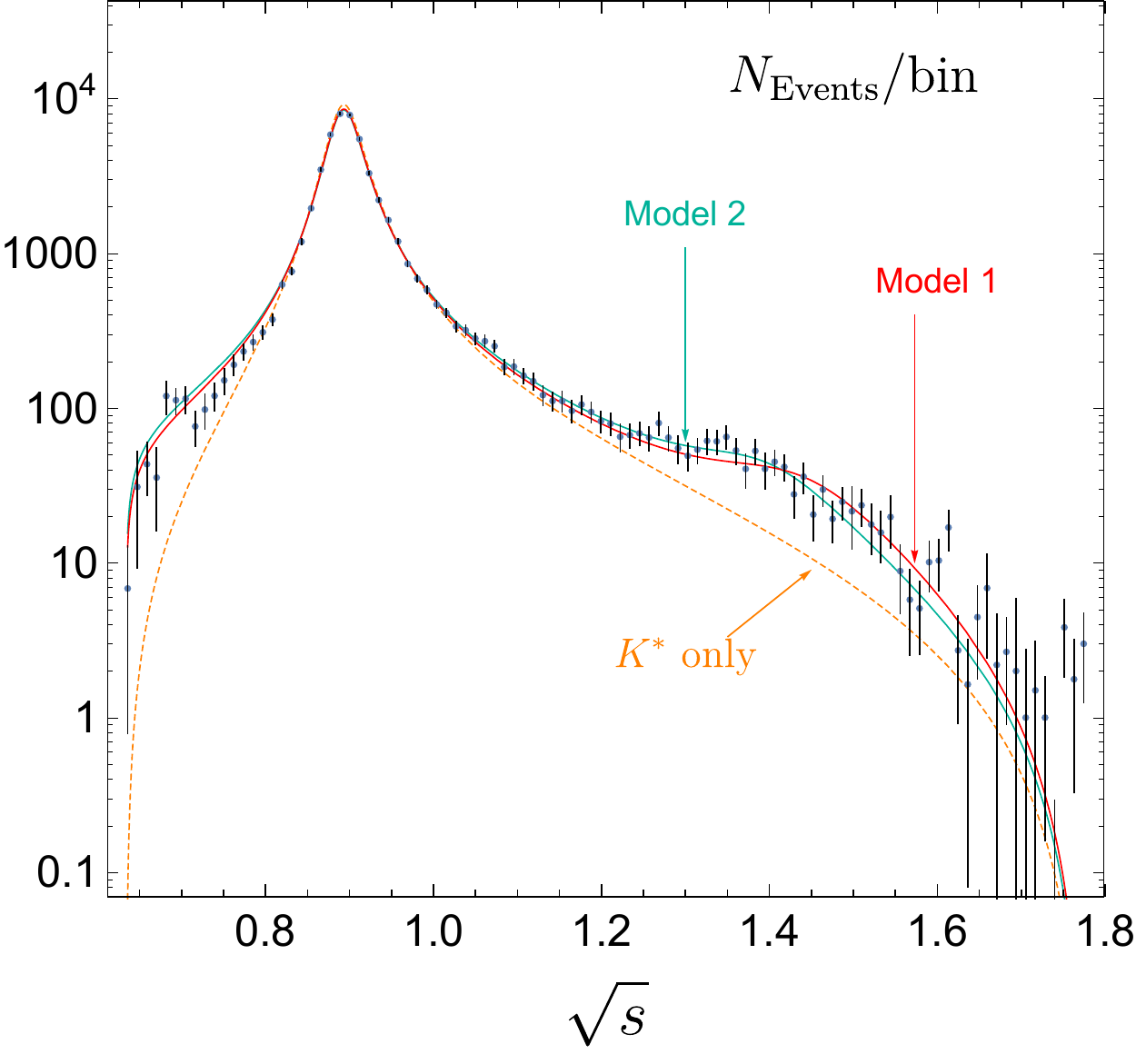}
\hspace{5mm}
\raisebox{0.6mm}{\includegraphics[height=8.2cm,width=7.6cm]{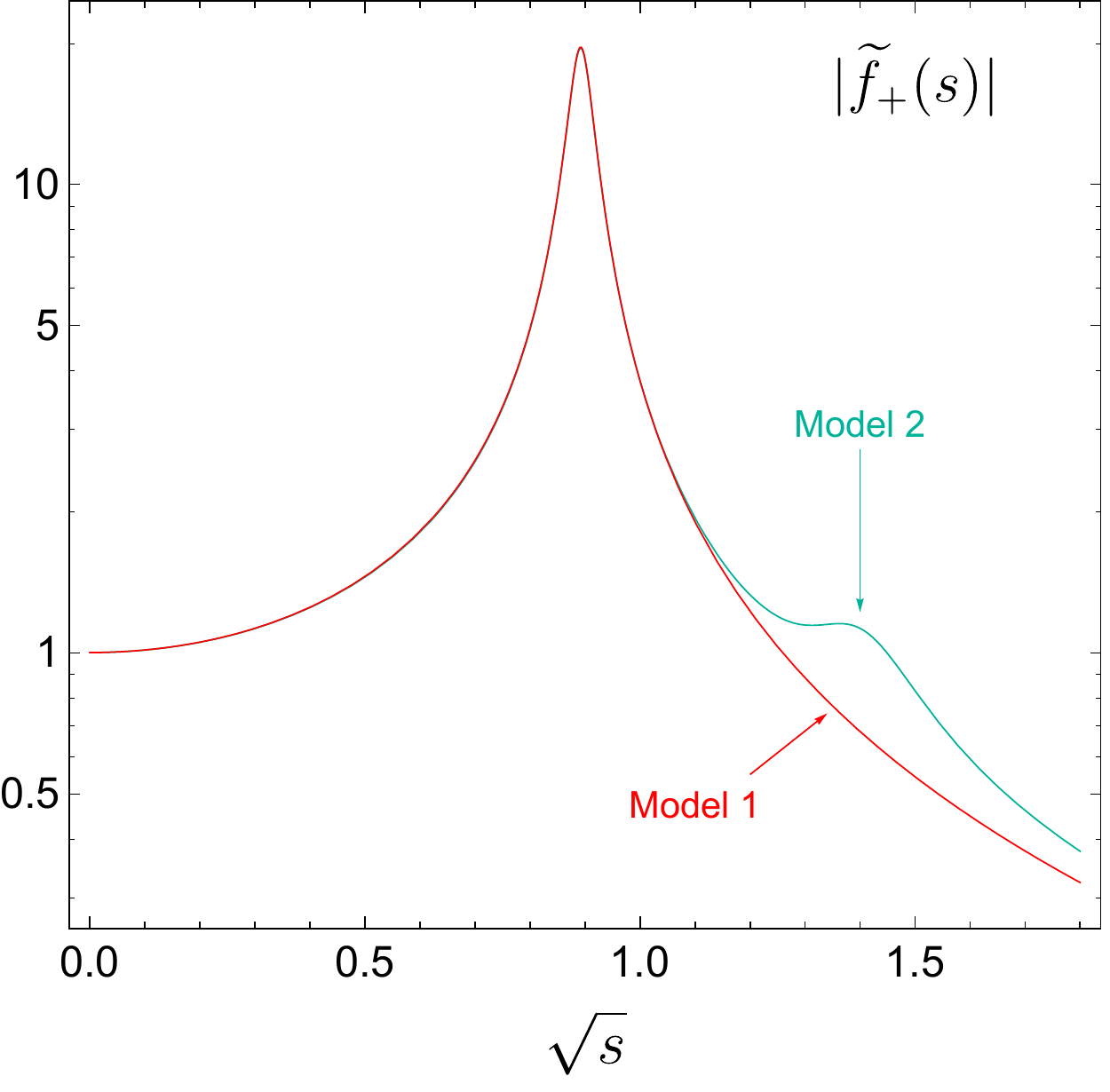}}
\caption{\it Left: $\tau \to K_S\pi^- \nu_\tau$ spectrum from Belle~\cite{Epifanov:2007rf}, and the corresponding curves from Model 1 and Model 2
(two solid lines) as well as the isolated contribution from the $K^*(892)$ (dashed).
Right: Normalized form factor $\widetilde f_+(s)$ in the two models that fit well the spectrum.}
\label{fig:tauspectrum}
\end{figure}

The spectrum of events given in~\Eq{eq:spectrum} does not depend on the normalization of the rate $N_\tau$. In~\Fig{fig:tauspectrum} (left panel) we show the Belle data
compared to the curves obtained from~Eqs.~(\ref{eq:spectrum}) and~(\ref{eq:dGammads}) in the two models for the form factors, and including a model
with only a $K^*(892)$ resonance. The normalized vector form factor $\widetilde f_+(s)$ is also shown in~\Fig{fig:tauspectrum} (right panel), within the
two models considered. 

In order to fix the normalization of the vector form factor $f_+(s)$ we consider the total branching fraction of $\tau \to K_S\pi^- \nu_\tau$.
Belle gives $\B(\tau \to K_S\pi^- \nu_\tau)_{\rm Belle} = 0.404 \pm 0.013\,\%$~\cite{Epifanov:2007rf}.
Integrating \Eq{eq:dGammads}, and using $\tau_\tau = 4.41\cdot 10^{11}\,\GeV^{-1}$ for the $\tau$ lifetime~\cite{PDG}, we find:
\eq{
\B(\tau \to K_S\pi^- \nu_\tau) = \left\{
\begin{array}{l}
0.409 \,|f_+(0)|^2\ \% \quad \text{(Model 1)} \\
0.411 \,|f_+(0)|^2\ \% \quad \text{(Model 2)}
\end{array}
\right.\ .
}
Thus, reproducing the central value of Belle requires $|f_+(0)| = 0.99$ (in both models), very close to unity,
and close to the central value of the lattice QCD average~\cite{FLAG}: $f^{\rm LQCD}_+(0) = 0.97$.
Keeping in mind the other sources of uncertainties affecting our computation, we find it simpler to approximate this number to unity and fix its phase such that $\phi_{K^*(892)}=0$ in~\Eq{eq:f+model}:
$f_+(0) = -1$ in Model 1 and $f_+(0) = -e^{i\,0.07}$ in Model 2.

\bigskip

All in all, the two models for the vector form factor $f_+(s)$ that we will use are given by~\Eq{eq:f+model}
with the following values for the resonance parameters: 
\eq{
f_R = \frac{m_R}{g_{RK\pi}} \ |\xi_R|\ , \quad
\phi_R = \arg{\xi_R} - \arg{\xi_{K^*(892)}}\ ,
}
which implies for each of the two models:
\eqa{
\text{\bf Model 1}: \quad	&& f_{K^*(892)}= 206\,\MeV \ ,\quad \phi_{K^*(892)}=0 \ , \\[2mm]
\text{\bf Model 2}:  \quad	&& f_{K^*(892)}= 203\,\MeV \ ,\quad \phi_{K^*(892)}=0 \ , \\
					&& f_{K^*(1410)}= 85\,\MeV \ ,\quad \phi_{K^*(1410)}= 1.44 \ ,
}
with $f_{K^*(1410)}=f_{K^*(1680)}=0$ in Model 1 and $f_{K^*(1680)}=0$ in Model 2.
In order to derive these numbers we have used the values for the strong couplings:
\eq{
g_{K^*(892)K\pi} = 4.36\ ,\quad
g_{K^*(1410)K\pi} = 1.24\ ,
}
which can be derived from~\Eq{eq:Gammatot} and the values in~\Tab{tab:parameters}.
We note that these values for $f_{K^*(892)}$ are somewhat lower than the narrow-width estimate $f_{K^*(892)}=217(5) \,\MeV$~\cite{Ball:2005vx}
obtained from the two-point QCD sum rule for vector currents with strangeness.
The reason for this difference will be discussed in the next subsection.
 
Having rewritten the models of Belle in the notation of~\Eq{eq:f+model} will be useful in order to check
the narrow-width limit.
Finally, we point out that in order to be able to distinguish the effects of the vector and scalar resonances $K^*(1410)$ and $K^*_0(1430)$,
a dedicated angular analysis of Belle and Belle-II data is necessary.

\subsection{Effective threshold from a two-point QCD sum rule}
\label{sec:2pt-s0}

In \Ref{Khodjamirian:2006st} the 
duality interval for the 
interpolating light-quark current was 
assumed the same as in the QCD (SVZ) sum rule
\cite{Shifman:1979BX,Shifman:1978by} 
for the two-point correlation function  
of these currents. 
In particular, in the LCSRs for  $B\to K^*$  form factors, the value $s_0=1.7$ GeV$^2$ \cite{Ball:2005vx} stems from  
the two-point QCD sum rule 
for vector currents with strangeness. In this sum rule the hadronic 
part in the duality interval was approximated by a single narrow $K^*$. We find  it more consistent to adopt for
this hadronic part the
$K\pi$ spectral density expressed
via the measured quantity $|f_+(s)|^2$, 
which includes the $K^*$ with a finite
width. This approach was already used
in \Ref{Cheng:2017smj} to set
the effective threshold for the 
dipion channel in the LCSR for the $B\to \pi\pi$ form factors.

The QCD  sum rule that we need  \cite{Shifman:1979BX,Shifman:1978by} is based on the two-point correlation function:
\eq{
\Pi_{\mu\nu}(k)=i\int d^4x e^{ikx}\langle 0 | {\rm T}\{\bar{d}(x)\gamma_\mu s(x),
\bar{s}(0)\gamma_\nu d(0)|0 \rangle =
(k_\mu k_\nu -k^2 g_{\mu\nu}) \,\Pi(k^2) + k_\mu k_\nu \,\widetilde{\Pi}(k^2) \,.
\label{eq:2ptcorr}
}
We focus specifically on the invariant function $\Pi(k^2)$ multiplying the transverse structure,
which receives no contributions from
the scalar form factor $f_0$.
Inserting the $K\pi$ intermediate states and performing the phase-space integrals, one finds
\eq{
\frac{1}{\pi}\im \Pi(s)=\frac{\lambda_{K\pi}^{3/2}(s)}{32 \pi^2 s^3} \,|f_+(s)|^2\,.
\label{eq:2ptim}
}
Using in this expression the model~(\ref{eq:f+model}) for $f_+(s)$ with one $K^*$ resonance and taking the narrow-width limit, $\Gamma_{K^*}\to 0$, 
one recovers the known result $\frac{1}{\pi} \im \Pi(s)= f_{K^*}^2 \delta(s-m_{K^*}^2)$.

Writing the dispersion relation for $\Pi(k^2)$, and performing the Borel transformation, we have
\eq{
\Pi(M^2,s_0) \equiv
\frac{1}{\pi}\int_{s_{\rm th}}^{s_0}\!\!ds\, e^{-s/M^2} \im\Pi(s)=
\int_{s_{\rm th}}^{s_0}\!\!ds\,e^{-s/M^2} \frac{\lambda_{K\pi}^{3/2}(s)}{32 \pi^2 s^3} \,|f_+(s)|^2 \,.
\label{eq:Pidisp}}
The above integral is equated to the Borel-transformed correlation function 
calculated in QCD and containing the perturbative
loop contribution (to NLO) and the vacuum condensate terms (up to $d=6$):

\eqa{
\Pi^{\rm OPE}(M^2,s_0) &=& 
\frac1{8\pi^2} \int_{m_s^2}^{s_0} \!\!ds\, e^{-s/M^2} \frac{(s-m_s^2)^2(2s+m_s^2)}{s^3} \nonumber\\
&&+ \frac{\alpha_s(M)}{\pi} \frac{M^2}{4\pi^2} \Big(1-e^{-s_0/M^2}\Big) +\frac{v_4}{M^2}+\frac{v_6}{2M^4}\ .
\label{eq:PiOPE}
}
Power-suppressed terms in the OPE with the coefficients
\eq{
v_4= m_d\langle 0 |\bar{s}{s}  | 0 \rangle + m_s\langle 0 |\bar{d}{d}  | 0 \rangle +
 \frac{1}{12}
\langle 0 |\frac{\alpha_s}{\pi}G^a_{\mu\nu}G^{a \,\mu\nu} | 0 \rangle\ ,
~~
v_6=-\frac{224\pi}{81} \alpha_s\langle 0 |\bar{q}{q}  | 0 \rangle ^2\ ,
\label{eq:cond}
}
include the contributions  from the quark ($d=3$), gluon ($d=4$)  
and four-quark ($d=6$) condensates, respectively. In the latter contribution, 
the approximation $\langle 0 |\bar{s}{s}|0\rangle(1 \GeV)=
\langle 0 |\bar{d}{d}|0\rangle$ is adopted and, following~\Ref{Shifman:1978by}, we rely on the vacuum saturation approximation to re-express the four-quark condensates in terms of the quark condensate.

In the numerical analysis of the above expressions, owing to the fact that 
$m_d\ll m_s$, we neglect the first term in the quark-condensate contribution.  
The input parameters are: 
$m_s(1 \GeV)=123\pm 14\MeV$~\cite{PDG},  $\langle 0 |\bar{d}{d}|0\rangle(1 \GeV)=
\langle 0 |\bar{q}{q}|0\rangle(1 \GeV)= (-250\pm 10 ~\MeV)^3 $~\cite{Leutwyler:1996qg,PDG} 
and $\langle 0 |\frac{\alpha_s}{\pi} G^a_{\mu\nu}G^{a \,\mu\nu} | 0 \rangle =0.012^{+ 0.006}_{-0.012}~\GeV^4$ \cite{Ioffe:2002ee}. Note that in the four-quark condensate contribution
a low-scale $\alpha_s(1\GeV) =0.46$ \cite{PDG} is taken.

The QCD sum rule obtained equating $\Pi(M^2,s_0)$ and $\Pi^{\rm OPE}(M^2,s_0)$ differs from the usual QCD sum rule~\cite{Shifman:1978by} in which the contribution of a narrow ground state is equated to the OPE result integrated over a duality interval, so that $s_0$ appears only on the OPE side. The threshold parameter is then usually fixed beforehand and the variation with $M^2$ is interpreted as an uncertainty of the sum rule prediction for the ground state parameters. Here we do not intend to calculate e.g., the ground state $K^*$ decay constant from the sum rule. On the contrary, we use data on the $K\pi$ form factor to fix the hadronic part and $s_0$ enters both sides of the sum rule equation. We determine the range of values of $s_0$ that best fit this equation and then use it in the LCSRs for the $B\to K\pi$ form factors. The extracted intervals of the effective threshold thus depend on the choice of Borel parameter. Hence, when taking a certain value of Borel parameter in the LCSR, we will use the corresponding fitted threshold.

Fitting the integral $\Pi(M^2,s_0)$ to its  QCD sum rule counterpart  $\Pi^{\rm OPE}(M^2,s_0)$ we find
the values for the effective threshold quoted in~\Tab{tab:EffectiveThreshold}.
These values depend on the Borel parameter $M^2$ and on the model used for the form factor $f_+(s)$.
We set the Borel parameter to the three different values $M^2=\{ 1.00, 1.25, 1.50\}\,\GeV^2$, and for
each of these values we calculate the resulting $s_0$ from a $\chi^2$ minimization of the difference $(\Pi - \Pi^{\rm OPE})$,
including the uncertainties in the OPE coefficients. This is done separately for each of the two models discussed in~\Sec{sec:taudata}
for $f_+(s)$. These results are shown in the second column of~\Tab{tab:EffectiveThreshold}. Finally, we combine both numbers
for each value of the Borel parameter to produce an average that accounts for the ``model dependence", as shown in the third column.
The estimated model dependence turns out to be small compared to the parametric uncertainty from the OPE coefficients.

\begin{table}
\centering
\setlength{\tabcolsep}{10pt}
\begin{tabular}{@{}ccc@{}}
\toprule[0.7mm]
Borel parameter $M^2$ & \multicolumn{2}{c}{Effective threshold $s_0$} \\
\midrule[0.7mm]
\multirow{2}{*}{$1.00\,\GeV^2$}		& $1.28\pm 0.18\,\GeV^2$  {\color{gray} (Model 1)}	&	\multirow{2}{*}{$1.26\pm 0.18\,\GeV^2$ {\color{gray} (Average)}} \\ 
 			  				& $1.25\pm 0.18\,\GeV^2$  {\color{gray} (Model 2)}	&\\
\midrule
\multirow{2}{*}{$1.25\,\GeV^2$}		& $1.33\pm 0.12\,\GeV^2$  {\color{gray} (Model 1)}	&      \multirow{2}{*}{$1.31\pm 0.12\,\GeV^2$ {\color{gray} (Average)}}\\
							& $1.31\pm 0.12\,\GeV^2$  {\color{gray} (Model 2)}	&\\
\midrule
\multirow{2}{*}{$1.50\,\GeV^2$}		& $1.36\pm 0.09\,\GeV^2$  {\color{gray} (Model 1)}	&      \multirow{2}{*}{$1.35\pm 0.09\,\GeV^2$  {\color{gray} (Average)}} \\
							& $1.34\pm 0.09\,\GeV^2$  {\color{gray} (Model 2)}	& \\
\bottomrule[0.7mm]
\end{tabular}
\caption{\it Values for the effective threshold $s_0$ extracted from the SVZ sum rules.}
\label{tab:EffectiveThreshold}
\end{table}

Our estimates for the effective threshold are relatively low as compared to the duality interval $s_0^{K^*}=1.7 \GeV^2$ 
adopted in the original SVZ sum rule for the $K^*$ meson \cite{Shifman:1978by}, and used again in \Ref{Ball:2005vx}. In the
latter, the sum rule was employed to 
obtain the value of the decay constant~$f_{K^*}$
(quoted at the end of \Sec{sec:taudata}) and 
the choice of the threshold was  
done in a seemingly qualitative way~\footnote{We can roughly 
reproduce this value assuming 
$s_0^{K^*}=(\sqrt{s_0^\rho}+m_s)^2$ where
$s_0^\rho=1.5$ GeV$^2$  is an established threshold in the $\rho$/dipion channel.}.
Following our procedure, we notice that
for $M^2\sim 1$ GeV$^2$ the integral on r.h.s. of \Eq{eq:Pidisp}  practically does not depend on $s_0$ starting from $s_0\sim 1.0\GeV^2$, which means that
the contribution of the $K\pi$ state to the hadronic spectral density weighted with the Borel exponent dominates for $s\lesssim 1$ GeV$^2$.
The OPE in \Eq{eq:PiOPE}, where the perturbative part dominates at $M^2\sim 1~ $GeV$^2$ is, on the contrary, sensitive
to increasing $s_0>$ 1 GeV$^2$ and fits the r.h.s.
of \Eq{eq:Pidisp} only at $s_0\sim 1.3$ GeV$^2$. 
We conclude that  
the OPE spectral density 
at $s>s_0$ is to a large extent dual to the hadronic states with
larger multiplicity
($K\pi\pi$, $K 3 \pi$, etc.)
including their resonance
contributions, and that the values of $s_0$ in \Tab{tab:EffectiveThreshold}
are truly reflecting the duality interval
for the $K\pi$ $P$-wave state.

Our second observation is that these values
are somewhat smaller
than $s_0^{2\pi} \simeq 1.5\GeV^2$ 
obtained in~\Ref{Cheng:2017smj}
for the dipion $P$-wave state. This might also seem unexpected 
but in reality it only
reflects the complexity of hadronic spectral functions in both $K\pi$ and $\pi\pi$ channels, including some diversity -- for instance,
three-body $K\pi\pi$ states are allowed in the former case
whereas the $3\pi$ ones are forbidden in the latter channel by isospin symmetry. These observations could open up a new
perspective for revisiting the ``classical"
two-point SVZ sum rules with a more accurate 
hadronic description, such as the one adopted here.

\subsection{Fitting the OPE to the $z$-expansion}
\label{sec:OPEz}

We now use the OPE expressions $\P_i^{(T),\rm OPE}$ in~\App{OPEexpressions} to determine the OPE coefficients of the $z$-expansion in~\Eq{eq:OPEz}:
\eq{
\P_i^{(T),\rm OPE}(q^2,\sigma_0,M^2) =
\frac{\kappa_{i}^{(T),\rm OPE}+\eta_{i}^{(T),\rm OPE} \ \zeta(q^2)}{1-q^2/m^2_{i}}  \ .
}
For this purpose, we first produce results for $\P_i^{(T)\rm OPE}(q^2,\sigma_0,M^2)$ for all seven form factors, for $q^2=\{ 0,1,2,3,4,5 \} \GeV^2$ and for $M^2=\{1.00, 1.25, 1.50\} \GeV^2$ (with the corresponding values of $s_0$ in~\Tab{tab:EffectiveThreshold}). This amounts to 18 determinations per form factor.
In addition, we consider all three models for the $B$-meson LCDAs discussed~\App{sec:modelsLCDAs}.
These results for $\P_i^{(T)\rm OPE}$ have central values and uncertainties that correspond to the mean and the standard deviation of a multivariate Gaussian scan over all input parameters. We have checked that these ensembles are approximately Gaussian and that the mean values are close to the most probable point, and also close to the result obtained from the central values of the input parameters.

From the results at $q^2=0$ we obtain directly the OPE parameters $\kappa_i^{(T),\rm OPE}$,
\eq{
\kappa_i^{(T),\rm OPE}(\sigma_0,M^2) = \P_i^{(T)\rm OPE}(0,\sigma_0,M^2)\ .
}
For the ``slope" OPE parameters $\eta_i^{(T),\rm OPE}$ we use the formula
\eq{
\eta_i^{(T),\rm OPE}(\sigma_0,M^2) =
\frac{(m_i^2-q^2)\P_i^{(T)\rm OPE}(q^2,\sigma_0,M^2)}{m_i^ 2\,\zeta(q^2)}
- \frac{\P_i^{(T)\rm OPE}(0,\sigma_0,M^2)}{\zeta(q^2)}\ ,
}
and taking into account that the left-hand side must be $q^2$-independent, we perform a fit using the determinations at $q^2=\{ 1,2,3,4,5 \} \GeV^2$ as pseudo-data.
Our final results for the OPE parameters $\kappa_i^{(T),\rm OPE}$ and $\eta_i^{(T),\rm OPE}$ are summarized in~\Tab{tab:OPEpars}. The uncertainties of this set of 42 numbers are strongly correlated among themselves.
The full $42\times 42$ correlation matrix in electronic format (in the form of a Mathematica file) is available from the authors upon request (see~\App{sec:OPEnumerics} for details).

\begin{table}
\centering
\setlength{\tabcolsep}{10pt}
\begin{tabular}{@{}clll@{}}
\toprule[0.7mm]
Form F. & $\ M^2=1.00\GeV^2$ & $\ M^2=1.25\GeV^2$ & $\ M^2=1.50\GeV^2$ \\
\midrule[0.7mm]
\multirow{2}{*}{$F_\perp$}
& $\kappa_\perp^{\rm OPE}=+0.007(4)(0)$ 
& $\kappa_\perp^{\rm OPE}=+0.008(5)(0)$ 
& $\kappa_\perp^{\rm OPE}=+0.009(5)(0)$ \\
& $\eta_\perp^{\rm OPE}=-0.010(14)(14)$ 
& $\eta_\perp^{\rm OPE}=-0.012(17)(17)$ 
& $\eta_\perp^{\rm OPE}=-0.013(19)(21)$ \\ 
\midrule
\multirow{2}{*}{$F_\|$}
& $\kappa_\|^{\rm OPE}=+0.100(58)(7)$ 
& $\kappa_\|^{\rm OPE}=+0.120(69)(8)$ 
& $\kappa_\|^{\rm OPE}=+0.137(78)(8)$ \\
& $\eta_\|^{\rm OPE}=+0.25(9)(16)$ 
& $\eta_\|^{\rm OPE}=+0.30(11)(20)$ 
& $\eta_\|^{\rm OPE}=+0.36(13)(24)$ \\ 
\midrule
\multirow{2}{*}{$F_-$}
& $\kappa_-^{\rm OPE}=-0.004(3)(1)$ 
& $\kappa_-^{\rm OPE}=-0.004(4)(1)$ 
& $\kappa_-^{\rm OPE}=-0.005(5)(1)$ \\
& $\eta_-^{\rm OPE}=-0.020(17)(3)$ 
& $\eta_-^{\rm OPE}=-0.025(21)(4)$ 
& $\eta_-^{\rm OPE}=-0.029(24)(5)$ \\ 
\midrule
\multirow{2}{*}{$F_t$}
& $\kappa_t^{\rm OPE}=-0.043(9)(2)$ 
& $\kappa_t^{\rm OPE}=-0.052(11)(3)$ 
& $\kappa_t^{\rm OPE}=-0.060(12)(3)$ \\
& $\eta_t^{\rm OPE}=+0.210(30)(10)$ 
& $\eta_t^{\rm OPE}=+0.249(34)(14)$ 
& $\eta_t^{\rm OPE}=+0.282(37)(19)$ \\ 
\midrule
\multirow{2}{*}{$F^T_\perp$}
& $\kappa_\perp^{T,\rm OPE}=+0.036(21)(2)$ 
& $\kappa_\perp^{T,\rm OPE}=+0.043(25)(2)$ 
& $\kappa_\perp^{T,\rm OPE}=+0.050(29)(2)$ \\
& $\eta_\perp^{T,\rm OPE}=-0.056(73)(66)$ 
& $\eta_\perp^{T,\rm OPE}=-0.065(85)(84)$ 
& $\eta_\perp^{T,\rm OPE}=-0.07(9)(10)$ \\ 
\midrule
\multirow{2}{*}{$F^T_\|$}
& $\kappa_\|^{T,\rm OPE}=+0.49(29)(3)$ 
& $\kappa_\|^{T,\rm OPE}=+0.59(35)(3)$ 
& $\kappa_\|^{T,\rm OPE}=+0.67(39)(3)$ \\
& $\eta_\|^{T,\rm OPE}=+1.35(46)(85)$ 
& $\eta_\|^{T,\rm OPE}=+1.7(6)(11)$ 
& $\eta_\|^{T,\rm OPE}=+1.9(7)(12)$ \\ 
\midrule
\multirow{2}{*}{$F^T_-$}
& $\kappa_-^{T,\rm OPE}=-0.021(19)(5)$ 
& $\kappa_-^{T,\rm OPE}=-0.025(23)(6)$ 
& $\kappa_-^{T,\rm OPE}=-0.028(26)(7)$ \\
& $\eta_-^{T,\rm OPE}=-0.10(10)(1)$ 
& $\eta_-^{T,\rm OPE}=-0.12(13)(2)$ 
& $\eta_-^{T,\rm OPE}=-0.14(15)(2)$ \\ 
\bottomrule[0.7mm]
\end{tabular}
\caption{\it Results for the OPE coefficients in the $z$-expansion in Model I for the $B$-meson LCDAs. The first error is parametric and the second one captures the model dependence from the LCDAs, as detailed in~\App{sec:OPEnumerics}.}
\label{tab:OPEpars}
\end{table}

These results correspond to Model I for the $B$-meson LCDAs as described in~\App{sec:modelsLCDAs}, which we regard as our default model. In order to estimate the model dependence of the OPE contributions, we look at the corresponding results in models IIA and IIB.
These results are collected in~\App{sec:OPEnumerics}. We use these results to produce the second set of errors in~\Tab{tab:OPEpars}, which capture the model dependence of the results. This estimate of model dependence does not imply that the three models discussed in~\App{sec:modelsLCDAs} and~\Ref{Braun:2017liq} must be regarded on the same footing. Models for LCDAs remain to be studied carefully and deserve further theoretical work (see e.g.~\Ref{Feldmann:2014ika}).
In relation to this, it has been determined that some invariant amplitudes in the correlation function relevant to $B\to \gamma\ell\nu$ are independent of the shape of some of the higher-twist LCDAs, within the types of models considered here~\cite{Beneke:2018wjp}. This correlation function is equal to the one in~\Eq{eq:corrV} up to flavor, and thus the invariant amplitudes considered in~\Ref{Beneke:2018wjp} are related to some of the invariant amplitudes in~\Eq{eq:corrV} in the limit $m_s\to 0$. 
To which extent the shape-independence of the $B$-meson LCDAs applies to the invariant amplitudes relevant to $B\to K\pi$ form factors is a question that we leave for future consideration.

\begin{sidewaystable}
\centering
\setlength{\tabcolsep}{8.2pt}
\begin{tabular}{@{}rrrrrrrrrrrrrr@{}}
\toprule[0.7mm]
$\F_\perp(0)$ & $b_\perp$ & $\F_\|(0)$ & $b_\|$ & $\F_-(0)$ & $b_-$ & $\F_t(0)$ & $b_t$ & $\F^T_\perp(0)$ & $b^T_\perp$ & $\F^T_\|(0)$ & $b^T_\|$ & $\F^T_-(0)$ & $b^T_-$ \\
\midrule[0.4mm]
$0.26$ & $-0.37$ & $0.20$ & $0.51$ & $0.14$ & $0.80$ & $0.30$ & $-1.44$ & $0.22$ & $-0.33$ & $0.22$ & $0.63$ & $0.13$ & $0.62$ \\
\midrule
$\pm0.15$ & $\pm0.53$ & $\pm0.12$ & $\pm0.19$ & $\pm0.13$ & $\pm0.68$ & $\pm0.07$ & $\pm0.20$ & $\pm0.13$ & $\pm0.44$ & $\pm0.13$ & $\pm0.24$ & $\pm0.12$ & $\pm0.66$ \\
\midrule[0.4mm]
$1.00$ & $-0.97$ & $1.00$ & $0.86$ & $1.00$ & $-0.68$ & $0.96$ & $-0.67$ & $1.00$ & $-0.97$ & $1.00$ & $0.87$ & $1.00$ & $-0.72$ \\
$-0.97$ & $1.00$ & $-0.97$ & $-0.72$ & $-0.98$ & $0.78$ & $-0.89$ & $0.56$ & $-0.97$ & $1.00$ & $-0.97$ & $-0.74$ & $-0.98$ & $0.81$ \\
$1.00$ & $-0.97$ & $1.00$ & $0.86$ & $1.00$ & $-0.68$ & $0.96$ & $-0.67$ & $1.00$ & $-0.97$ & $1.00$ & $0.87$ & $1.00$ & $-0.72$ \\
$0.86$ & $-0.72$ & $0.86$ & $1.00$ & $0.83$ & $-0.25$ & $0.91$ & $-0.84$ & $0.85$ & $-0.70$ & $0.85$ & $1.00$ & $0.83$ & $-0.30$ \\
$1.00$ & $-0.98$ & $1.00$ & $0.83$ & $1.00$ & $-0.70$ & $0.93$ & $-0.64$ & $1.00$ & $-0.98$ & $1.00$ & $0.84$ & $1.00$ & $-0.74$ \\
$-0.68$ & $0.78$ & $-0.68$ & $-0.25$ & $-0.70$ & $1.00$ & $-0.55$ & $-0.06$ & $-0.68$ & $0.79$ & $-0.68$ & $-0.27$ & $-0.70$ & $1.00$ \\
$0.96$ & $-0.89$ & $0.96$ & $0.91$ & $0.93$ & $-0.55$ & $1.00$ & $-0.74$ & $0.96$ & $-0.88$ & $0.95$ & $0.92$ & $0.94$ & $-0.60$ \\
$-0.67$ & $0.56$ & $-0.67$ & $-0.84$ & $-0.64$ & $-0.06$ & $-0.74$ & $1.00$ & $-0.67$ & $0.55$ & $-0.67$ & $-0.83$ & $-0.64$ & $-0.01$ \\
$1.00$ & $-0.97$ & $1.00$ & $0.85$ & $1.00$ & $-0.68$ & $0.96$ & $-0.67$ & $1.00$ & $-0.97$ & $1.00$ & $0.87$ & $1.00$ & $-0.72$ \\
$-0.97$ & $1.00$ & $-0.97$ & $-0.70$ & $-0.98$ & $0.79$ & $-0.88$ & $0.55$ & $-0.97$ & $1.00$ & $-0.97$ & $-0.72$ & $-0.98$ & $0.82$ \\
$1.00$ & $-0.97$ & $1.00$ & $0.85$ & $1.00$ & $-0.68$ & $0.95$ & $-0.67$ & $1.00$ & $-0.97$ & $1.00$ & $0.87$ & $1.00$ & $-0.72$ \\
$0.87$ & $-0.74$ & $0.87$ & $1.00$ & $0.84$ & $-0.27$ & $0.92$ & $-0.83$ & $0.87$ & $-0.72$ & $0.87$ & $1.00$ & $0.85$ & $-0.32$ \\
$1.00$ & $-0.98$ & $1.00$ & $0.83$ & $1.00$ & $-0.70$ & $0.94$ & $-0.64$ & $1.00$ & $-0.98$ & $1.00$ & $0.85$ & $1.00$ & $-0.74$ \\
$-0.72$ & $0.81$ & $-0.72$ & $-0.30$ & $-0.74$ & $1.00$ & $-0.60$ & $-0.01$ & $-0.72$ & $0.82$ & $-0.72$ & $-0.32$ & $-0.74$ & $1.00$ \\
\bottomrule[0.7mm]
\end{tabular}
\caption{\it Results for the form factors in the narrow-width limit. The first row contains the central values, the second row the uncertainties, and the rest the correlation coefficients.}
\label{tab:ResultsFFs}
\end{sidewaystable}

\subsection{$B\to K^*$ form factors in the narrow-width limit}
\label{sec:BtoK*NWL}

\begin{table}
\centering
\setlength{\tabcolsep}{10pt}
\begin{tabular}{@{}cclllll@{}}
\toprule[0.7mm]
Form Factor && This work&\Ref{Khodjamirian:2006st}&\Ref{Khodjamirian:2010vf}&\Ref{Gubernari:2018wyi}&\Ref{Straub:2015ica} \\
\midrule[0.4mm]
$\F_{K^*,\perp}(0)= V^{BK^*}(0)$ && 0.26(15) & 0.39(11) & 0.36(18) & 0.32(11) & 0.34(4) \\
$\F_{K^*,\|}(0)= A_1^{BK^*}(0)$ && 0.20(12) & 0.30(8) & 0.25(13) & 0.26(8) & 0.27(3) \\
$\F_{K^*,-}(0)= A_2^{BK^*}(0)$ && 0.14(13) & 0.26(8) & 0.23(15) & 0.24(9) & 0.23(5) \\
$\F_{K^*,t}(0)= A_0^{BK^*}(0)$ && 0.30(7) & -- & 0.29(8) & 0.31(7) & 0.36(5) \\
$\F^T_{K^*,\perp}(0)= T_1^{BK^*}(0)$ && 0.22(13) & 0.33(10) & 0.31(14) & 0.29(10) & 0.28(3) \\
$\F^T_{K^*,\|}(0)= T_2^{BK^*}(0)$ && 0.22(13) & 0.33(10) & 0.31(14) & 0.29(10) & 0.28(3) \\
$\F^T_{K^*,-}(0)= T_3^{BK^*}(0)$ && 0.13(12) & -- & 0.22(14) & 0.20(8) & 0.18(3) \\
\bottomrule[0.7mm]
\end{tabular}
\caption{\it Results for the form factors at $q^2=0$ in the narrow-width limit,compared to corresponding results in the literature. The approach in~\Ref{Straub:2015ica} is a completely different LCSR approach, in terms of $K^*$ DAs.}
\label{tab:FFcomparison}
\end{table}

Having studied the OPE side of the sum rules in the previous section, we can move to the hadronic side. This is the part of the sum rules that has been generalized in this article to go beyond the Narrow-Width Limit (NWL). In \Sec{sec:NWL}, we have demonstrated explicitly that in the limit $\Gamma_{K^*} \to 0$ the integrand of the integral over the $K\pi$ invariant mass becomes a delta function, and thus the usual LCSRs for $B\to {K^*}$ form factors are recovered from our sum rules, analytically. Furthermore, we have also checked that the limit $\Gamma_{K^*} \to 0$ works also numerically, and that making $\Gamma_{K^*}$ smaller and smaller the results for the form factors from the full LCSRs converge to the results from the $B\to K^*$ sum rules in~\App{app:BtoVLCSRs}.

In this section we thus study the $B\to {K^*}$ form factors in the NWL and compare our results to those in the literature. For this purpose we take the formulae for the form factors in~\App{app:BtoVLCSRs} and the numerical determination of the OPE functions in~\Tab{tab:OPEpars}.
In this way, we have:
\eqa{
V^{BK^*}(0) &=&
\frac{m_B+m_{K^*}}{2 f_{K^*} m_{K^*}} e^{m_{K^*}^2/M^2} \kappa_\perp^{\rm OPE}(\sigma_0,M^2)\ ,\\[2mm]
b^{BK^*}_V &=&
\frac{m_B+m_{K^*}}{2 f_{K^*} m_{K^*}} e^{m_{K^*}^2/M^2} \eta_\perp^{\rm OPE}(\sigma_0,M^2)\ ,
}
and similarly for the other form factors.

We calculate the form factors $\F^{BK^*}_i(0)$ and the slope parameters $b^{BK^*}_i$ by performing a Gaussian scan over all input parameters, including the three different values for $M^2$.
Our results are collected in~\Tab{tab:ResultsFFs}, together with the correlation coefficients.

In~\Tab{tab:FFcomparison} we compare our results for the form factors at $q^2=0$ with the analogous results in~Refs.\cite{Khodjamirian:2006st,Khodjamirian:2010vf,Gubernari:2018wyi,Straub:2015ica}. 
We see that our results are consistent with all the other determinations within uncertainties, but with central values that are somewhat lower. We ascribe this difference to four factors: the difference in the numerical input, the effect of twist-four two-particle contributions from $g_+(\omega)$ and $\bar G_\pm(\omega)$, the substantially lower value of the effective threshold parameter $s_0$ as described in~\Sec{sec:2pt-s0}, and the effect of three-particle contributions, which in our case reduce the form factors by around $10\%$, while they are negligible and excluded from the numerical analysis in~\Ref{Gubernari:2018wyi}\,\footnote{Using the same inputs as in~\Ref{Gubernari:2018wyi}, we agree with the results of that reference very precisely. The reason that three-particle contributions are negligible in~\Ref{Gubernari:2018wyi} is due to the use of the numerical values in~\Eq{eq:lambdaElambdaHnum} from~\Ref{Nishikawa:2011qk} without the use of the EOM (see~\App{sec:modelsLCDAs}).}.
These effects are summarized in~\Tab{tab:deconstruction}, where we show the central values for the form factors corresponding to:
\Ref{Khodjamirian:2006st} (first row);
our calculation but with the numerical inputs of \Ref{Khodjamirian:2006st}: $f_B=180\MeV$, $f_{K^*}=217\MeV$, $s_0=1.7\GeV^2$, $m_s=130\MeV$, and excluding the twist-four two-particle contributions (second row);
the same but including the twist-four contributions (third row);
the calculation with $M^2=1$ and our inputs in~\Tab{tab:parameters}, but with the effective threshold at $s_0=1.7\GeV$ (fourth row);
all our inputs but excluding the twist-four two-particle contributions (fifth row);
and our final central values with the value of $s_0$ from~\Sec{sec:2pt-s0}, which coincide with the values quoted in~\Tab{tab:FFcomparison} (sixth row).
Our higher input value for $f_B$ and lower value for $f_{K^*}$ increase the values of the form factors,
but this cancels approximately the decrease from the substantially lower value of $s_0$.
The effect of $g_+$ is ultimately responsible for the low values of our form factors (albeit consistent with other determinations within errors).
The parametrical hierarchy of twists in the OPE deserves further careful study, which we postpone to a future work.

To close this subsection, we mention a very recent LCSR calculation of the ``soft overlap" $B\to V$ form factors (in the narrow-width limit), done at NLO in SCET and including twist-6 contributions~\cite{Gao:2019lta}.
The results for $B\to K^*$ form factors are given in Tables~6 and 7 there.

\begin{table}
\centering
\setlength{\tabcolsep}{10pt}
\begin{tabular}{@{}lllllll@{}}
\toprule[0.7mm]
$\F^{BK^*}(q^2=0)$ & $V^{BK^*}$ & $A_1^{BK^*}$ & $A_2^{BK^*}$ & $A_0^{BK^*}$ & $T_{1,2}^{BK^*}$ & $T_3^{BK^*}$ \\
\midrule[0.4mm]
\Ref{Khodjamirian:2006st} & $0.39$ & $0.30$ & $0.26$ & -- & $0.33$ & -- \\
\midrule
Inputs~\cite{Khodjamirian:2006st}, no $g_+$ & $0.38$ & $0.29$ & $0.26$ & $0.31$ & $0.33$ & $0.25$  \\
Inputs~\cite{Khodjamirian:2006st}, with $g_+$ & $0.27$ & $0.21$ & $0.14$ & $0.31$ & $0.24$ & $0.14$ \\
Our inputs, but $s_0=1.7\GeV^2$& $0.33$ & $0.26$ & $0.17$ & $0.38$ & $0.29$ & $0.17$ \\
Our inputs, our $s_0$, no $g_+$ & $0.36$ & $0.28$ & $0.25$ & $0.30$ & $0.31$ & $0.23$ \\
Our inputs, our $s_0$, with $g_+$ & $0.26$ & $0.20$ & $0.14$ & $0.30$ & $0.22$ & $0.13$ \\
\bottomrule[0.7mm]
\end{tabular}
\caption{\it Deconstruction of the different effects explaining the difference between our results for the form factors at $q^2=0$ and those in~\Ref{Khodjamirian:2006st}. The difference stems mainly from the inclusion of the twist-four two-particle contributions. See the text for more details.}
\label{tab:deconstruction}
\end{table}

\subsection{Finite-width effects in $B\to K^*$ form factors}
\label{sec:finitewidtheffects}

The LCSRs in the form of~\Eq{eq:LCSRszpar} imply that, in the one-resonance approximation, each $B\to K^*$ form factor normalized to its narrow-width limit is a constant that does not depend on the form factor type. 
To see this we consider the LCSRs~(\ref{eq:LCSRsmodel}) with a single $K^*$:
\eq{
\F_{K^*,i}^{(T)}(q^2)
\ d_{K^*,i}^{(T)}\ I_{K^*}(s_0,M^2)  
= \P_i^{(T)\rm OPE}(q^2,\sigma_0,M^2)\ .
}
The key observation is that the only quantity here that depends on both $i$ and the width $\Gamma_{K^*}$ is the form factor itself $\F_{K^*,i}^{(T)}(q^2)$. The other quantity that depends on $\Gamma_{K^*}$
is the function $I_{K^*}(s_0,M^2)$, which is universal for all form factors and independent of $q^2$.
Therefore, defining the ratio $\W_{K^*}$ of any of the form factors to its NWL, we find:
\eq{
\label{eq:R}
\W_{K^*} \equiv \frac{\F_{K^*,i}^{(T)}(q^2)}{\F_{K^*,i}^{(T)}(q^2)_{\rm NWL}}
= \frac{I_{K^*}(s_0,M^2)\big|_{\Gamma_{K^*}\to 0}}{I_{K^*}(s_0,M^2)}
= \frac{2 m_{K^*} f_{K^*} e^{-m_{K^*}^2/M^2}}{I_{K^*}(s_0,M^2)}\ ,
}
where the NWL of the numerator has been performed as described in~\Sec{sec:NWL}.
Assuming $K^*$ dominance, the quantities $\F_{K^*,i}^{(T)}(q^2)_{\rm NWL}$ are precisely the ones determined in the previous subsection, and thus the form factors $\F_{K^*,i}^{(T)}(q^2)$ can be obtained by multiplying the results in Tables~\ref{tab:ResultsFFs} and \ref{tab:FFcomparison} by $\W_{K^*}$.

\begin{figure}
\begin{center}
\includegraphics[width=11cm]{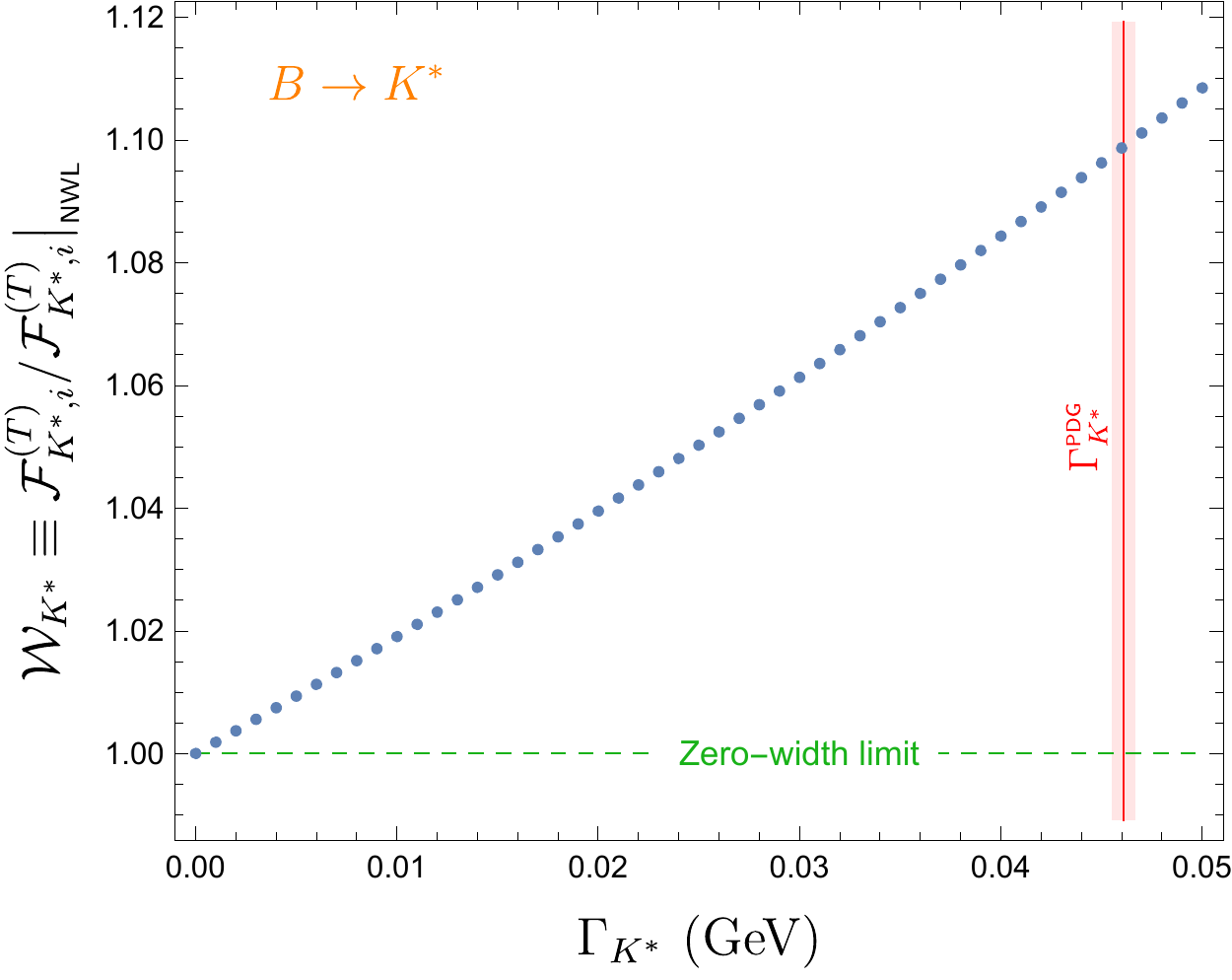}
\end{center}
\caption{\it Ratio $\W_{K^*}$ quantifying the finite-width correction to the $B\to K^*$ form factors, as a function of the $K^*$ width $\Gamma_{K^*}$, for $M^2=1\GeV^2$. The red vertical band corresponds to the physical width (PDG). This finite-width correction is universal and $q^2$-independent.}
\label{fig:R}
\end{figure}

The ratio $\W_{K^*}$ as a function of the $K^*$ width $\Gamma_{K^*}$ is shown in~\Fig{fig:R}. We can see that for $\Gamma_{K^*}\to 0$ we recover smoothly the NWL $\W_{K^*}\to 1$ (as discussed above for the form factors themselves). The dependence is very approximately linear:
\eq{
\W_{K^*} \simeq 1+ 1.9\,\frac{\Gamma_{K^*}}{m_{K^*}}\ ,
}
with a coefficient ($\approx 2$) of order one multiplying the expected
$\Gamma_{K^*}/m_{K^*}$ correction.
For the measured width $\Gamma_{K^*} \simeq 46\MeV$ (see \Tab{tab:parameters}) we find that the finite-width correction in $B\to K^*$ form factors is of order of $\sim 10\%$, which is similar to the corresponding corrections to the $B\to \rho$ form factors investigated in~\Ref{Cheng:2017smj}. More precisely, from~\Eq{eq:R} and the numerical inputs in~\Tab{tab:parameters} we find:
\eq{
\W_{K^*} = 1.09 \pm 0.01\ .
\label{eq:RKsresult}
}
We have used the Model 1 for $f_{K^*(892)} = 206\MeV$, since in this model the $\tau$ data is consistent with the absence of a $K^*(1410)$ resonance (see \Sec{sec:taudata}). This number is an average of the results $\W_{K^*}=\{1.099(16),1.091(9),1.085(7) \}$ for $M^2=\{ 1.00,1.25,1.50 \}\GeV^2$, where the corresponding values for $s_0$ in~\Tab{tab:EffectiveThreshold} -- for Model 1 -- have been used. The parametric uncertainties in $\W_{K^*}$ are negligible compared to those arising from the uncertainty in $s_0$.
The variation of $s_0$ and $M^2$ in $\W_{K^*}$ will be correlated with the one in the calculation of the form factors,
and therefore the separate determinations of $\W_{K^*}$ for different values of $M^2$ may result in more accurate estimates of the finite-width effect. Nevertheless, the three values quoted for $\W_{K^*}$ are consistent among themselves within errors, and the average in~\Eq{eq:RKsresult} is meaningful.

The robustness of the previous results is also supported by the expansion at $\op(\Gamma)$ discussed at the end of \Sec{sec:NWL}. From \Eq{eq:beyondNWL3resonancessumrules}, we see that
\eq{
\W_{K^*}=1-\frac{\Gamma_{K^*}}{m_{K^*}}\, \Delta_{K^*}(s_0,M^2)+\op(\Gamma_{K^*}^2)\ .
}
This linearised expression depends only on the mass and the width of the resonance as well as the sum rule parameters $s_0$ and $M^2$, and it is thus less dependent on the details of the hadronic model used. With the same inputs as above, we obtain 
\eq{
\W_{K^*} = 1.08 \pm 0.01 \qquad [{\rm linearised}]\ .
}
The range is obtained by varying the sum rule parameters in the same way as in \Eq{eq:RKsresult}. The slight difference 
with the central value of \Eq{eq:RKsresult} can be attributed to higher orders in the expansion in powers of $\Gamma_{K^*}$, indicated by the slight curvature of the function $\W_{K^*}(\Gamma_{K^*})$ shown in~\Fig{fig:R}~\footnote{Interestingly, a similar effect (in size and direction) was found in \Ref{Khodjamirian:1997tk} in the case of
the form factor for the $\gamma^*\rho\to\pi$ transition. Indeed, taking into account the measured width of the $\rho$-meson leads to an increase by 12\% of the form factor compared to the LCSR prediction in the narrow-width limit.}.

The fact that the ratio $\W_{K^*}$ is independent of the form factor helicity has two consequences, which will be discussed further in~\Sec{sec:applications}. First, corrections to the NWL in branching fractions of exclusive $B\to K^*$ observables will be proportional to $|\W_{K^*}|^2\simeq 1.20$. This $20\%$ increase in the theory predictions with respect to the narrow-width limit is very relevant phenomenologically in view of the systematically low experimental determinations of branching ratios in $b\to s\mu\mu$ modes reported by the LHCb collaboration (see for instance \Refs{Descotes-Genon:2015uva,Capdevila:2017bsm,Alguero:2019ptt,Bifani:2018zmi} and references therein). The correction to the NWL discussed here would tend to increase the discrepancy between the SM predictions and the LHCb measurements. Second, normalized observables such as $P'_5$~\cite{DescotesGenon:2012zf,Descotes-Genon:2013vna}, which depend only on ratios of form factors, are insensitive to finite-width corrections. Technically, these considerations apply only to {\it factorizable} decay modes; it remains to be determined if non-local contributions to $b\to s\ell\ell$ also have this property.

\subsection{Beyond the $K^*$ window and the $K^*(1410)$ contribution}
\label{sec:K*1410effects}

It is evident from~\Eq{eq:fitrel} that the LCSRs can only constrain one particular combination of the $K^*(892)$ and $K^*(1410)$ contributions. As mentioned earlier, this is due to the fact that the sum rules only depend on an integral of the $B\to K\pi$ form factors over the $K\pi$ invariant mass, weighted by the $K\pi$ form factor $f_+(s)$. Since this form factor is peaked strongly around the $K^*(892)$ resonance, the LCSRs are mostly sensitive to the $B\to K\pi$ form factors in the region $s\simeq m_{K^*} \pm \Gamma_{K^*}$.
This is the reason why the traditional LCSRs for $B\to K^*$ form factors work well. But this also means that the $K^*(1410)$ and other ``non-resonant" contributions will be only weakly constrained by the LCSRs.
In addition, the two models for $f_+(s)$ considered in~\Sec{sec:taudata} (consistent with the $\tau$ decay), will presumably provide a slightly different sensitivity to the $K^*(1410)$ contribution, as the form factor differs by a factor of two on the vicinity of this resonance.

More quantitatively, the contributions from $K^*(892)$ and $K^*(1410)$ to the sum rules~(\ref{eq:fitrel}) are proportional to the factors $I_{K^*(892)}$ and $I_{K^*(1410)}$. The numerical values for these factors are collected in~\Tab{tab:IR}, where one can see that $I_{K^*(1410)}/I_{K^*(892)} \simeq 0.03$. Thus, for the $K^*(1410)$ to have a significant weight in the sum rule for a given form factor, $\F_{K^*(1410)}$ must be at least an order or magnitude larger than  $\F_{K^*(892)}$.

\begin{table}
\centering
\setlength{\tabcolsep}{10pt}
\begin{tabular}{@{}ccccc@{}}
\toprule[0.7mm]
 && $\ M^2=1.00\GeV^2$ & $\ M^2=1.25\GeV^2$ & $\ M^2=1.50\GeV^2$ \\
\midrule[0.7mm]
\multirow{2}{*}{Model 1}
& $I_{K^*(892)}$  & $0.1506(23)$ & $0.1781(16)$ & $0.1992(13)$ \\
& $I_{K^*(1410)}$ & $0.0050(07)$ & $0.0062(07)$ & $0.0072(06)$ \\ 
\midrule
\multirow{2}{*}{Model 2}
& $I_{K^*(892)}$  & $0.1491(22)$ & $0.1766(20)$ & $0.1975(16)$ \\
& $I_{K^*(1410)}$ & $0.0048(07)$ & $0.0061(06)$ & $0.0070(06)$ \\
\bottomrule[0.7mm]
\end{tabular}
\caption{\it Values for the quantities $I_R$ for $R=\{K^*(892),K^*(1410)\}$ for the different values of the Borel parameter $M^2$ and for the two models for the $K\pi$ form factor. The $K^*(1410)$ contribution is very suppressed in the sum rules, with $I_{K^*(1410)}/I_{K^*(892)} \simeq 0.03$ in all cases.}
\label{tab:IR}
\end{table}

This issue was also pointed out in~\Ref{Cheng:2017smj} where various alternative assumptions were adopted in order to estimate the $\rho'$ and $\rho''$ contributions to $B\to \pi\pi$ form factors. As a first approach, the LCSRs with $\rho$-meson LCDAs were used to fix the $\rho$ contribution, and thus the $\rho'$ contribution could be estimated from our LCSRs for $B\to \pi\pi$ form factors. This assumed that the LCSRs with $\rho$-meson LCDAs are insensitive to the presence of the $\rho'$~\cite{Straub:2015ica}. The corresponding results for $B\to \rho'$ form factors were relatively imprecise, given the insensitivity of the LCSRs to the region outside the $\rho$ window. As a second approach, it was assumed that the relative contribution from each resonance is the same in the $B\to \pi\pi$ form factors as in the time-like pion form factor.
This is a bolder assumption but provides relatively precise predictions. One may see this as a model which is consistent with the LCSRs.

\begin{figure}[t]
\begin{center}
\includegraphics[width=15cm]{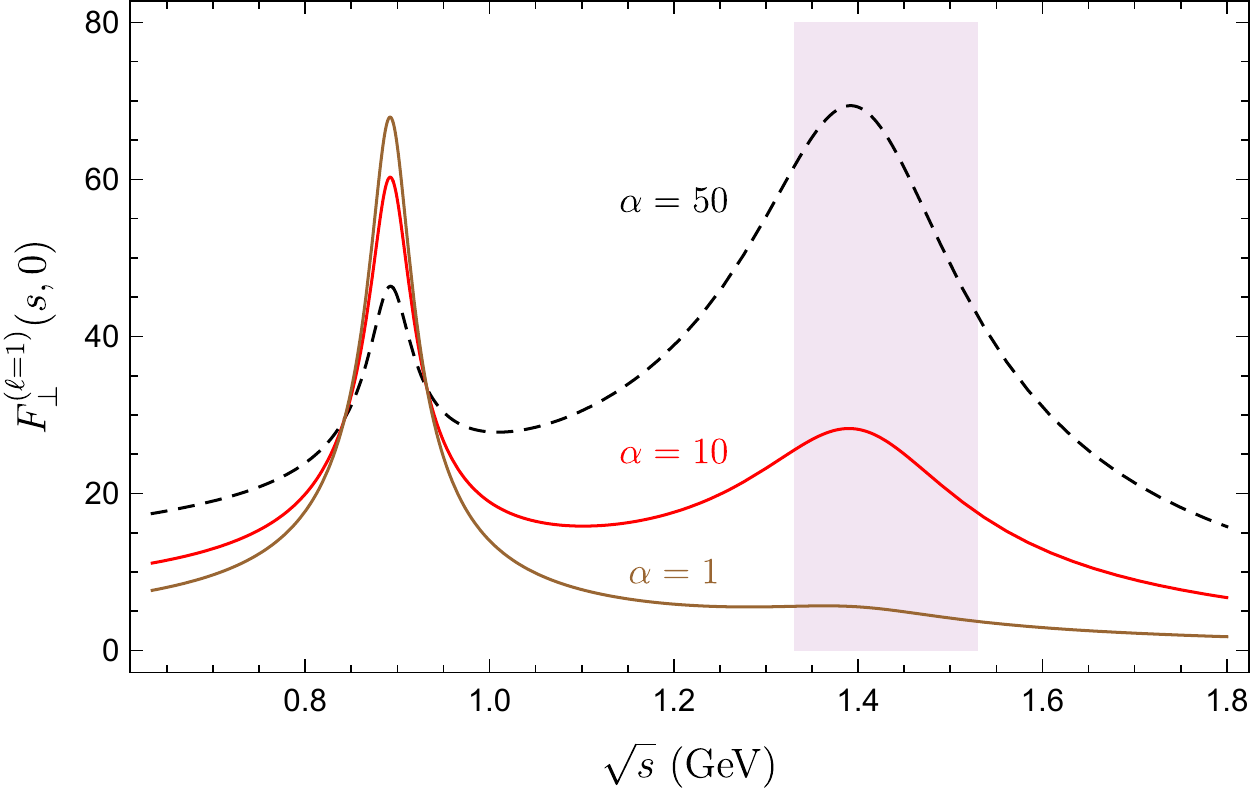}
\end{center}
\caption{\it Form factor $F^{(\ell=1)}_\perp(s,0)$ at $q^2=0$, as a function of the $K\pi$ invariant mass, for three different values of the parameter $\alpha$, describing the relative size of $K^*(892)$ and $K^*(1410)$ contributions. All the curves are consistent with the Light-Cone Sum Rule. The vertical band indicates the region of the measurements in~\Ref{Aaij:2016kqt}.}
\label{fig:Fperp}
\end{figure}

As a more pragmatic and model-independent alternative, one may attempt to constrain the $B\to K\pi$ form factors in the $K^*(1410)$ region from data. Once this is done, our LCSRs can be used to determine the $B\to K^*$ form factors in a way which takes into account the contributions beyond the $K^*$ window.
Note that this data-driven determination of the form factors in the $K^*(1410)$ region needs not be very precise. In order to have a significant impact on the $K^*$ region, these would need to be huge, but this is a possibility that has not been discarded.

In order to illustrate this point, we consider the form factors in the form of~\Eq{eq:FFmodelsFinal}, and set $\F_{K^*(1410)} = \alpha\,\F_{K^*(892)}$ with $\alpha$ a floating parameter. For different values of $\alpha$, we can use the sum rules~(\ref{eq:fitrel}) to fix the parameters $\F^{(T)}_{K^*,i}$ and $b^{(T)}_{K^*,i}$ as in~\Sec{sec:BtoK*NWL}, and plug these results into~\Eq{eq:FFmodelsFinal} to predict the $B\to K\pi$ form factors $F^{(T)(\ell=1)}_i(s,q^2)$. In~\Fig{fig:Fperp} we show the outcome of this exercise for the form factor $F^{(\ell=1)}_\perp(s,0)$, choosing the values $\alpha=\{1,10,50\}$. One can see that for $\alpha=1$, the presence of the $K^*(1410)$ is barely noticeable, but for $\alpha=50$ it dominates the form factor. These two extremes are perfectly allowed by the LCSRs. But there is a competition between both contributions. Higher values of $\alpha$ suppress the $B\to K^*$ form factor in order to maintain the sum rule constraint~\footnote{Note that the value $0.28$ corresponding to $\alpha=1$ is equal to the NWL result in~\Tab{tab:FFcomparison} corrected by $\W_{K^*}$.}:
\eq{
\alpha=1:\ \F_{K^*,\perp}(0)=0.28\ ; \quad
\alpha=10:\ \F_{K^*,\perp}(0)=0.22\ ; \quad
\alpha=50:\ \F_{K^*,\perp}(0)=0.11\ .
}
Therefore, a suppression of the $B\to K^*$ form factors favoured by $B\to K^*\mu\mu$ data could be the result of a very large $B\to K^*(1410)$ form factor, being all consistent with the LCSRs. However, this would at the same time produce a huge enhancement of the $B\to K\pi\mu\mu$ rate around $\sqrt{s}=1.4\GeV$, as can be seen in~\Fig{fig:Fperp}, which would impact significantly the measurements in this region performed by the LHCb collaboration~\cite{Aaij:2016kqt}. Thus these measurements can be used to constrain the predictions for $B\to K^*$ form factors. We shall discuss this in more detail in~\Sec{sec:moments}.

As a side note, we point out that the exact interplay between both contributions depends on the relative phase $\phi_R$ in the form factor. \Fig{fig:Fperp} and the above discussion correspond to the phase choice in~\Eq{eq:phasecondition}. All in all, it would also be important to study these phases in more depth.


\section{Applications to rare decays}
\label{sec:applications}

We have derived the sum rules for the $B\to K\pi$ form factors and we have determined the constraints set on models based on a series of resonances. We want now to exploit these sum rules for $B\to K\pi\ell\ell$. We consider first a toy example to illustrate the connection between the general $K\pi$ case and the narrow-width $K^*$ case at the level of the differential decay rates. Then we consider the $B\to K\pi\ell\ell$ decay: we derive the expression of the differential decay rate using the $B\to K\pi$ form factors, we discuss the connection with the narrow-width limit around the $K^*$ peak, and we exploit experimental information obtained for a $K\pi$ invariant mass around the $K^*(1410)$ in order to further constrain our model for the $B\to K\pi$ form factors.

\subsection{A toy example}\label{sec:toyexample}

We start with a toy example that captures the essence of the generalization beyond the narrow-width limit at the level of decay rates. We consider a new scalar particle $\Phi$ with mass-squared $m_\Phi^2=q^2$ that couples to the pseudoscalar current $\bar s \gamma_5 b$:
\eq{
\cL_{sb\Phi} = -g \,\bar s \gamma_5 b\, \Phi + {\rm h.c.}\ ,
}
and study the decay $B\to \Phi K^-\pi^+$. The amplitude of the process to leading order in $g$ is
\eq{
i\A = -\frac{g\,\sqrt{q^2}}{m_b+m_s} F_t(k^2,q^2,q\cdot \bar k)\ ,
}
and the differential decay rate is given by
\eq{
\frac{d\Gamma}{dk^2\, d\cos\theta_K} = \frac{1}{(2\pi)^3 32 m_B^3} \frac{\sqrt{\lambda \lambda_{K\pi}}}{2k^2} |\A|^2\ .
}
Expanding the form factor in partial waves, the squared amplitude is
\eq{
|\A|^2 = \frac{g^2\,q^2}{(m_b+m_s)^2}
\sum_{\ell,m} \sqrt{(2\ell+1)(2m+1)} F_t^{(\ell)}(k^2,q^2) F_t^{(m)*}(k^2,q^2)
P_\ell^{(0)}(\cos\theta_K) P_m^{(0)}(\cos\theta_K)\ .
}
Therefore, integrating over the angle $\theta_K$ and using the orthogonality of Legendre polynomials we find
\eq{
\frac{d\Gamma}{dk^2} = \frac{1}{(2\pi)^3 32 m_B^3}
\frac{g^2 q^2 \sqrt{\lambda \lambda_{K\pi}}}{(m_b+m_s)^2 k^2} \sum_{\ell=0}^\infty |F_t^{(\ell)}(k^2,q^2)|^2\ .
}
We now consider the $K^*$ contribution to this decay rate, which means taking only the $\ell=1$ term in the sum, and using the parametrization of~\Eq{eq:FFmodels} with only one resonance:
\eq{
|F_t^{(\ell=1)}|^2 = \frac{32 \pi^2 s \lambda}{3q^2 \lambda_{K\pi}^{1/2}}\,|\F_{K^*,t}(q^2)|^2\, \Delta(s,m_{K^*})\ ;\quad
\Delta(s,m_{K^*})\equiv \frac1{\pi}\frac{\sqrt{s}\,\Gamma_{K^*}(s)}{(m_{K^*}^2-s)^2+s\Gamma_{K^*}(s)}\ .
\label{eq:FtmodelDelta}
}
The function $\Delta(s,m_{K^*})$ has an integral equal to $1$ and goes to $\delta(s-m_{K^*})$ in the narrow-width limit. We use the notation $s=k^2$. Thus, if we integrate the decay rate around a window that contains the $K^*$ resonance, we find~\footnote{Here we assume that the prefactor multiplying $\Delta(s,m_{K^*})$ in~\Eq{eq:FtmodelDelta} varies slowly in the resonance region. This is certainly true if the width is small. A more careful description of this effect is given in~\App{app:beyondNWL}.}:
\eq{
\Gamma(B\to \Phi K^*[\to K^-\pi^+]) = \frac{g^2 \lambda^{3/2}(m_K^*)}{24\pi m_B^3 (m_b+m_s)^2}  |\F_{K^*,t}(q^2)|^2\ .
\label{eq:GammaPhiKpi}
}

We need to compare this result with its narrow-width approximation.
In this case the amplitude would be:
\eq{
i \A_{K^*} = -\frac{2 g\, m_{K^*}\, \epsilon^*_\eta \cdot q}{m_b+m_s} A^{BK^*}_0(q^2)\ ,
}
where $\eta$ is the polarization of the $K^*$ (with $\epsilon^{*\mu}_\eta$ its polarization vector), and $A^{BK^*}_0$ is the timelike-helicity $B\to K^*$ form factor (see~\Eq{eq:BVFFdef}).
Squaring the amplitude and summing over polarizations:
\eq{
\sum_\eta |\A_{K^*}|^2 = \frac{g^2\, m^2_{K^*}\, \lambda(m_{K^*})}{(m_b+m_s)^2 k^2}\, |A^{BK^*}_0(q^2)|^2\ ,
}
where we have used that $\sum_\eta 4 q_\mu q_\nu \epsilon_\eta^\mu \epsilon_\eta^{*\nu} = \lambda(m_{K^*})/k^2$. The decay rate is then:
\eq{
\Gamma(B\to \Phi K^*) = \frac{g^2 \lambda^{3/2}(m_K^*)}{16\pi m_B^3 (m_b+m_s)^2}  |A^{BK^*}_0(q^2)|^2\ .
\label{eq:GammaPhiK*}
}
Since the $K^*$ decays with probability 2/3 to $K^-\pi^+$, \Eqs{eq:GammaPhiKpi} and~(\ref{eq:GammaPhiK*}) coincide if we identify $\F_{K^*,t}(q^2) = A^{BK^*}_0(q^2)$.
Thus, when we integrate the decay rate around a resonance in a region wide enough to contain it, the finite-width-corrected result is obtained multiplying the rate by the squared of the ratio $\W$ discussed in~\Sec{sec:finitewidtheffects}. In the case of the $K^*$, where $\W_{K^*}=1.09$, the impact is $20\%$:
\eq{
\Gamma(B\to \Phi K^*[\to K^-\pi^+]) = 1.2\times \Gamma(B\to \Phi K^*[\to K^-\pi^+])_{\rm NWL}\ . 
}
As a final note, since the ratio $\W_K^*$ is independent of the form factor helicity (see~\Sec{sec:finitewidtheffects}), the correction $\W_{K^*}^2 = 1.19\,$ factorizes in the decay rate of any (factorizable) decay mode, even if the amplitude depends on all 7 form factors. Such is the case in $B\to K^*\nu\bar\nu$ or $B_s\to K^* \ell\nu$, and in the factorizable part of more complicated decay modes such as the non-leptonic decay $B\to M K^*$, and the rare decay $B\to K^*\ell\ell$ discussed in the following section.

\subsection{Angular distribution of the non-resonant $B\to K\pi\,\ell \ell$ decay}\label{sec:nonresonantdecay}

After the study of this toy example, we can move to the more realistic case of the rare decay $B\to K\pi\ell\ell$.
The amplitude $\A \equiv \A(\bar B^0 \to K^-(k_1) \pi^+(k_2) \ell^-(q_1) \ell^+(q_2) )$ in the SM is given by \cite{Altmannshofer:2008dz,Khodjamirian:2010vf,Bobeth:2017vxj}:
\eq{
i\A = g_F \frac{\alpha}{4\pi} \bigg[ (C_9 \,L_{V\mu} + C_{10} \,L_{A\mu})\  \F_L^\mu 
+  \frac{L_{V\mu}}{q^2} \Big\{  2 m_b C_7\, \F^{T\mu}_R  -i\, 16\pi^2 \,\H^\mu \Big\}   \bigg]
\label{AmplitudeBKpill}
}
with $q=q_1+q_2$, $g_F \equiv 4G_F/\sqrt{2} \,V^*_{ts} V_{tb}$, $L_{V(A)}^\mu \equiv \bar u_\ell(q_1) \gamma^\mu(\gamma_5) v_\ell(q_2)$,
and the local and non-local hadronic matrix elements:
\eqa{
\F_L^\mu &\equiv& i\, \langle K^-(k_1) \pi^+(k_2)|\bar{s}\gamma^\mu P_L\, b|\bar{B}^0(p)\rangle
= \frac12 \big( F_\perp \, k_\perp^\mu + F_\| \, k_\|^\mu + F_0 \, k_0^\mu +F_t \, k_t^\mu \big) \ , \\
\F^{T\mu}_R &\equiv& \langle K^-(k_1) \pi^+(k_2)|\bar{s}\sigma^{\mu\nu} q_\nu P_R\, b|\bar{B}^0(p)\rangle
= \frac12 \big( F^T_\perp \, k_\perp^\mu + F^T_\| \, k_\|^\mu + F^T_0 \, k_0^\mu \big) \ , \\
\H^\mu  &\equiv& i \int dx\,e^{i\,q\cdot x } \langle K^-(k_1) \pi^+(k_2)| T\{ j_{\rm em}^\mu(x) \,\op_{\rm 4q}(0) \} |\bar{B}^0(p)\rangle
= \H_\perp  k_\perp^\mu + \H_\|  k_\|^\mu + \H_0 \, k_0^\mu \ ,\ \quad
}
with $p=q+k$. Besides the form factors $F^{(T)}_i$ discussed in this article, we have introduced here the functions $\H_i(k^2,q^2,q\cdot \bar k)$ describing the
non-local effects which appear when the lepton pair couples to the electromagnetic current, through a penguin contraction of four-quark $\op_{\rm 4q} \sim \bar s b \bar q q$ operators.
In complete analogy with the form factors in~Eqs.\,(\ref{eq:PWE0t}$-$\ref{eq:PWEperppar}), the functions $\H_i(k^2,q^2,q\cdot \bar k)$ can be expanded in partial waves,
resulting in the corresponding functions $\H^{(\ell)}_i(k^2,q^2)$.

The Lorentz decomposition and the structure of the leptonic currents define the different
{\it transversity amplitudes} $\A^{L,R}_i$ by:
\eq{
i \A = \frac{\alpha \,g_F}{8\pi \N} \bigg\{
L_\mu \big( \A^L_\perp \, k_\perp^\mu + \A^L_\| \, k_\|^\mu + \A^L_0 \, k_0^\mu + \A^L_t \, k_t^\mu \big) +
R_\mu \big( \A^R_\perp \, k_\perp^\mu + \A^R_\| \, k_\|^\mu + \A^R_0 \, k_0^\mu + \A^R_t \, k_t^\mu \big) 
\bigg\} \ ,
\label{eq:iA1}
}
where $L^\mu \equiv \bar u_\ell(q_1) \,\gamma^\mu P_L\, v_\ell(q_2)$ and $R^\mu \equiv \bar u_\ell(q_1) \,\gamma^\mu P_R\, v_\ell(q_2)$,
with $P_{L,R} = (1\mp \gamma_5)/2$, and $\N$ is a normalization constant which is introduced for convenience and will be fixed later.
Comparing with \Eq{AmplitudeBKpill} one sees that
\eq{
\label{eq:calABtoKll}
\A_i^{L,R} = \N \bigg[
(C_9 \mp C_{10}) F_i + \frac{2m_b}{q^2} \Big\{  C_7 F_i^T - i \frac{16\pi^2}{m_b} \H_i  \Big\}
\bigg]\ ,
\quad i=\{\perp,\|,0,t\}\ ,
}
keeping in mind that $\A_i^{L,R} \equiv \A_i^{L,R}(k^2,q^2,q\cdot \bar k)$, etc. For $\A_t^{L,R}$ only the first term exists ($F_t^T = \H_t = 0$).
In addition, when $m_1=m_2$ one has $L_\mu k_t^\mu = - R_\mu k_t^\mu$ and the timelike-helicity amplitude depends only on the combination
$\A_t\equiv \A_t^L-\A_t^R$, which is independent of $C_9$. However this is not true if the two leptons have different mass. The transversity amplitudes $\A_i^{L,R}$ in~\Eq{eq:calABtoKll} can be expanded in partial waves $\A_i^{L,R(\ell)}(k^2,q^2)$ in the same way as the form factors, i.e.~\Eqs{eq:PWE0t}-(\ref{eq:PWEperppar}).

We follow the same approach as in \Ref{Altmannshofer:2008dz},
exploiting the fact that we use the same definitions for the various angles (the link with the experimental kinematics from the LHCb experiment~\cite{Aaij:2013iag} is given by \Ref{Gratrex:2015hna}).
We consider the decay as the chain $B\to V^*(\to \ell\ell) K\pi$, and
we introduce the polarisations of the virtual intermediate gauge boson $V^*$ defined in the $B$-meson rest frame,
\eq{
\epsilon_\pm^\mu=(0,1,\mp i,0)/\sqrt{2}\quad 
\epsilon_0^\mu=(-q_z,0,0,-q_0)/\sqrt{q^2}\quad
\epsilon_t^\mu=(q_0,0,0,q_z)/\sqrt{q^2}\ ,
}
where $q^\mu=(q_0,0,0,q_z)$. We can then use the completeness relation for this basis of polarisation vectors to write
\eq{
\label{eq:Mpolarisations}
i \A=\frac{\alpha g_F}{8\pi {\mathcal {N}}}\sum_i \sum_\lambda g_{\lambda\lambda}
   (L_\lambda {\cal A}_i^L+R_\lambda {\cal A}_i^R) (\epsilon_\lambda^{*}\cdot k_i)
   =\frac{\alpha g_F}{8\pi {\mathcal {N}}}
      \sum_\lambda g_{\lambda\lambda} (L_\lambda H^L_\lambda + R_\lambda H^R_\lambda)\ ,
}
with $i=\{\perp,\|,0,t\}$, $\lambda=\{0,t,+,-\}$ and $g_{tt}=1$, $g_{00}=g_{++}=g_{--}=-1$.
We have also defined $L_\lambda = \epsilon_\lambda \cdot L$ and  $R_\lambda = \epsilon_\lambda \cdot R$, with $R^\mu, L^\mu$ given after~\Eq{eq:iA1}. The quantities $H_\lambda^{L,R}$ are called {\it helicity amplitudes}.

We can define transversity amplitudes similar to the $\bar{B}\to \bar{K}^*\ell^+\ell^-$ case, by considering the $B$-meson rest frame described in \App{app:kinematics}, and performing the partial-wave expansion of the various amplitudes up to the $P$ wave:
\begin{align}
H_+^{L,R}&=\sqrt{3}\,\frac{\widehat{A}_{\|}^{L,R}+\widehat{A}_\perp^{L,R}}{\sqrt{2}}(-\sin\theta_K) + \cdots\ ,
&
H_-^{L,R}&=\sqrt{3}\,\frac{\widehat{A}_{\|}^{L,R}-\widehat{A}_\perp^{L,R}}{\sqrt{2}}(-\sin\theta_K) + \cdots\ ,
\\[2mm]
H_0^{L,R}&=\sqrt{2} \sqrt{3}(\widehat{S}_0^{L,R}+\widehat{A}_0^{L,R} \cos\theta_K + \cdots)\ ,
&
H_t&=-\sqrt{2} \sqrt{3}(\widehat{S}_t+\widehat{A}_t \cos\theta_K + \cdots)\ ,
\end{align}
with $H_t \equiv H_t^L - H_t^R$. Here $\widehat{S}$ and $\widehat{A}$ denote $\ell_{K\pi}=0$ and $\ell_{K\pi}=1$ amplitudes, respectively, and the ellipsis indicates $D$ and higher partial waves.
The normalisations have been chosen to make the connection between ${\cal A}_i$ and $\widehat{A}_i$ amplitudes easier, 
taking into account the partial-wave decompositions and the powers of $\sin\theta_K$ stemming from $\epsilon(\lambda)\cdot k_{\|}$ and $\epsilon(\lambda)\cdot k_{\perp}$ in \Eq{eq:Mpolarisations}:
\begin{align}
\widehat{A}_{\perp}^{L,R} &= \frac{\sqrt{\lambda_{K\pi}}}{k^2}\A_{\perp}^{L,R(1)}\ ,
&
\widehat{A}_{\|}^{L,R} &= \frac{\sqrt{\lambda_{K\pi}}}{k^2}\A_{\|}^{L,R(1)}\ ,
\nonumber\\[2mm]
\widehat{A}_0^{L,R}&= -\A_0^{L,R(1)}/\sqrt{2}\ , 
&
\widehat{A}_t &= -\A_t^{(1)}/\sqrt{2}\ .
\label{eq:helampl}
\end{align}
The differential decay rate is given by
\begin{equation}
\frac{d\Gamma}{dq^2 dk^2 d\cos\theta_\ell d\cos\theta_K d\phi}
  =\frac{1}{2^{15}\pi^6m_B}
  \frac{\sqrt{\lambda\lambda_q\lambda_{K\pi}}}{m_B^2q^2k^2}
   \sum_{s_1,s_2} |\A|^2\ ,
\end{equation}
where $|\A|^2$ is a product of the hadronic amplitudes $\widehat{A}_i^{L,R}$ (known in terms of the form factors $F_i,F_i^{T}$) and the leptonic amplitudes $L_\lambda$ and $R_\lambda$ (which can be easily evaluated in the $B$-meson rest frame) and $\lambda_q \equiv \lambda(q^2,m_{\ell}^2,m_{\ell}^2)$.
We can then perform the summation over the spins of the outgoing leptons to obtain the final expression
\begin{equation} \label{eq:differentialdecayrate}
\frac{d\Gamma}{dq^2 dk^2 d\cos\theta_\ell d\cos\theta_K d\phi}
  =\frac{9}{32\pi}\bar{I}(q^2,k^2,\theta_\ell,\theta_K,\phi)\ .
\end{equation}
If we choose the normalisation
\begin{equation}
{\cal N}= \alpha G_F V_{tb} V_{ts}^* \sqrt{\frac{ \sqrt{\lambda\lambda_{K\pi}\lambda_q}}{3\cdot 2^{13}\pi^7 m_B^3\,k^2}}
\end{equation}
the expression of \Eq{eq:differentialdecayrate} is indeed very simple.
$\bar{I}$ is formally the same expression obtained in Eqs~(3.10) and (3.21) of \Ref{Altmannshofer:2008dz}  with the angular coefficients $\bar{I}_{1s,1c,2s,2c,3,4,5,6s,6c,7,8,9}$ given by Eqs.~(3.34)-(3.45) of the same reference. The main difference comes from the transversity amplitudes $A_i$, which are not be given by Eqs.~(3.28)-(3.31) of \Ref{Altmannshofer:2008dz} but should be replaced by the transversity amplitudes $\widehat{A}_i$ given in \Eq{eq:helampl}.

\subsection{Finite-width effects in $B\to K^*\ell\ell$}

Following the same arguments as in \Sec{sec:toyexample}, we expect the results of \Sec{sec:nonresonantdecay} to be 
compatible with \Ref{Altmannshofer:2008dz} if we assume that the $K^-\pi^+$ pair comes only from the decay of a narrow $K^*$. In order to prove this agreement, we can take the expressions of \Sec{sec:nonresonantdecay} and determine the
expressions of the amplitudes $\widehat{A}_i$ assuming that they are dominated by a narrow $K^*$ contribution.
We can connect $\widehat{A}_i$ to $\A_i$ using \Eq{eq:helampl}, express them in terms of $F_i^{(1)}$ and $F_i^{T(1)}$ using \Eq{eq:calABtoKll}, and describe the latter
using the model in \Eq{eq:FFmodels}. One can then use the narrow-width limit expression of ${\cal F}^{(T)}_{R,i}$ in terms of the $B\to K^*$ form factors described in \App{app:BtoVLCSRs}. 

The resulting expressions can be  related to the $B\to K^*$ amplitudes $A_i$ given in Eqs.~(3.28)-(3.31) of \Ref{Altmannshofer:2008dz}:
\eq{
\widehat{A}_i^{L,R}=
-\frac{1}{4\pi\sqrt{3}}
\frac{\lambda_{K\pi}^{3/4}}{m_R^2}\frac{g_{RK\pi}e^{i\phi_R(s)}}{m_R^2-s-i\sqrt{s}\Gamma_{K^*}(s)}
\left[A_i^{L,R}+{\cal O}(\Gamma_{K^*})\right]
}
where we have already considered the narrow-width limit for the form factors, but we have still to take this limit for the propagator.

Interferences between these amplitudes will thus become
\eq{
\widehat{A}_i^{L,R}(\widehat{A}_j^{L,R})^*
 = \frac{1}{48\pi^2}\frac{\lambda_{K\pi}^{3/2}}{m_R^4}
    \frac{g_{RK\pi}^2}{(m_R-s)^2+s\Gamma_R^2}
    A_i^{L,R}(A_j^{L,R})^* + {\cal O}(\Gamma_{K^*})\ .
}
In the narrow-width limit, the squared propagator becomes
\begin{equation}
\frac{1}{(k^2-m^2_{K^*})^2+(m_{K^*}\Gamma_{K^*})^2}\xrightarrow{\Gamma_{K^*}\to 0} \frac{\pi}{m_{K^*}\Gamma_{K^*}}\delta(k^2-m_{K^*}^2)
\end{equation}
whereas the width can be reexpressed using \Eqs{eq:Gamma(s)} and (\ref{eq:Gammatot})
\begin{equation}
\Gamma_{K^*}=\frac{g_{K^*K\pi}^2 \lambda_{K\pi}^{3/2}}{48\pi m_{K^*}^5} {\cal B}(K^*\to K^-\pi^+)
\end{equation}
so that we have
\eq{
\widehat{A}_i^{L,R}(\widehat{A}_j^{L,R})^*\xrightarrow{\Gamma_{K^*}\to 0}
    A_i^{L,R}(A_j^{L,R})^* \delta(k^2-m_{K^*}^2) {\cal B}(K^*\to K^-\pi^+)\ ,
}
proving that the narrow-width limit of the differential decay rate in \Eq{eq:differentialdecayrate} agrees with the results of \Ref{Altmannshofer:2008dz}.

As discussed in \Sec{sec:NWL}, the $\op(\Gamma)$ correction to the narrow-width limit can be determined quite easily. If we take finite-width effects into account, the sum-rule determination of the form factors entering 
the amplitudes $\widehat{A}_i$ are enhanced by a factor ${\cal W}_K^*$, leading to an enhancement of all angular coefficients $I_i$ by a factor $\simeq 1.2$
compared to the value obtained using the $B\to K^*$ form factors in the narrow-width limit.

\subsection{High $K\pi$-mass moments of the $B\to K\pi\ell\ell$ angular distribution}
\label{sec:moments}

We can combine the analysis of the $B\to K\pi\ell\ell$ angular distribution in terms of $B\to K\pi$ form factors with the sum rules derived in \Sec{sec:LCSRs} to constrain the $K^*(1410)$ contribution to the models in \Sec{sec:LCSRs@Work}, thanks to recent experimental measurements. Indeed, in~\Ref{Aaij:2016kqt}
the LHCb experiment has analysed 
the moments $\Gamma_i$ ($i=1\ldots 41$) of the angular distribution of $B\to K\pi\mu\mu$ in the region of $K\pi$ and dilepton invariant masses $\sqrt{k^2}\in [1.33,1.53]\GeV$ and $q^2\in [1.1,6]\GeV^2$, respectively.
This region of $K\pi$ masses contains contributions from $K^*$ resonances in the $S$, $P$ and $D$ waves, and the moments analysed in \Ref{Aaij:2016kqt} contain contributions from all partial waves, following the analysis in \Ref{Dey:2015rqa}. The corresponding expansion can be written as
\eq{
\frac{d\Gamma}{dq^2 dk^2 d\Omega}
=\frac{1}{4\pi} \sum_{i=1}^{41} f_i(\Omega)\tilde{\Gamma}_i(q^2,k^2) 
}
where $d\Omega=d\cos\theta_\ell\,d\cos\theta_K\,d\phi$. Since the decomposition takes into account the possibility of $S$, $P$ and $D$-wave contributions, it features many different angular structures $f_i(\Omega)$.

We can compare these results with our predictions using our $B\to K\pi$ form factors for the combinations of moments that depend only on the $P$-wave contributions. The normalisations chosen are such that
\begin{equation}\label{eq:BRBtoKpill}
 \frac{d\Gamma}{dq^2dk^2}=
    \tilde{\Gamma}_1=|\widehat{A}_{\|}^L|^2+|\widehat{A}_{\|}^R|^2+|\widehat{A}_{\perp}^L|^2+|\widehat{A}_{\perp}^R|^2+|\widehat{A}_{0}^L|^2+|\widehat{A}_{0}^R|^2+
    \ldots
\end{equation}
where the ellipsis denote other partial-waves.
The other moments can be obtained from Table~5 of \Ref{Aaij:2016kqt} in a similar way. 
The experimental value integrated over the ranges in $k^2$ and $q^2$ can be obtained from Table~3 of the same reference using $\tilde\Gamma_i=\bar\Gamma_i \tilde\Gamma_1$.

One should be careful that \Ref{Aaij:2016kqt} uses the same definition of the kinematics as in \Ref{Dey:2015rqa}, whereas we follow a prescription for the angles in agreement with \Ref{Altmannshofer:2008dz}: we have thus to perform the redefinition\,\footnote{As indicated in \Ref{Gratrex:2015hna},
the definitions in the LHCb analyses (see \Ref{Aaij:2015oid})  and the theoretical analyses (e.g. \Ref{Altmannshofer:2008dz}) for the decay of $B\to K^*(\to K\pi)\mu\mu$ can be related by the changes $\theta_\ell \to \pi-\theta_\ell$ and $\phi\to -\phi$. However, the LHCb analysis at higher invariant $K\pi$ mass \Ref{Aaij:2016kqt} uses a different convention for $\phi$ from the LHCb analysis of the $B\to K^*(\to K\pi)\mu\mu$ decay \Ref{Aaij:2015oid}, which explains that we only have to change the definition of $\theta_\ell$ here.}
\begin{equation}
    \theta_\ell \to \pi-\theta_\ell\ ,
\end{equation}
leading to a change of sign for $\Gamma_i$ for $i$ from 11 to 18 and 29 to 33 between our definition and the one used in \Ref{Aaij:2016kqt}.

We could try to compute the various moments $\Gamma_i$ using our model for the $B\to K\pi$ form factors. Although possible, this is probably not the best use that can be made of the LHCb measurements. Indeed, by construction, our sum rules involve the $K\pi$ form factor, which yields a better sensitivity to the low-energy resonances (and most prominently to the $K^*(892)$ resonance). As 
discussed in~\Sec{sec:K*1410effects}
the sensitivity to the parameters of the excited $K^*$ resonances is limited, which would lead to predictions for the moments $\Gamma_i$ with large uncertainties. On the contrary, one can think of using the LHCb measurements to constrain the parameters describing the contributions of the higher resonances. 

\begin{figure}
\begin{center}
\includegraphics[height=7.2cm,width=9.6cm]{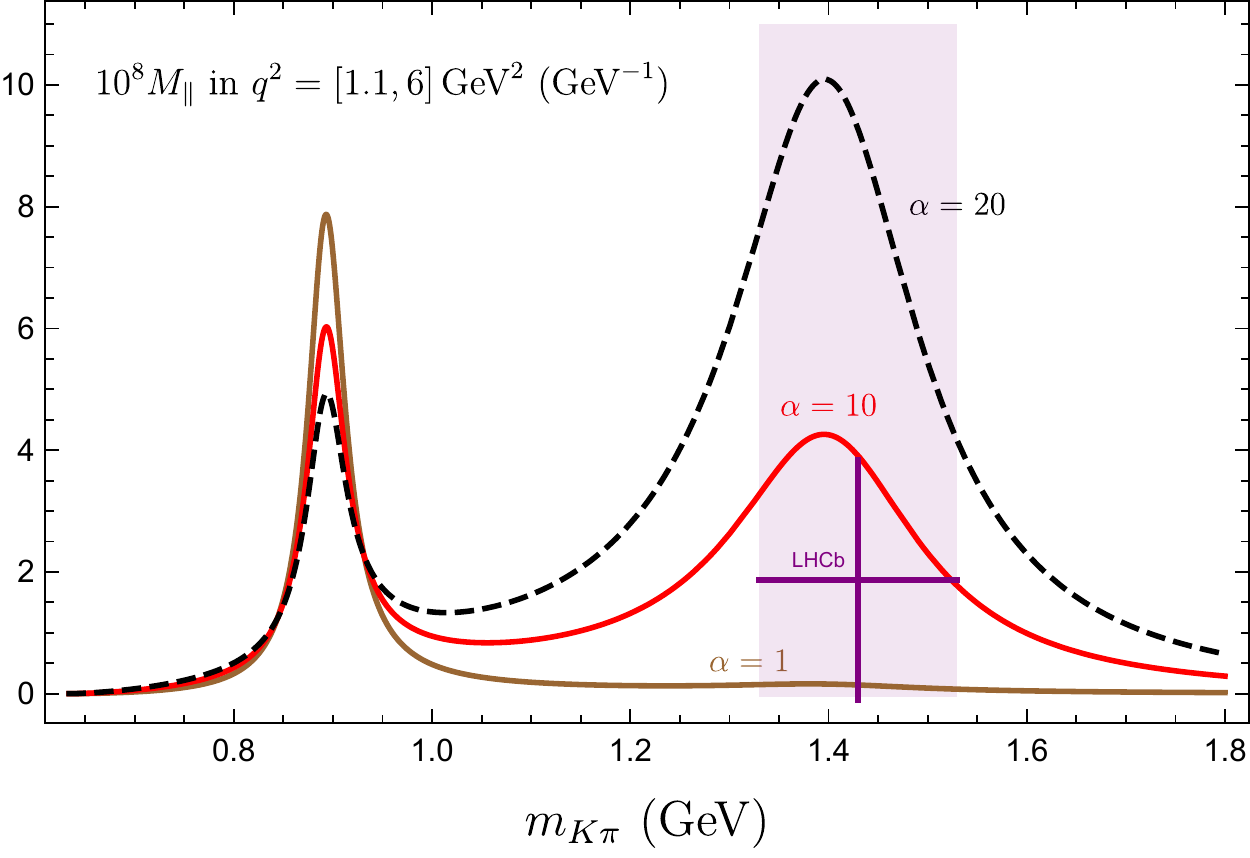}
\hspace{5mm}
\raisebox{0.5mm}{\includegraphics[height=7.16cm,width=6.4cm]{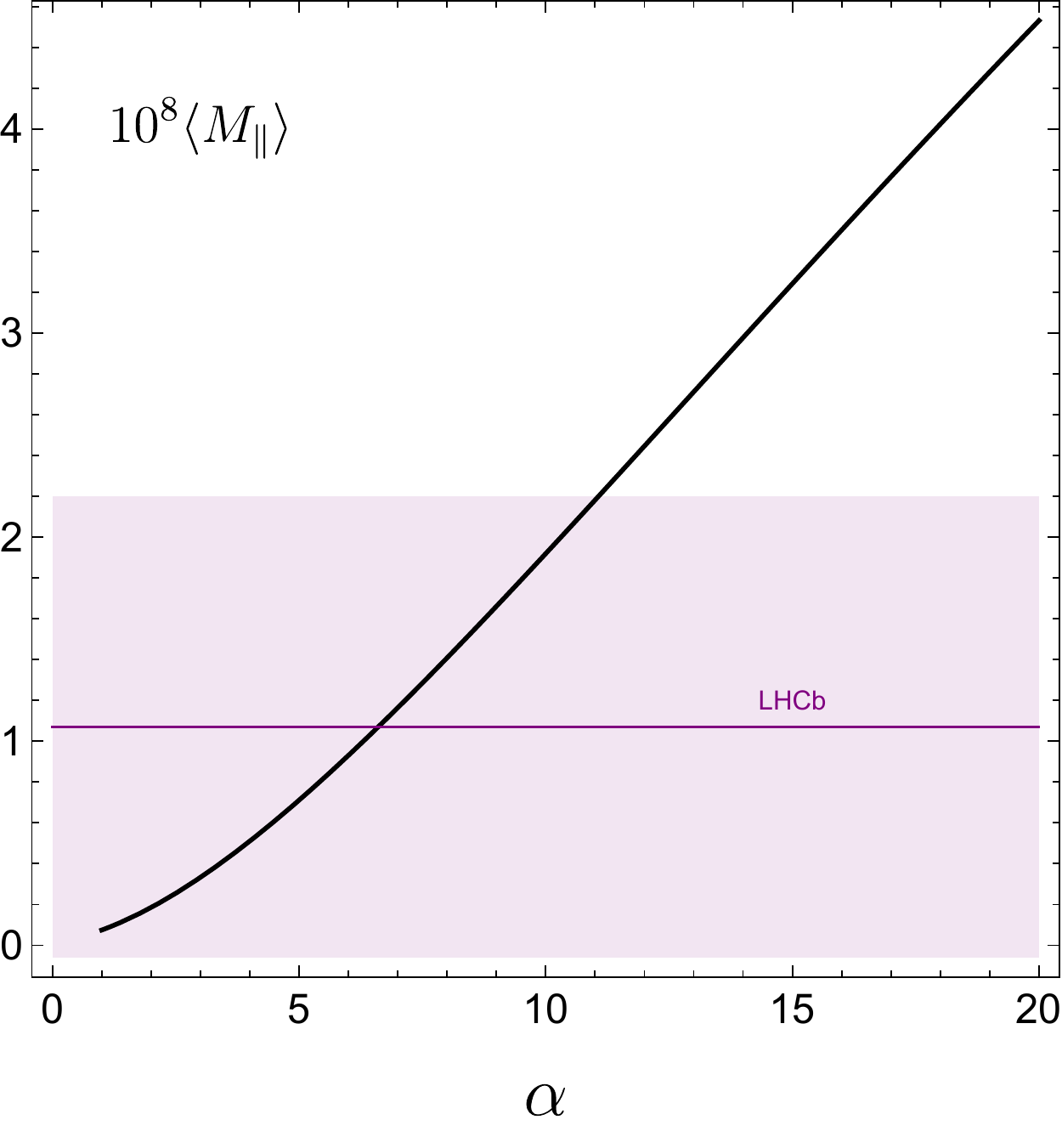}}\\
\vspace{-5mm}
\end{center}
\caption{\it Left: The moment $\av{M_\|}$ differentially in $m_{K\pi}$ for three values of $\alpha=\{1,10,20\}$, compared to the LHCb measurement in the bin $m_{K\pi}^2\in [1.33^2,1.53^2]\GeV^2$. For easy comparison, the LHCb binned measurement has been divided by the bin size.
Right: The integrated moment $\av{M_\|}$ as a function of $\alpha$, compared to the LHCb measurement (horizontal band).}
\label{fig:Mpar}
\end{figure}

In order to perform this analysis, we have to isolate combinations of the moments that are only dependent on the $P$-wave contributions. Using Appendix~A of \Ref{Aaij:2016kqt},
we find the following combinations free from $S$ and $D$-wave contributions
\begin{eqnarray}\label{eq:moments1}
|\widehat{A}_{\|}^L|^2+|\widehat{A}_{\|}^R|^2 &=&
\frac{1}{36} (5 \tilde\Gamma_{1} - 7 \sqrt{5} \tilde\Gamma_{3} + 5 \sqrt{5} \tilde\Gamma_{6} - 
   35 \tilde\Gamma_{8} - 5 \sqrt{15} \tilde\Gamma_{19} + 35 \sqrt{3} \tilde\Gamma_{21})\ ,
\quad\\
|\widehat{A}_{\perp}^L|^2+|\widehat{A}_{\perp}^R|^2 &=&
\frac{1}{36} (5 \tilde\Gamma_{1} - 7 \sqrt{5} \tilde\Gamma_{3} + 5 \sqrt{5} \tilde\Gamma_{6} - 
    35 \tilde\Gamma_{8} + 5 \sqrt{15} \tilde\Gamma_{19} - 35 \sqrt{3} \tilde\Gamma_{21})\ ,
\\
{\rm Im}(\widehat{A}_\perp^L \widehat{A}_{\|}^{L*}+\widehat{A}_\perp^R \widehat{A}_{\|}^{R*}) &=&
\frac{5}{36} (\sqrt{15} \tilde\Gamma_{24} - 7 \sqrt{3} \tilde\Gamma_{26})\ ,
\\
{\rm Re}(\widehat{A}_\perp^L \widehat{A}_{\|}^{L*}-\widehat{A}_\perp^R \widehat{A}_{\|}^{R*})
&=& \frac{1}{36} (-5 \sqrt{3} \tilde\Gamma_{29} + 7 \sqrt{15} \tilde\Gamma_{31})\ .
\label{eq:moments2}
\end{eqnarray}
There is some ambiguity in the previous expressions due to the following degeneracies among the moments\,\footnote{One has also the following degeneracies, of no use here:
\begin{equation}
0=\tilde\Gamma_2+\sqrt\frac{7}{3}\tilde\Gamma_4+\sqrt{5}\tilde\Gamma_7+\sqrt\frac{35}{3}\tilde\Gamma_9=
\tilde\Gamma_{20}+\sqrt{\frac{7}{3}}\tilde\Gamma_{22}=\tilde\Gamma_{25}+\sqrt{\frac{7}{3}}\tilde\Gamma_{27}=\tilde\Gamma_{30}+\sqrt{\frac{7}{3}}\tilde\Gamma_{32}
\end{equation}
The current LHCb data~\cite{Aaij:2016kqt} obeys all the degeneracy relations at 1.4 $\sigma$ or less (taking into account the experimental correlations).
}:
\begin{eqnarray}
0&=&\tilde\Gamma_1+\sqrt{5}\tilde\Gamma_3+3\tilde\Gamma_5+\sqrt{5}\tilde\Gamma_6+5\tilde\Gamma_8+3\sqrt{5}\tilde\Gamma_{10} \nonumber\\
&=&\tilde\Gamma_{19}+\sqrt{5}\tilde\Gamma_{21}+3\tilde\Gamma_{23}=\tilde\Gamma_{24}+\sqrt{5}\tilde\Gamma_{26}+3\tilde\Gamma_{28}=\tilde\Gamma_{29}+\sqrt{5}\tilde\Gamma_{31}+3\tilde\Gamma_{33}\ .
\end{eqnarray}
Using the experimental values and correlations of the moments,
we obtain from Eqs.~(\ref{eq:moments1}$-$\ref{eq:moments2}):
\begin{eqnarray}
\tau_B\, \av{|\widehat{A}_{\|}^L|^2+|\widehat{A}_{\|}^R|^2} & \equiv& \av{M_\|} =
(1.07\pm 1.13)\times 10^{-8}\ ,
\label{Mpar}\\
\tau_B\, \av{|\widehat{A}_{\perp}^L|^2+|\widehat{A}_{\perp}^R|^2} & \equiv& \av{M_\perp} =
(0.94 \pm 1.06)\times 10^{-8}\ ,
\\
\tau_B\, \av{{\rm Im}(\widehat{A}_\perp^L \widehat{A}_{\|}^{L*}+\widehat{A}_\perp^R \widehat{A}_{\|}^{R*})}
& \equiv& \av{M_{\rm im}} =
(-0.75 \pm 0.79)\times 10^{-8}\ ,
\\
\tau_B\, \av{{\rm Re}(\widehat{A}_\perp^L \widehat{A}_{\|}^{L*}-\widehat{A}_\perp^R \widehat{A}_{\|}^{R*})}
& \equiv& \av{M_{\rm re}} =
(0.27\pm 0.50)\times 10^{-8}\ ,
\label{Mre}
\end{eqnarray}
where $\langle X\rangle$ denotes the integration with respect to $q^2$ and $k^2$ over the experimental ranges, and $\tau_B\simeq 2.3\cdot 10^{12}\GeV^{-1}$ is the lifetime of the $B$ meson.

As an illustration we show in \Fig{fig:Mpar} the comparison between our predictions for the moment $M_\|$ and the LHCb measurement in~\Eq{Mpar}.
From these measurements we obtain the following bounds on the parameter $\alpha$:
\begin{align}
&\text{From}\ \av{M_\|}: \hspace{-4cm} & \alpha \lesssim 11\ ,\\
&\text{From}\ \av{M_\perp}: \hspace{-4cm} & \alpha \lesssim 17\ ,\\
&\text{From}\ \av{M_{\rm re}}: \hspace{-4cm} & \alpha \lesssim 18\ ,
\end{align}
by imposing that the corresponding observable is lower than the central value plus the uncertainty quoted in~Eqs.~(\ref{Mpar})-(\ref{Mre}).
From the measurement of $\av{M_{\rm im}}$ we do not obtain any meaningful bound since within our phase ansatz the product $\av{(\widehat{A}_\perp^L \widehat{A}_{\|}^{L*}+\widehat{A}_\perp^R \widehat{A}_{\|}^{R*})}$ is real.
For the non-local contributions $\H_\perp$ and $\H_\|$ in~\Eq{eq:calABtoKll} we have used the leading-order OPE approximation\,\footnote{The OPE coefficients at dimension three are known to NLO~\cite{Asatryan:2001zw,Asatrian:2019kbk}.}, where they are proportional to the form factors $F_\perp$ and $F_\|$, by absorbing them into the ``effective" Wilson coefficients $C_9^{\rm eff}$ and $C_7^{\rm eff}$~\cite{Beneke:2001at}.
For the values of these Wilson Coefficients we have used their SM values (see e.g. Table~1 of~\Ref{Matias:2012xw}), as the bounds given here on $\alpha$ should be understood as ballpark estimates. 
But in more refined analyses one should correlate the determination of the high-mass moments with the analysis of $B\to K^*\ell\ell$.

We can also use the branching ratio in~\Eq{eq:BRBtoKpill} directly to obtain an upper bound on the $P$-wave contribution, since the contributions from the other partial waves are necessarily positive. LHCb has measured this branching ratio within the considered $k^2$ bin and in several bins of $q^2$, and resulting in the following upper bounds on the parameter $\alpha$:
\eqa{
10^8\cdot\av{\B}_{[0.10,0.98]} &=& 1.41 \pm 0.27 \rightarrow\ \alpha\lesssim 5\ ,\\
10^8\cdot\av{\B}_{[1.10,2.50]} &=& 1.60 \pm 0.29 \rightarrow\ \alpha\lesssim 6\ ,\\
10^8\cdot\av{\B}_{[2.50,4.00]} &=& 1.37 \pm 0.26 \rightarrow\ \alpha\lesssim 5\ ,\\
10^8\cdot\av{\B}_{[4.00,6.00]} &=& 1.12 \pm 0.26 \rightarrow\ \alpha\lesssim 4\ ,\\
10^8\cdot\av{\B}_{[6.00,8.00]} &=& 0.98 \pm 0.23 \rightarrow\ \alpha\lesssim 3\ .
}
which are obtained as described above, and turn out to be somewhat stronger than the bounds from the angular moments.
Again, the SM has been assumed here.

The ballpark bounds obtained here, while still rather loose, already constrain the very large values for the $B\to K^*(1410)$ form factors that would affect significantly the LCSR predictions for the $B\to K^*$ form factors. 
In particular, for $\alpha \lesssim 3$ the reduction of the $B\to K^*$ form factors is at most of order $\sim 10\%$.
This simple analysis presented here should be refined in future studies, which together with more precise experimental measurements of the moments will provide more solid and stringent constraints on both the $B\to K^*(1410)$ and the $B\to K^*$ form factors.
Moreover, further generalization of the sum rules to the $S$-wave $K\pi$ system~\cite{WIPSwave} will make it possible to include other combination of moments beyond those considered here, and depending on the $S$-wave amplitudes.

\section{Conclusions and Perspectives}
\label{sec:conclusions}

Accurate theoretical predictions for exclusive $B$-meson decay observables are of utmost importance for studies of the the Standard Model and New Physics, but require a careful assessment of the theoretical inputs used. LCSRs are particularly prominent tools to compute the form factors involved. In this article, we  focused on the determination of $B\to K^*$ form factors from LCSRs with $B$-meson LCDAs. We  extended the framework to consider $B\to K\pi$ form factors for the $K\pi$ $P$ wave, which include effects such as the non-resonant $K\pi$ production and the width of the $K^*$ meson. We  analysed all vector, axial and tensor form factors needed for the phenomenological analysis of $B\to K\pi\ell\ell$ in the $P$ wave in the Standard Model.

We first derived light-cone sum rules with $B$-meson LCDAs for these form factors, generalising the results of \Ref{Cheng:2017sfk} for the case of two pseudoscalar mesons of different masses in the final state.
These sum rules provide relationships between
the convolution of the $K\pi$ vector form factor and a  $B\to K\pi$ form factor on one side, and the OPE of a well-chosen correlation function expressed in terms of $B$-meson LCDAs on the other side.
On the OPE side of the sum rules, we computed higher-twist two- and three-particle contributions (see also \Ref{Gubernari:2018wyi}), and we proceeded to a new,
well motivated, determination of the  threshold parameter $s_0$, somewhat lower than earlier determinations.
On the hadronic side,
we  introduced a resonance model for the $B\to K\pi$ including the $K^*(980)$ and the $K^*(1410)$, so that our sum rules can be used to constrain the parameters describing the two contributions. This resonance model is related to the model used by the Belle collaboration to determine the $K\pi$ vector form factor from the $\tau\to K\pi\nu$ differential decay rate, and which describes the spectrum very well.

We  then exploited our set-up phenomenologically with the following results: 

\begin{itemize}

\item We  considered the narrow-width limit of our LCSRs to check that we recover the known $B\to K^*$ sum rules, and to assess the correction due to the finite width of the $K^*$ meson. We find that this correction is universal for all form factors and amounts to a multiplicative factor ${\cal W}_K^*\simeq 1.1$, corresponding to a 20\,\% enhancement of the decay rate. This result does not depend strongly on the details of our model and it should be taken into account when computing observables with $B\to K^*$ form factors obtained from LCSRs derived in the narrow-width limit.

\item Comparing with earlier determinations, we find that our results for the form factors in the narrow width-limit are consistent but with lower central values. Several competing effects compensate each other (choice of the $B$-meson decay constant, the $K^*(980)$ coupling constant, the threshold parameters $s_0$) so that the difference can be mainly attributed to the contribution from the twist-four two-particle contribution $g_+(\omega)$.

\item We then considered the impact of an additional $K^*(1410)$ contribution to our model. This contribution is small in the case of the $K\pi$ vector form factor,  but in principle it could be large for the $B\to K\pi$ case. Our sum rules provide a combined constraint on the $K^*(980)$ and $K^*(1410)$ coupling constants describing the height of the two resonances in the $B\to K\pi$ form factors (with a much smaller weight for $K^*(1410)$). An increased $K^*(1410)$ contribution would correspond to a smaller $K^*(980)$ contribution, and thus a lower value for $B\to K^*$ form factors.

\item We turned to the $B\to K\pi\mu\mu$ differential decay rate in the $K^*(1410)$ region, which has been measured recently by the LHCb collaboration. The branching ratio and the angular decay distribution in this $K\pi$ window provide bounds on the contribution of the $K^*(1410)$. If a huge contribution is ruled out, the current data leave room for a $K^*(1410)$ contribution of moderate size, which could lead to a decrease of the $K^*(890)$ contribution (and thus the values of the $B\to K^*$ form factors) of around 10\%.

\end{itemize}

Considering the more general $B\to K\pi$ form factors and constraining them using LCSRs with $B$-meson LCDAs, we have identified three effects of rather different nature that affect the determination of the $B\to K^*$ form factors: a universal effect related to the finite-width of the $K^*$ increasing by 1.1 the value compared to the narrow-width limit, an effect related to the inclusion of the two-particle twist-four $B$-meson LCDAs leading to smaller values in our case compared to earlier calculations, and an effect of the $K^*(1410)$ contribution that could decrease the $K^*(980)$ peak up to 10\% according to the data currently available. All three effects have a direct impact on the prediction of the $B\to K^*\mu\mu$ branching ratio and thus on the importance of the current discrepancy between SM expectations and LHCb measurements of $b\to s\mu\mu$ branching ratios. Let us add that $B\to K^*\mu\mu$ angular observables (such as $P_5'$) are less affected by this discussion as they are insensitive to the overall normalisation of the form factors.

One may wonder whether previous determinations or other approaches are sensitive to these issues. Previous light-cone sum rule determinations using the $B$-meson LCDAs have worked under the assumption of the narrow-width limit, so that any correction related to a finite width is missed. The $K^*(1410)$ contribution is included through quark-hadron duality in the choice of the threshold parameter, but this contribution is obviously hard to disentangle from the continuum contribution and it suffers from significant uncertainties (suppressed after Borel transformation). A huge contribution from the $K^*(1410)$ would require
a significant failure of the quark-hadron duality, but the moderate contribution discussed here could be included in uncertainties associated with the continuum contribution.

Other LCSRs exploit different correlation functions, so that the OPE contribution is expressed in terms of light-meson LCDAs (either $K^*$ or $K\pi$). They have the advantage of providing an expression of the form factors directly in terms of the OPE part, and not through a convolution with the $K\pi$ form factor as in our case (which required us to design a resonance model for the $B\to K\pi$ form factors). However, this simplicity prevents these sum rules from providing corrections to the narrow-width limit (the LCSRs with a vector resonance are not smooth limits of the LCSRs of the dimeson DAs). The size of the contribution of the $K^*(1410)$ is in principle partially encoded in the dimeson LCDAs, which are however poorly known, whereas the $K^*$ LCDAs are essentially unable to probe this question.

Finally, lattice QCD simulations investigate a different kinematic range for the transfer momentum. The first results on $B\to K^*$ form factors~\cite{Horgan:2013hoa} include configurations where the $K^*$ meson may decay into $K\pi$, but there was not enough data to investigate the impact of finite-width effects in this kinematic regime. This could in principle be studied more extensively using a lattice set-up dedicated to the analysis of unstable resonances~\cite{Agadjanov:2016fbd}. 
Concerning the $K^*(1410)$ contribution to the $B\to K\pi$ form factors, a huge effect
would require an usually large $q^2$-dependence of the form factors to bridge the LCSR and the lattice results and 
it could have led to difficulties in extracting the $K^*(980)$ signal from lattice data under a very large background of excited states, but the moderate contribution discussed here does not seem to contradict
the current results from lattice QCD on these form factors.

Our study of $B\to K\pi$ form factors through an extension of well-known sum rules has highlighted several effects that may impact the current determination of the form factors used to analyse $B\to K^*\ell\ell$ and other decays of the form $B\to K^*X$. 
Several improvements would help assessing these effects more precisely.
The models for the $B$-meson LCDAs should be investigated and constrained more tightly to assess the role played by higher-twist contributions.
The LCSRs could be exploited with additional models for the $K\pi$ and $B\to K\pi$ form factors to consolidate our results.
Finally, the differential decay rate for $B\to K\pi\mu\mu$ at high $K\pi$ invariant mass could be measured more precisely to provide tighter constraints on the $K^*(1410)$ contributions. This would also require a better knowledge of the other partial waves that are contributing in this invariant mass window. If the $D$ wave seems small~\cite{Aaij:2016kqt}, the $S$ wave interferes significantly with the $P$ wave analysed here. The relevant $B\to K\pi$ form factors could be investigated with similar LCSRs to the ones considered here~\cite{WIPSwave}. This should lead to a consistent picture of the contributions from higher resonances to the $B\to K\pi\ell\ell$ decay, and to a deeper understanding of the anomalies currently observed in $b\to s\ell\ell$ transitions.

\section*{Acknowledgments}
We thank Marcin Chrzaszcz, Nico Gubernari, Pablo Roig, Danny van Dyk, Keri Vos and Yuming Wang for useful discussions.
This work is supported by the DFG Research Unit FOR 1873
``Quark Flavour Physics and Effective Theories'', 
contract No KH 205/2-2. This project has received funding from the European Union’s Horizon 2020 research and innovation programme under the Marie Sklodowska-Curie grant agreements No 690575 and No 674896.
J.V. acknowledges funding from the European Union's Horizon 2020 research and innovation programme under the Marie Sklodowska-Curie grant agreement No 700525 `NIOBE', and support from SEJI/2018/033 (Generalitat Valenciana).

\newpage


\appendix

\numberwithin{equation}{section}


\section{Light-Cone Sum Rules for $B\to V$ Form Factors}
\label{app:BtoVLCSRs}

Here we rederive the LCSRs for the $B\to V$ form factors, assuming that the vector $V$ is stable in QCD and leads to a pole in the correlation functions.
The original derivation of these LCSRs was given in \Ref{Khodjamirian:2006st}, but the sum rules for $T_{2,3}$ were not explicitly given there.

The $B\to V$ form factors are defined here as\,\footnote{
This definition agrees with \Refs{Cheng:2017smj,Khodjamirian:2006st} but differs by a factor of $i$ with respect to \Ref{Beneke:2000wa}:
$F_{\!\!\text{\cite{Beneke:2000wa}}} = i F_{\!\!\text{\cite{Cheng:2017smj}}}$.
The reason for this factor of $i$ is that we want to define the phase of the $B\to K\pi$ form factors such that it is real below threshold,
it reaches $\pi/2$ {\it on} the vector resonance and $\pi$ once the resonance is left behind.
}:
\eqa{
&&
\hspace{-4mm}
i\,\langle V(k,\varepsilon) |\bar{s}\gamma^\mu(1-\gamma_5) b |\bar{B}(p)\rangle=
i\,\epsilon^{\mu\nu\alpha\beta}\varepsilon^*_{\nu} q_\alpha k_\beta\ \frac{2V^{BV}(q^2)}{m_B+m_V}
+2 m_V \frac{\varepsilon^*\!\cdot q}{q^2} q^\mu\,A_0^{BV}(q^2) 
\label{eq:BVFFdef}\\
&&
+(m_B+m_V) A_1^{BV}(q^2)\, \bigg[  \varepsilon^{*\mu} -  \frac{\varepsilon^*\!\cdot q}{q^2} q^\mu\bigg]
-\frac{\varepsilon^*\!\cdot q}{m_B+m_V} A_2^{BV}(q^2)\, \bigg[  (p+k)^\mu -  \frac{m_B^2-m_V^2}{q^2} q^\mu\bigg]
\,,\quad
\nonumber\\[3mm]
&&
\hspace{-4mm}
i\,\langle V(k,\varepsilon) |\bar{s}\sigma^{\mu\nu} q_\nu(1+\gamma_5) b |\bar{B}(p)\rangle=
-2 \epsilon^{\mu\nu\alpha\beta}\varepsilon^*_{\nu} q_\alpha k_\beta\ T_1^{BV}(q^2)
\label{eq:BTFFdef}\\
&&
+i\,T_2^{BV}(q^2)\,\Big[ (m_B^2-m_V^2) \varepsilon^{*\mu} - (\varepsilon^*\!\cdot q) (p+k)^\mu \Big]
+i\,(\varepsilon^*\!\cdot q) \,T_3^{BV}(q^2)\,\bigg[ q^\mu - \frac{q^2(p+k)^\mu}{m_B^2-m_V^2} \bigg]
\ .
\nonumber
}

We start from the correlation functions in Eqs~(\ref{eq:corrV}), (\ref{eq:corrt}) and~(\ref{eq:corrT}), but in calculating the imaginary part of the
invariant amplitudes through the unitarity relation we include the single-particle states $\sum_\lambda |V_\lambda \rangle \langle V_\lambda |$
(with $\lambda$ the polarization of the vector particle) instead of the
two-particle states $|K\pi \rangle \langle K\pi |$. Using the same invariant functions and conventions for the OPE functions as for the $B\to K\pi$ form factors, we find:
\eqa{
V^{BV}(q^2) &=& \F_{V,\perp}(q^2)= \frac{m_B + m_V}{2 f_V m_V} e^{m_V^2/M^2} \cdot \P_\perp^{\rm OPE}(q^2,\sigma_0, M^2)\ ,
\label{eq:LCSRBVV}\\
A_1^{BV}(q^2) &=& \F_{V,\|}(q^2)= \frac1{f_V m_V(m_B + m_V)} e^{m_V^2/M^2} \cdot \P_\|^{\rm OPE}(q^2,\sigma_0, M^2)\ ,\\
A_2^{BV}(q^2) &=& \F_{V,-}(q^2)= -\frac{m_B + m_V}{2 f_V m_V} e^{m_V^2/M^2} \cdot \P_-^{\rm OPE}(q^2,\sigma_0, M^2)\ ,\\
A_0^{BV}(q^2) &=& \F_{V,t}(q^2)= -\frac1{2 f_V m_V^2} e^{m_V^2/M^2}
\cdot \P_t^{\rm OPE}(q^2,\sigma_0, M^2)\ ,\\
T_1^{BV}(q^2) &=& \F^T_{V,\perp}(q^2)= \frac1{2f_V m_V} e^{m_V^2/M^2}
\cdot \P_\perp^{T, {\rm OPE}}(q^2,\sigma_0, M^2)\ ,\\
T_2^{BV}(q^2) &=& \F^T_{V,\|}(q^2)= \frac1{f_V m_V(m_B^2 - m_V^2)} e^{m_V^2/M^2}
\cdot \P_\|^{T, {\rm OPE}}(q^2,\sigma_0, M^2)\ ,\\
T_3^{BV}(q^2) &=& \F^T_{V,-}(q^2)= -\frac1{2 f_V m_V} e^{m_V^2/M^2}
\cdot \P_-^{T, {\rm OPE}}(q^2,\sigma_0, M^2)\ ,
\label{eq:LCSRBVT3}
}
where the functions $\P_i^{T, {\rm OPE}}$ are the same as the ones for the $B\to K\pi$ sum rules,
and are given in Appendix~\ref{OPEexpressions}.


\newpage

\section{$B$-meson distribution amplitudes up to twist 4}
\label{BLCDAs}

\subsection{Definition of $B$-meson distribution amplitudes}

We consider the two- and three-particle $B$-meson light-cone distribution amplitudes (LCDAs) up to twist four as recently discussed in \Ref{Braun:2017liq}\footnote{
For earlier articles discussing $B$-meson LCDAs,
see~\cite{Grozin:1996pq,Beneke:2000wa,DescotesGenon:2002mw,Khodjamirian:2006st,DescotesGenon:2009hk,Kawamura:2010tj,Knodlseder:2011gc,Bell:2013tfa,Beneke:2018wjp,Wang:2016qii,Wang:2018wfj}.
}.

\subsubsection{Two-particle LCDAs}

The two-particle LCDAs up to twist-four are given by~\cite{Braun:2017liq}: 
\begin{eqnarray}
\hspace{-3mm} \D_\Gamma^{(2)}(x) &\equiv& \langle 0| \bar q(x) \Gamma h_v(0) |\bar B(v)\rangle =
-\frac{i  f_B m_B}2 \Tr\Big[\gamma_5 \Gamma P_+ \Big] \int_0^\infty d\omega \, e^{-i\omega v\cdot x}
\Big\{\phi_+(\omega) + x^2 g_+(\omega)\Big\}
\nonumber\\
&&\hspace{-16mm} +\frac{i  f_B m_B}4  \Tr\Big[\gamma_5 \Gamma P_+ \slashed{x} \Big] \frac{1}{v\cdot x}
\int_0^\infty d\omega \, e^{-i\omega v\cdot x} \Big\{[\phi_+-\phi_-](\omega) + x^2 [g_+-g_-](\omega)\Big\} + \op(x^4)
\label{def:g+g-}
\end{eqnarray}
for an arbitrary Dirac matrix $\Gamma$. Here  $P_+ \equiv (1+\slashed v)/2$ and we have omitted the collinear Wilson line $W(0,x)$ that
makes the matrix element gauge invariant.
The functions $\phi_+(\omega)$, $\phi_-(\omega)$, $g_+(\omega)$ and $g_-(\omega)$ are respectively twist-two, three, four and five LCDAs.
We will therefore neglect $g_-(\omega)$. We use the same normalization conventions for the LCDAs as in \Ref{Khodjamirian:2006st} (see also \cite{Beneke:2000wa}), with the $B$-meson decay constant $f_B$ defined in full QCD and with the relativistic normalization of the state $|B(v)\rangle$. These conventions  differ from the ones used in \Ref{Braun:2017liq}, where the LCDAs are normalized to the scale-dependent decay constant $F_B$ defined in the Heavy Quark Effective Theory (HQET) and the HQET normalization for the $|B(v)\rangle$ state is adopted. This explains an extra factor $m_B$ in \Eq{def:g+g-} as compared to the definition in \cite{Braun:2017liq}. Since we will not include gluon radiative corrections in the correlation function, the difference between $F_B$ and $f_B$ -- which starts at $\op(\alpha_s)$ -- remains within the considered accuracy.

We write \Eq{def:g+g-} as:
\eq{
\D_\Gamma^{(2)}(x) = -\frac{i}2 f_B m_B \Tr\Big[ \gamma_5 \Gamma P_+ \Big] \A(x)
+ \frac{i}4  f_B m_B \Tr\Big[\gamma_5 \Gamma P_+ \slashed{x} \Big] \B(x)
}
where:
\eqa{
\A(x) &=& \int_0^\infty d\omega \, e^{-i\omega v\cdot x} \Big\{\phi_+(\omega) + x^2 g_+(\omega)\Big\} \ ,
\label{eq:A(x)}\\
\B(x) &=& \int_0^\infty d\omega \, e^{-i\omega v\cdot x} \Big\{i\bar\Phi_\pm(\omega) + x^2 i \bar G_\pm(\omega)\Big\} \ ,\label{eq:B(x)}\\[3mm]
\bar\Phi_\pm(\omega) &=& \int_0^\omega d\tau\,\big(\phi_+(\tau) - \phi_-(\tau)\big) \ ,
\quad
\bar G_\pm(\omega) = \int_0^\omega d\tau\,g_+(\tau) \ \ .
\label{eq:barred}
}
The functions $\bar\Phi_\pm(\omega)$, $\bar G_\pm(\omega)$ satisfy the normalization condition
$\bar\Phi_\pm(\infty)=\bar G_\pm(\infty)=0$, in order for $\D_\Gamma^{(2)}(x)$ to be well-defined as $v\cdot x\to 0$.
The appearance of these functions in $\B(x)$ follows from integration by parts:
\eq{
0=\int_0^\infty d\omega \frac{d}{d\omega}(e^{-i\omega v\cdot x} F_\pm(\omega))
= \int_0^\infty d\omega e^{-i\omega v\cdot x}  \bigg[ (-iv\cdot x) F_\pm(\omega) + f_\pm(\omega)  \bigg] 
}
where $\{ f_\pm,F_\pm\}$ denote $\{ \phi_+-\phi_-,\bar \Phi_\pm \}$ or $\{g_+,\bar G_\pm\}$
and the first equality follows from the aforementioned normalization conditions $F_\pm(\infty)=0$.
This implies that under the integral one may perform the substitution
\eq{
\frac{f_\pm(\omega)}{v\cdot x} \to i F_\pm(\omega)\ ,
\label{eq:subsbarred}
}
which is what we have done in~\Eq{eq:B(x)}.

\subsubsection{Three-particle LCDAs}

The three-particle quark-gluon LCDAs up to twist four are given by~\cite{Braun:2017liq}:
\begin{eqnarray}
&&
\D_\Gamma^{(3)}(x,u) \equiv
\lefteqn{\langle 0| \bar q(x) G_{\mu\nu}(ux)\Gamma^{\mu\nu} h_v(0) |\bar B(v)\rangle =
\frac{f_B m_B}2 \int_0^\infty d\omega_1\, d\omega_2\ e^{-i (\omega_1 + u \omega_2) v\cdot x}}
\nonumber\\
&&
\Tr\biggl\{\gamma_5 \Gamma^{\mu\nu} P_+
\biggl[
(v_\mu\gamma_\nu-v_\nu\gamma_\mu)  \phi_3
+\frac{i}2\sigma_{\mu\nu} \big[ \phi_3 - \phi_4 \big]
+ \frac{x_\mu v_\nu-x_\nu v_\mu}{2v\cdot x} \big[ \phi_3 + \phi_4 - 2\psi_4 \big]
\nonumber\\
&&
- \frac{x_\mu \gamma_\nu-x_\nu \gamma_\mu}{2v\cdot x}\big[\phi_3 + \widetilde\psi_4 \big]
+ \frac{i\epsilon_{\mu\nu\alpha\beta} x^\alpha v^\beta}{2v\cdot x}  \gamma_5\,
\big[ \phi_3 - \phi_4 +2 \widetilde\psi_4 \big]
- \frac{i\epsilon_{\mu\nu\alpha\beta} x^\alpha}{2v\cdot x}  \gamma^\beta\gamma_5\, \big[ \phi_3 - \phi_4 + \widetilde\psi_4 \big]
\nonumber\\
&&
- \frac{(x_\mu v_\nu-x_\nu v_\mu)\slashed{x}}{2(v\cdot x)^2} \, \big[ \phi_4 - \psi_4 - \widetilde\psi_4 \big]
- \frac{(x_\mu \gamma_\nu-x_\nu \gamma_\mu)\slashed{x}}{4(v\cdot x)^2} \, \big[ \phi_3 - \phi_4 + 2 \widetilde\psi_4 \big]
\biggr]\biggr\}(\omega_1,\omega_2)
\ ,
\label{def:three}
\end{eqnarray}
where the functional dependence $\phi_3=\phi_3(\omega_1,\omega_2)$, etc., is indicated outside the curly bracket.
Again, $\Gamma^{\mu\nu}$ denotes an arbitrary Dirac matrix and Wilson lines have been suppressed.
This form of the matrix element has been obtained from the one in~\cite{Braun:2017liq} by defining $x_\mu = z_1 n_\mu$ and $u = z_2/z_1$.
In analogy to~\Eq{eq:barred} we define~\cite{Khodjamirian:2006st}
\eq{
\overline \Phi (\omega_1,\omega_2) = \int_0^{\omega_1} d\tau\,\Phi (\tau,\omega_2)\ , \qquad
\overline{\overline \Phi} (\omega_1,\omega_2) = \int_0^{\omega_1} d\tau\, \overline\Phi (\tau,\omega_2)\ ,
}
for $\Phi = \{ \phi_3,\phi_4,\psi_4, \widetilde \psi_4 \}$.
This allows us to make the substitutions such as \Eq{eq:subsbarred} in \Eq{def:three} and get rid of the
$v\cdot x$ denominators. In the last two terms one needs to do this twice (thus appearing `double-barred' LCDAs),
but the final results for the correlation functions will only contain `single-barred' LCDAs.
In addition we define the variable
\eq{
\sigma(u) \equiv (\omega_1 + u \,\omega_2)/m_B
}
such that
\eq{
\D_\Gamma^{(3)}(x,u) =  \frac{f_B m_B}2 \int_0^\infty d\omega_1\, d\omega_2\ e^{-i \sigma m_B v\cdot x} \, 
\Tr \Big[ \gamma_5\, \Gamma^{\mu\nu}\, P_+\, \Psi_{\mu\nu}(x,\omega_1,\omega_2) \Big]\ ,
\label{def:three3}
}
with
\eqa{
&&\Psi_{\mu\nu}(x,\omega_1,\omega_2) =
\biggl[
(v_\mu\gamma_\nu-v_\nu\gamma_\mu)  \phi_3
+\frac{i}2\sigma_{\mu\nu} \big[ \phi_3 - \phi_4 \big]
+ \frac{i}2 (x_\mu v_\nu-x_\nu v_\mu) \big[ \overline \phi_3 + \overline \phi_4 - 2 \overline \psi_4 \big]
\nonumber\\
&&
- \frac{i}2 (x_\mu \gamma_\nu-x_\nu \gamma_\mu) \big[ \overline \phi_3 + \overline{\widetilde\psi}_4 \big]
- \frac12 \epsilon_{\mu\nu\alpha\beta} x^\alpha v^\beta  \gamma_5\,
\big[ \overline \phi_3 - \overline \phi_4 +2 \overline {\widetilde\psi}_4 \big]
+ \frac12 \epsilon_{\mu\nu\alpha\beta} x^\alpha  \gamma^\beta\gamma_5\, \big[ \overline\phi_3 - \overline\phi_4 + \overline{\widetilde\psi}_4 \big]
\nonumber\\
&&
+ \frac12 (x_\mu v_\nu-x_\nu v_\mu)\slashed{x} \, \big[ \overline{\overline\phi}_4 -  \overline{\overline\psi}_4 - \overline{\overline{\widetilde\psi}}_4 \big]
+ \frac14 (x_\mu \gamma_\nu-x_\nu \gamma_\mu)\slashed{x} \, \big[ \overline{\overline\phi}_3 - \overline{\overline\phi}_4 + 2 \overline{\overline{\widetilde\psi}}_4 \big]
\biggr](\omega_1,\omega_2)\ .
}

\subsection{Models and numerics for LCDAs}
\label{sec:modelsLCDAs}

We will use the three models for the two- and three-particle $B$-meson LCDAs considered in~\Ref{Braun:2017liq},
called {\bf Model I}, {\bf Model IIA} and {\bf Model IIB} in that reference, and we keep these names. 
All models are characterized by two input parameters: $\lambda_B$ and $R$.
Our numerical values for these two parameters are collected in \Tab{tab:parameters}, and will be justified at the end of this section.

\bigskip

\noindent {\bf Model I :} The first model is based on the exponential model of \Refs{Grozin:1996pq,Khodjamirian:2006st}:
\eqa{
\phi^{\bf I}_+(\omega) &=& \frac{\omega}{\lambda_B^2} e^{-\omega/\lambda_B}  \\
\phi^{\bf I}_-(\omega) &=& \frac1{\lambda_B} e^{-\omega/\lambda_B}
+ \frac1{2\lambda_B} \bigg[ 1 - \frac{2\omega}{\lambda_B} + \frac{\omega^2}{2\lambda_B^2} \bigg] \frac{1-R}{1+2R}\, e^{-\omega/\lambda_B} \\
g^{\bf I}_+(\omega) &=& \frac{3 \omega^2}{16\lambda_B} \frac{3+4R}{1+2R} \,e^{-\omega/\lambda_B}  \\
\phi^{\bf I}_3(\omega_1,\omega_2) &=&  - \frac{3\,\omega_1 \omega_2^2}{4 \lambda_B^3} \frac{1-R}{1+2R}  \,e^{- \overline \omega/\lambda_B} \\
\phi^{\bf I}_4(\omega_1,\omega_2) &=&  \frac{3\,\omega_2^2}{4 \lambda_B^2} \frac{1+R}{1+2R}  \,e^{- \overline \omega/\lambda_B} \\
\psi^{\bf I}_4(\omega_1,\omega_2) &=& R \ \widetilde \psi^{\bf I}_4(\omega_1,\omega_2)
= \frac{3\,\omega_1 \omega_2}{2 \lambda_B^2} \frac{R}{1+2R} \,e^{- \overline \omega/\lambda_B} \ ,
}
where $\overline \omega \equiv \omega_1 + \omega_2$.

\bigskip

\noindent {\bf Model IIA :} The second model is based on the `local duality' assumption~\cite{Braun:2017liq} with linear behaviour $\sim(3\lambda_B-\omega)$
for all LCDAs (and constant for $\phi_3$):
\eqa{
\phi^{\bf IIA}_+(\omega) &=& \frac{2\omega}{9\lambda_B^3} (3\lambda_B-\omega) \ \theta(3\lambda_B-\omega)  \\
\phi^{\bf IIA}_-(\omega) &=&
\bigg[  \frac{(3\lambda_B-\omega)^2}{9\lambda_B^3}
+\frac{2\omega^2 - 6 \omega \lambda_B + 3 \lambda_B^2}{24 \lambda_B^3} \frac{1-R}{1+2R} \bigg] \ \theta(3\lambda_B-\omega)  \\
g^{\bf IIA}_+(\omega) &=& \frac{\omega^2 (3\lambda_B-\omega)^2}{192\lambda_B^3} \frac{7+12R}{1+2R} \ \theta(3\lambda_B-\omega)  \\
\phi^{\bf IIA}_3(\omega_1,\omega_2) &=&
- \frac{\omega_1 \omega_2^2}{24 \,\lambda_B^3} \frac{1-R}{1+2R} \ \theta(3\lambda_B-\overline\omega) \\
\phi^{\bf IIA}_4(\omega_1,\omega_2) &=&
\frac{\omega_2^2 (3\lambda_B - \overline \omega)}{24 \,\lambda_B^3} \frac{1+R}{1+2R} \ \theta(3\lambda_B-\overline\omega) \\
\psi^{\bf IIA}_4(\omega_1,\omega_2) &=& R\ \widetilde \psi^{\bf IIA}_4(\omega_1,\omega_2) 
= \frac{\omega_1 \omega_2 (3\lambda_B - \overline \omega)}{12 \,\lambda_B^3} \frac{R}{1+2R} \ \theta(3\lambda_B-\overline\omega)  \ .
}

\bigskip

\noindent {\bf Model IIB :} The third model is the same type as IIA, but with cubic behaviour $\sim(5\lambda_B-\omega)^3$
for all LCDAs (and quadratic for $\phi_3$):
\eqa{
\phi^{\bf IIB}_+(\omega) &=& \frac{4\omega}{625\lambda_B^5} (5\lambda_B-\omega)^3 \ \theta(5\lambda_B-\omega)  \\
\phi^{\bf IIB}_-(\omega) &=&
\frac{(5\lambda_B-\omega)^4}{125\lambda_B^5}  \bigg[  \frac{1}{5}
+\frac{3\omega^2 - 10 \omega \lambda_B + 5 \lambda_B^2}{4 (5\lambda_B-\omega)^2} \frac{1-R}{1+2R} \bigg] \ \theta(5\lambda_B-\omega)  \\
g^{\bf IIB}_+(\omega) &=& \frac{\omega^2 (5\lambda_B-\omega)^4}{20000\lambda_B^5} \frac{13+20R}{1+2R} \ \theta(5\lambda_B-\omega)  \\
\phi^{\bf IIB}_3(\omega_1,\omega_2) &=&
- \frac{9\,\omega_1 \omega_2^2 \,(5\lambda_B- \overline \omega)^2}{1250 \,\lambda_B^5} \frac{1-R}{1+2R} \ \theta(5\lambda_B-\overline\omega) \\
\phi^{\bf IIB}_4(\omega_1,\omega_2) &=&  
\frac{3\,\omega_2^2 (5\lambda_B - \overline \omega)^3}{1250 \,\lambda_B^5} \frac{1+R}{1+2R} \ \theta(5\lambda_B-\overline\omega) \\
\psi^{\bf IIB}_4(\omega_1,\omega_2) &=& R\ \widetilde \psi^{\bf IIB}_4(\omega_1,\omega_2)
=\frac{3\,\omega_1 \omega_2 (5\lambda_B - \overline \omega)^3}{625 \,\lambda_B^5} \frac{R}{1+2R} \ \theta(5\lambda_B-\overline\omega) \ .
}

\bigskip

All these expressions for the LCDAs up to twist-four in the three models have been obtained from the expressions in~\Ref{Braun:2017liq} by enforcing
the Equation-Of-Motion (EOM) constraints derived in that article.  
The expressions given here for $g_+(\omega)$ correspond to the simplified expressions in~\Ref{Braun:2017liq}.
We have checked that they reproduce numerically the full expressions in~\cite{Braun:2017liq} for a very wide range of parameters with a very high precision.

We now discuss briefly the numerical values chosen here for the two input parameters $\lambda_B$ and~$R$ on  which the models depend.
These values are also collected for reference in~\Tab{tab:parameters}.
For $\lambda_B$ we follow~\cite{Cheng:2017smj} and take:
\eq{
\lambda_B = \lambda_B(1\GeV) = 460 \pm 110 \,\MeV,
\label{eq:lambdaBnum}
}
which is the value derived from the QCD sum rules in~\Ref{Braun:2003wx}.
This value is consistent with the experimental bound obtained in~\Ref{Heller:2015vvm}
and with the determination from the $B\to \pi$ form factor LCSRs~\cite{Wang:2015vgv}.

The parameter $R$ is defined as the ratio $R = \lambda_E^2/\lambda_H^2$,
where $\lambda_{E,H}^2$ parametrize the local correlation function $\D_\Gamma^{(3)}(0,0)$ (see~\Eq{def:three}), for example
$\lambda_E^2 = 3 \int d\omega_1 d \omega_2 \,\psi_4(\omega_1,\omega_2)$ and 
$\lambda_H^2 = 3 \int d\omega_1 d \omega_2 \,\widetilde\psi_4(\omega_1,\omega_2)$.
\Ref{Braun:2017liq} considers the following two determinations of the local matrix element from QCD sum rules:
\eq{
\arraycolsep=5mm
\begin{array}{llr}
\lambda_E^2 = 110 \pm 60 \MeV^2\ ,  & \lambda_H^2 = 180 \pm 70 \MeV^2\ ,  &  \text{\cite{Grozin:1996pq}}  \\[2mm]
\lambda_E^2 =   30 \pm 20 \MeV^2\ ,  & \lambda_H^2 =   60 \pm 30 \MeV^2\ .  &  \text{\cite{Nishikawa:2011qk}}
\end{array}
\label{eq:lambdaElambdaHnum}
}
These results are only marginally compatible with each other, and in addition when combined with the EOM they lead to values of $\lambda_B$
significantly lower than \Eq{eq:lambdaBnum}. However, as pointed out in~\Ref{Braun:2017liq}, the ratio $R$ is expected to be more reliable:
 indeed both determinations are consistent with each other with $R\sim 0.5$.
Here we average both determinations:
\eq{
R = 0.4^{+0.5}_{-0.3}\ .
}
This number is obtained in the following way. We assume the errors in~\Eq{eq:lambdaElambdaHnum} are Gaussian and uncorrelated.
We generate a large and equal number of normally distributed pairs $\{ \lambda_E^2, \lambda_H^2 \}$ for each of the two determinations,
and calculate the ratio $R$ for each pair. This leads to a set of values of $R$ that are not normally distributed but have a tail to the positive side.
From this distribution we take the central value as the most probable one (the maximum of the distribution), and for the ``$1\,\sigma$''
region we take the most probable region containing the $68\%$ of the probability.
This error bar has to be regarded as a conservative estimate, given the fact that the errors in~\Eq{eq:lambdaElambdaHnum} are certainly correlated. A reevaluation of the QCD sum rules leading to~\Eq{eq:lambdaElambdaHnum} should be performed, with a simultaneous determination of the ratio $R$ taking into account such correlations.

\begin{figure}
\begin{center}
\includegraphics[width=5cm,height=4cm]{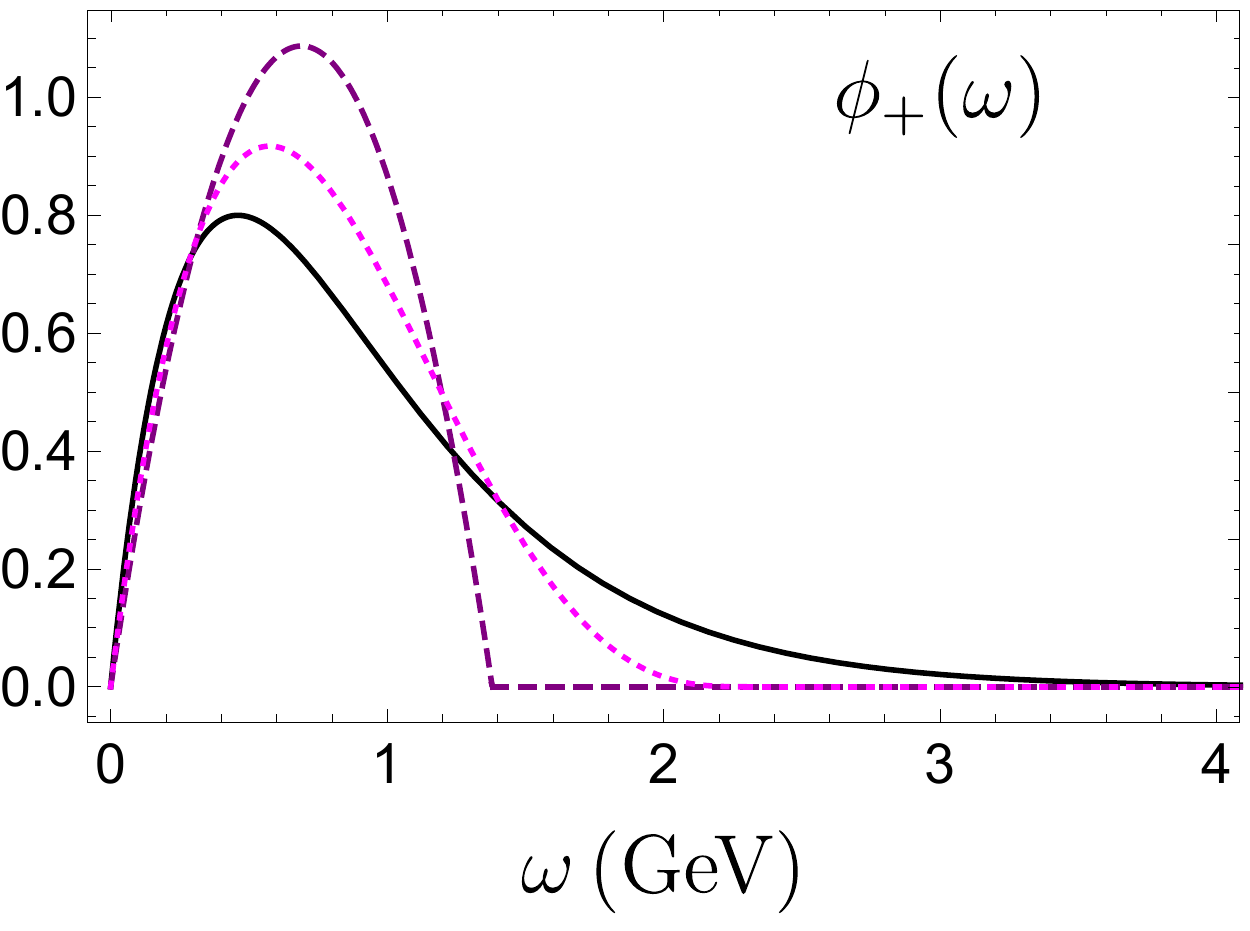}
\hspace{5mm}
\includegraphics[width=5cm,height=4cm]{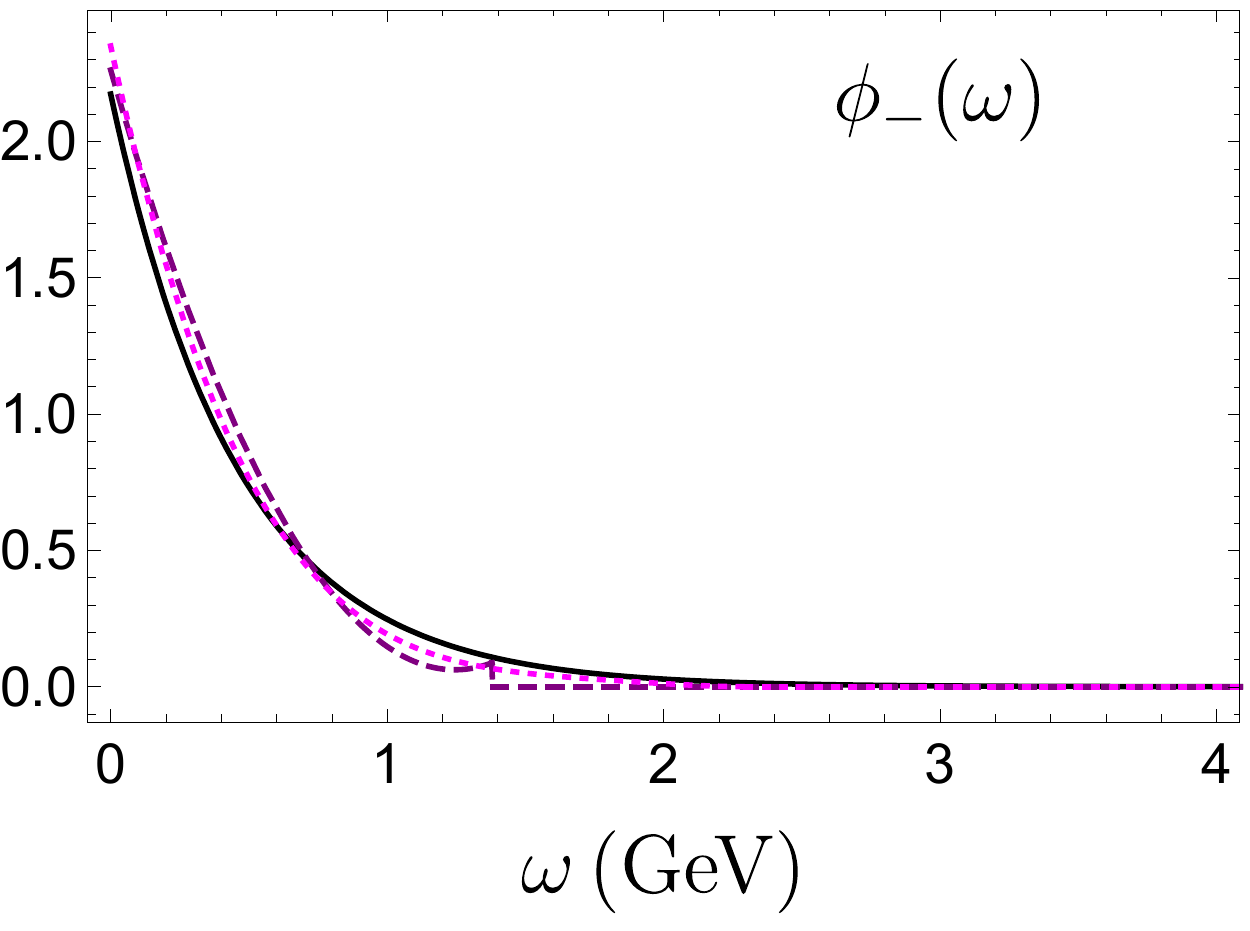}
\hspace{5mm}
\includegraphics[width=5.1cm,height=4cm]{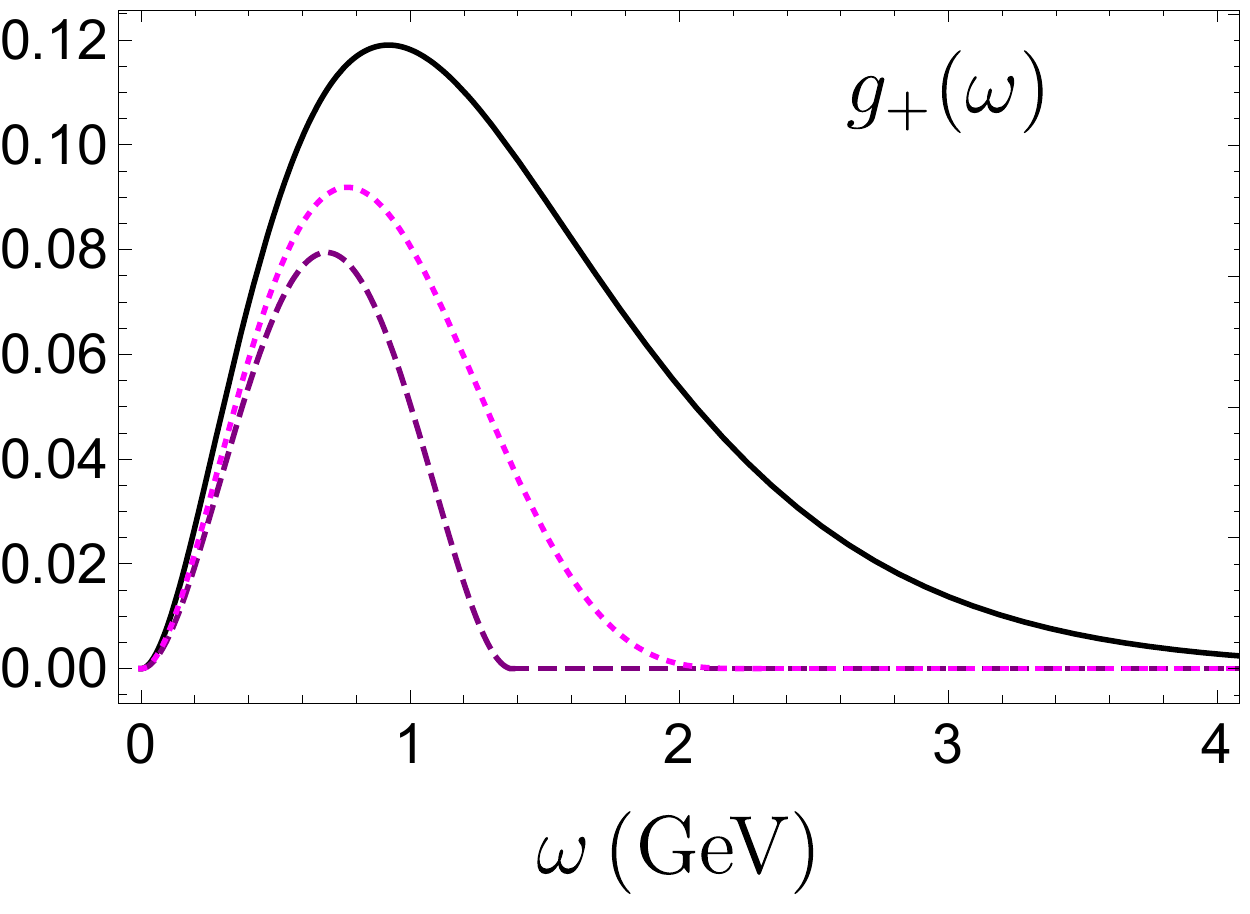}\\[2mm]
\includegraphics[width=5cm,height=4cm]{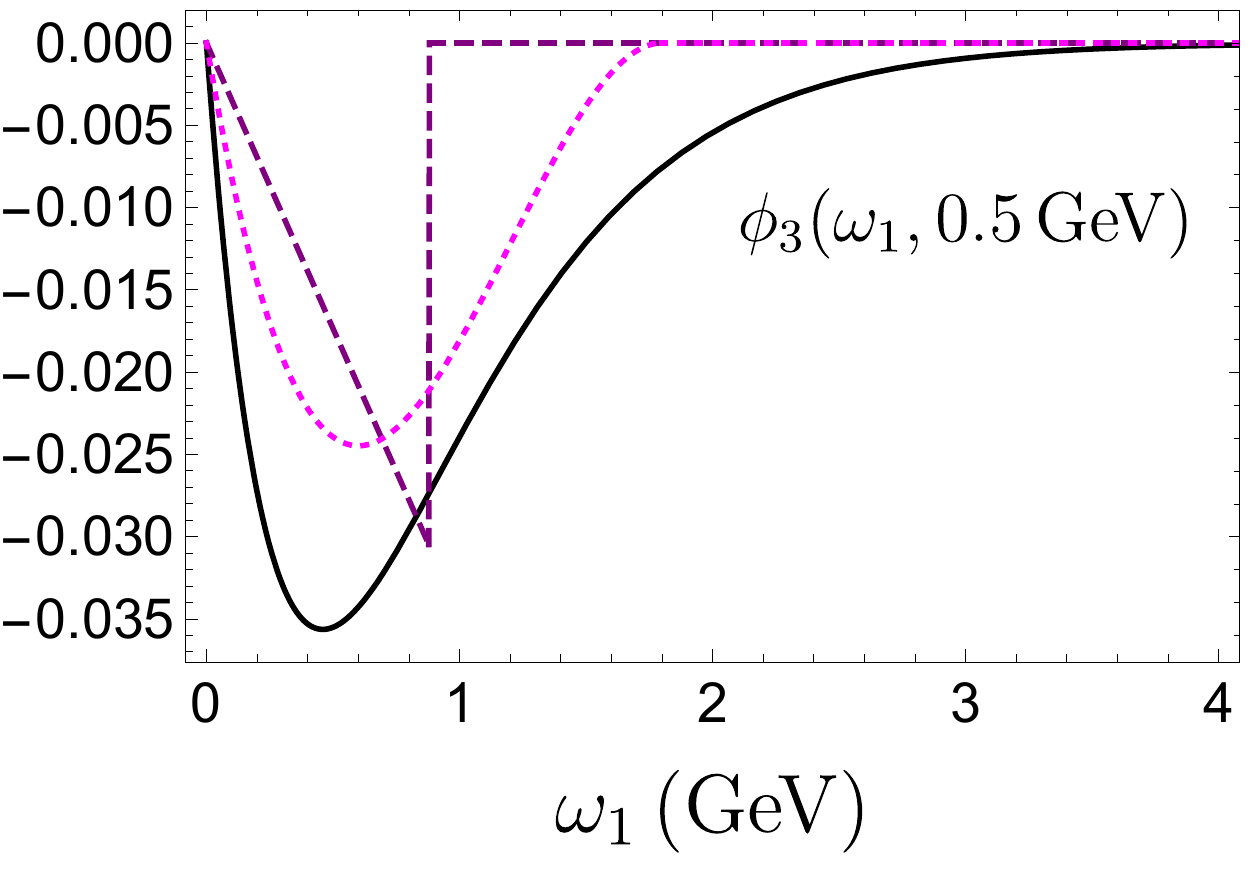}
\hspace{5mm}
\includegraphics[width=5cm,height=4cm]{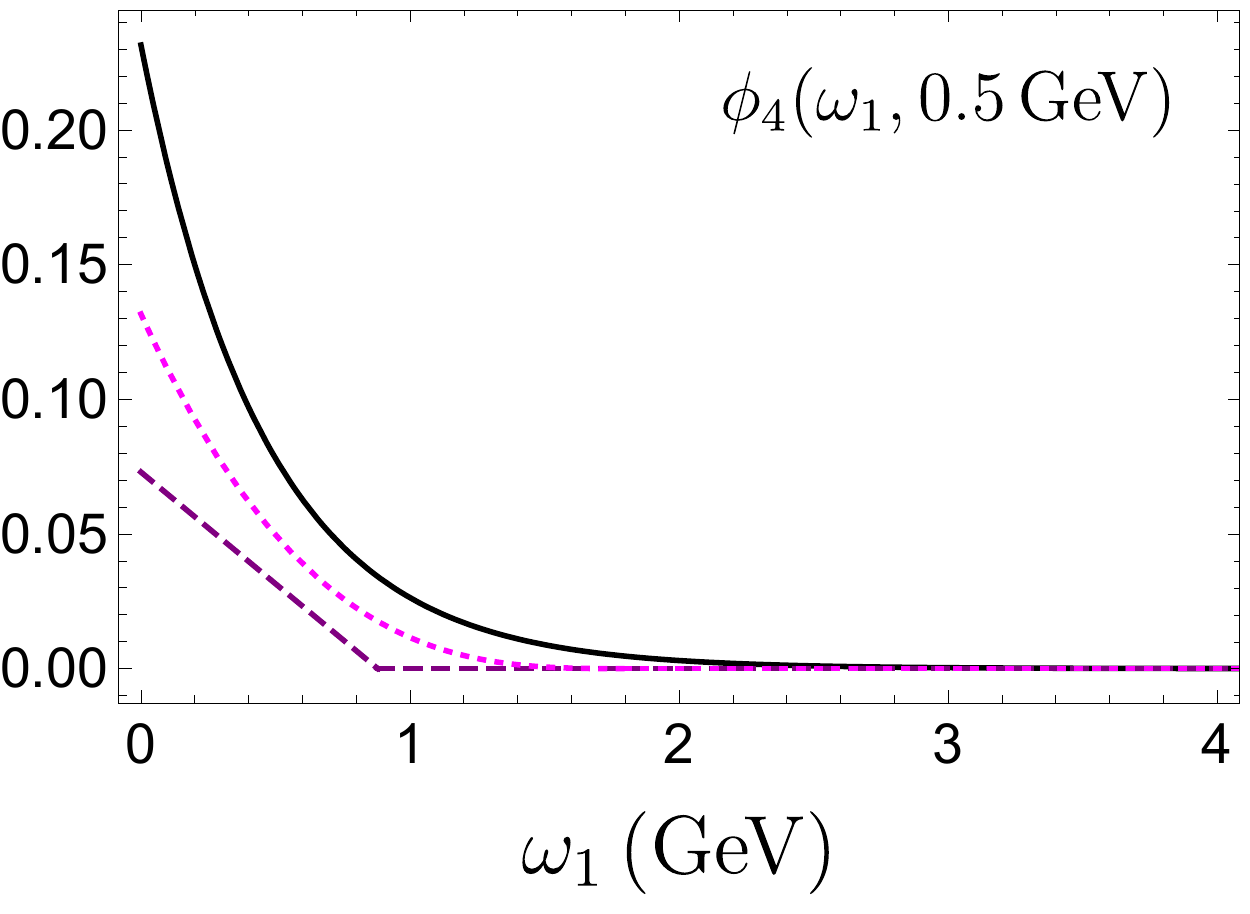}
\hspace{5mm}
\includegraphics[width=5.1cm,height=4cm]{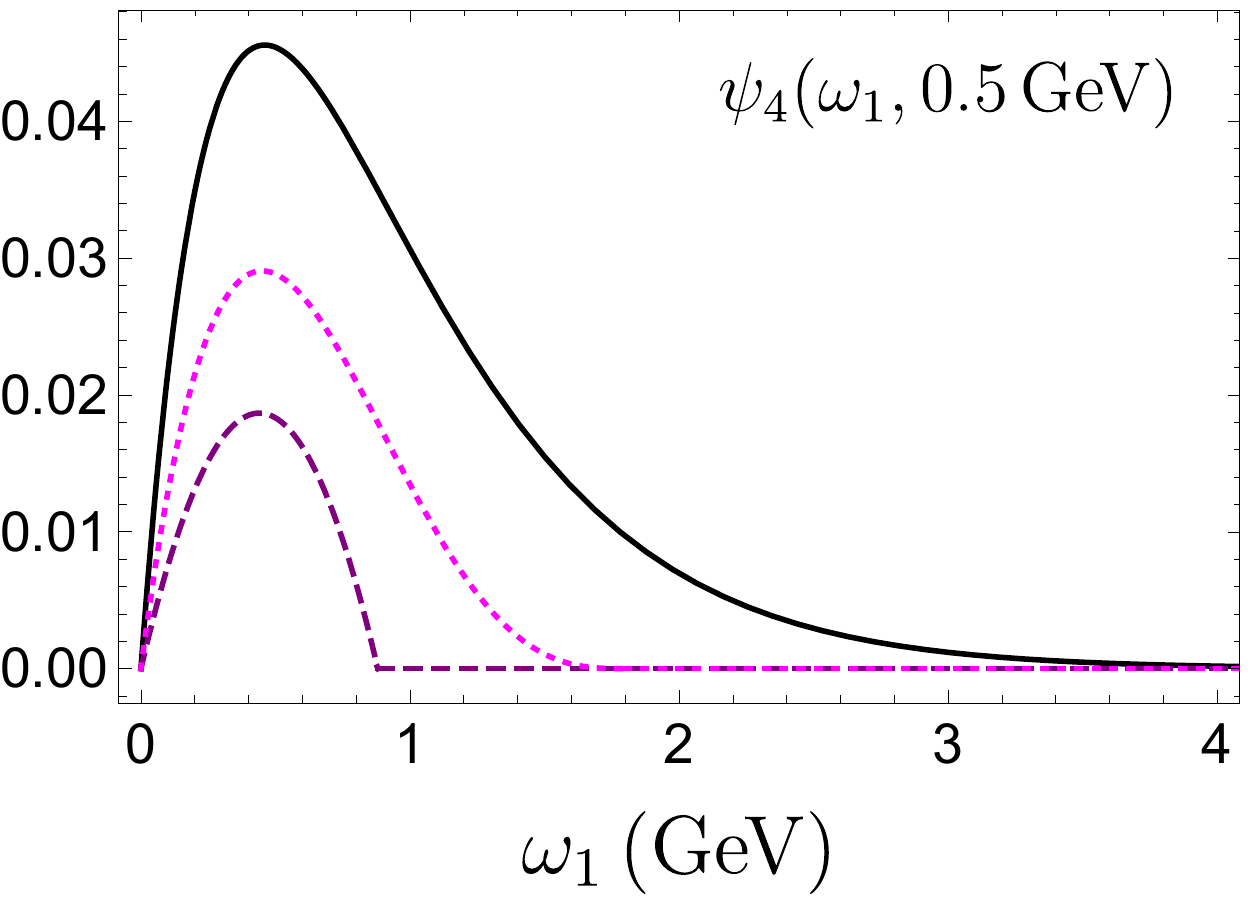}
\end{center}
\caption{\it Two- and three-particle LCDAs within the three different models considered: Model~I (solid), Model~IIA (dashed) and Model~IIIA (dotted).
These plots are obtained fixing $\lambda_B$ and $R$ to their central values, and $\omega_2= 0.5\,\GeV$.}
\label{fig:LCDAs}
\end{figure}

In \Fig{fig:LCDAs} we show plots for the different LCDAs in the different models.

\section{Calculation of Correlation Functions}
\label{app:OPEcalc}

We need to calculate the correlation functions:
\eq{
\P_{ab}(k,q) = i\int d^4x\, e^{i k \cdot x} \langle 0 | {\rm T} \{\bar{d}(x) \Gamma_a\, s(x),\bar{s}(0)\Gamma_b\, b(0) \} | \bar{B}^0(q+k) \rangle
}
with generic Dirac structures $\Gamma_{a,b}$. Contracting the $s$ fields, we write:
\eq{
\P_{ab}(k,q) = i\int d^4x\, e^{i k \cdot x} \langle 0 | \bar{d}(x) \Gamma_a \hat S(x) \Gamma_b\, b(0) | \bar{B}^0(q+k) \rangle
}
where $\hat S(x) = \hat S^{(2)}(x) + \hat S^{(3)}(x) + \cdots$ is the light-cone expansion of the quark propagator, with~\cite{Balitsky:1987bk}
\eqa{
\hat S^{(2)}(x) &=& \int \frac{d^4 p}{(2\pi)^4} e^{-i p\cdot x} \frac{i (\slashed p + m_s)}{p^2-m_s^2} \ ,\\
\hat S^{(3)}(x) &=& - i  \int_0^1 du \,G_{\alpha\beta}(ux) \,
\int \frac{d^4 p}{(2\pi)^4}\, e^{-i p\cdot x}
\,\frac{\big[\bar u\, (\slashed{p}+m_s) \sigma^{\alpha\beta} +u \,\sigma^{\alpha\beta} (\slashed{p}+m_s) \big]}{2 (p^2 - m_s^2)^2}  \ .
}
The contributions from $\hat S^{(2)}(x)$ and $\hat S^{(3)}(x)$ to the correlation functions are called two- and three-particle contributions, respectively:
\eqa{
\P_{ab}(k,q) &=& \P_{ab}^{(2)}(k,q) + \P_{ab}^{(3)}(k,q) + \cdots\\
\P_{ab}^{(j)}(k,q) &=& i\int d^4x\, e^{i k \cdot x} \langle 0 | \bar{d}(x) \Gamma_a \hat S^{(j)}(x) \Gamma_b\, b(0) | \bar{B}^0(q+k) \rangle\ .
}
In the following two subsections we outline the calculation of these two- and three-particle contributions.

\subsection{Two-particle contributions}

Collecting the results in the previous subsections we have, for the two-particle contribution:
\eqa{
\P_{ab}^{(2)}(k,q) &=& i\int d^4x\, e^{i k \cdot x} \ \D_{[\Gamma_a \hat S^{(2)}(x) \Gamma_b]}^{(2)} (x)
\label{eq:T*fi}\\
&&\hspace{-6mm} = \frac{f_B m_B}2 \int d^4x\, e^{i k \cdot x} \bigg\{  \Tr\Big[ \gamma_5 \Gamma_a \hat S^{(2)}(x) \Gamma_b P_+ \Big] \A(x)
- \frac12 \Tr\Big[\gamma_5 \Gamma_a \hat S^{(2)}(x) \Gamma_b P_+ \slashed{x} \Big] \B(x)  \bigg\}  .\nonumber
}
The $d^4x$ integrals can be performed by applying the general formula
\eq{
\int d^4x \frac{d^4 p}{(2\pi)^4} \ e^{i(k_\omega-p)\cdot x} \,f(x) \,S(p) = f(-i\partial_p) S(p) \big|_{p=k_\omega}\ ,
\label{eq:d4xd4p}
}
leading to
\eqa{
\int d^4x \ e^{i k_\omega\cdot x} \, \hat S^{(2)}(x) &=& \frac{i(\slashed k_\omega + m_s)}{k_\omega^2 - m_s^2} \ ,\\
\int d^4x \ e^{i k_\omega\cdot x} \, x^2 \hat S^{(2)}(x) &=&
\frac{4i \slashed k_\omega}{(k_\omega^2 - m_s^2)^2}  - \frac{8im_s^2 (\slashed k_\omega + m_s)}{(k_\omega^2 - m_s^2)^3} \ ,\\
\int d^4x \ e^{i k_\omega\cdot x} \, x^\mu \hat S^{(2)}(x) &=&
\frac{\gamma^\mu}{k_\omega^2 - m_s^2}   - \frac{2 k_\omega^\mu(\slashed k_\omega + m_s)}{(k_\omega^2 - m_s^2)^2} \ ,\\
\int d^4x \ e^{i k_\omega\cdot x} \, x^\mu x^2 \hat S^{(2)}(x) &=&
\frac{4 \gamma^\mu}{(k_\omega^2 - m_s^2)^2} - \frac{8 m_s^2 \gamma^\mu+16k_\omega^\mu \slashed k_\omega}{(k_\omega^2 - m_s^2)^3}
+ \frac{48 m_s^2 k_\omega^\mu(\slashed k_\omega + m_s)}{(k_\omega^2 - m_s^2)^4} \ . \quad 
}
In our case,
\eq{
k_\omega = k - \omega v = \bar \sigma k - \sigma q\ ,
}
where $\sigma \equiv \omega/m_B$ and $\bar\sigma = 1-\sigma$.

Inserting these expressions in \Eq{eq:T*fi} we reduce the latter to integrals over the variable~$\omega$ -- see Eqs.~(\ref{eq:A(x)})-(\ref{eq:B(x)}) -- 
which is then traded for the variable $\sigma$, with a support in the range~$[0,1]$.
Performing the Lorentz algebra and the Dirac traces, we can now identify the invariant amplitudes such as in~\Eq{eq:corrV}.
Each invariant amplitude can now be written as:
\eq{
\P^{(2)}_{ab,(i)}(k^2,q^2) = f_B m_B \sum_n  
\int_0^1 \! d\sigma\, \frac{\bar \sigma^{n+1}\, \hat I^{(i)}_n(\sigma)}{(k_\omega^2-m_s^2)^{n+1}}\ .
\label{eq:terms}
}
where the functions $\hat I^{(i)}_n(\sigma)$ are linear combinations of the distribution amplitudes with $\sigma$-dependent coefficients. 
In order to implement duality, we cast the OPE expression of each invariant amplitude as a dispersion relation, with a subsequent subtraction above threshold and the Borel transformation:
\eq{ \P(k^2,q^2)=\frac{1}{\pi}\int\limits_{m_s^2}^{\infty} ds\frac{\mbox{Im}\,\P(s,q^2)}{s-k^2}
\quad \longrightarrow \quad
\P(q^2,s_0,M^2)=\frac{1}{\pi}\int\limits_{m_s^2}^{s_0} ds\, \mbox{Im}\,\P(s,q^2)e^{-s/M^2}\ .
\label{transf}
}
To write the terms in~\Eq{eq:terms} in the form of a dispersion relation, it is useful to define the variable $s(\sigma)$:
\eq{
s(\sigma) \equiv \sigma m_B^2 - \frac{\sigma q^2 - m_s^2}{\bar\sigma}\ ,
\label{eq:s(sigma)}
}
such that $k_\omega^2-m_s^2 = - \bar\sigma \,[s(\sigma)-k^2]$, before changing the integration variable from $\sigma$ to $s$.
This works directly for the terms with $n=0$ in~(\ref{eq:terms}).
For the terms with $n\ge 1$ we perform the integration by parts sequentially, in order to express them in terms of terms with $n=0$
plus ``boundary terms". Once all the terms have been written in dispersive form, we apply duality by cutting out the $s$ integrals
above $s>s_0$, after which we work backwards to the original non-dispersive form of the $d\sigma$ integrals with a cut-off at
$\sigma_0 \equiv \sigma(s_0)$.
Then we perform the Borel transformation in the variable $k^2$:
\eq{
\B_{M^2} \Bigg[ \frac1{(s-k^2)^N} \Bigg] =\frac{1}{(N-1)!} \frac{e^{-s/M^2}}{M^{2(N-1)}}
}
for generic $(s,N)$ -- and vanishing for $N\le 0$.

At the end of the day, we find that the whole procedure can be implemented by performing the following substitution in
the terms of the r.h.s of~\Eq{eq:terms}:
\eq{
\int_0^1 \! d\sigma \frac{\bar \sigma^{n+1}\, \hat I_n^{(i)}(\sigma)}{(k_\omega^2-m_s^2)^{n+1}} \longrightarrow
\bigg\{
\int_0^{\sigma_0} \!\! d\sigma \, \frac{I_n^{(i)}(\sigma)\,e^{-s(\sigma)/M^2}}{(M^2)^{n}}
+ e^{-s_0/M^2} \sum_{\ell=0}^{n-1} \frac{\eta(\sigma_0)\, \D_\eta^\ell\big[I_n^{(i)}\big](\sigma_0)}{(M^2)^{n-\ell-1}}
\bigg\}
\label{eq:MasterDuality}
}
where the second term in the r.h.s. (the boundary term) arises only for $n\ge 1$.
We have defined $s_0\equiv s(\sigma_0)$
and $\hat I_n^{(i)}(\sigma) =(-1)^{n+1}\, n!\, I_n^{(i)}(\sigma)$.
The operator $\D_\eta$ is defined by:
\eq{
\D_\eta [F](\sigma_0) = \frac{d}{d\sigma} \big[\eta(\sigma) F(\sigma)\big]_{\sigma=\sigma_0} \ ,
\label{eq:Deta}
}
which is applied $\ell$ times in each term of the above formula, so that:
\eq{
\D_\eta^0 [F](\sigma) = F(\sigma)\ ;\quad
\D_\eta^2 [F](\sigma) = \frac{d}{d\sigma} \bigg[\eta(\sigma) \frac{d}{d\sigma} [\eta(\sigma) F(\sigma)]\bigg]\ ; \quad \text{etc.}
}
The function $\eta(\sigma)$ is the Jacobian of the change of integration variable $\sigma(s)$:
\eq{
\eta(\sigma) \equiv \frac{d\sigma}{ds} = 
\frac{\bar \sigma^2}{ \bar \sigma^2 m_B^2 - (q^2-m_s^2)}\ .
}
Collecting everything leads to the structure in~\Eq{eq:FOPE} for each invariant amplitude.
All the results for the choices of currents $\Gamma_{a,b}$ relevant for $P$-wave form factors are collected in~\App{OPEexpressions}.

\subsection{Three-particle contributions}

Collecting the different pieces, we have:
\eq{
\P_{ab}^{(3)}(k,q) = \frac{f_B m_B}2 \int_0^1 \! du \int_0^\infty d \omega_1\, d\omega_2 \ \F_{ab}(u,\omega_1\omega_2) 
}
with
\eq{
\F_{ab}(u,\omega_1\omega_2) \equiv
\int d^4x\, \ e^{i k_\omega \cdot x} \ 
\Tr \Big[  \gamma_5 \Gamma_a \, \widetilde S_{(3)}^{\mu\nu}(x,u) \, \Gamma_b P_+ \Psi_{\mu\nu}(x,\omega_1,\omega_2) \Big]\ ,
\label{eq:Fab}
}
where $k_\omega = k - \sigma m_B v = \bar\sigma k - \sigma q$,\footnote{Note that the variable $\sigma$ here is different from the 2-particle case, but since it becomes an integration variable in both cases, it will not matter.}
and
\eq{
\widetilde S_{(3)}^{\mu\nu}(x,u) \equiv 
\int \frac{d^4 p}{(2\pi)^4}\, e^{-i p\cdot x}
\,\frac{\big[\bar u\, (\slashed{p}+m_s) \sigma^{\mu\nu} +u \,\sigma^{\mu\nu} (\slashed{p}+m_s) \big]}{2 (p^2 - m_s^2)^2} \ .
}
We perform the change of variable $u\to \sigma(u) \equiv (\omega_1 + u \omega_2)/m_B$, so that:
\eq{
\int_0^1\!\! du \int _0^\infty \!\!d\omega_1 \, d\omega_2 \ \F_{ab}(u,\omega_1 ,\omega_2) = 
\int_0^1 \!\!m_B\,d\sigma \int _0^{m_B \sigma}  \!\!d\omega_1 \int_{m_B\sigma -\omega_1}^\infty \!\! \frac{d\omega_2}{\omega_2}
\  \F_{ab}(u(\sigma),\omega_1,\omega_2)
}
with $u(\sigma) = (\sigma m_B - \omega_1)/\omega_2$. With this, we write:
\eq{
\P_{ab}^{(3)}(k,q) = \frac{f_B m_B^2}{2} \int_0^1 \!\!\,d\sigma \int _0^{m_B \sigma}  \!\!d\omega_1 \int_{m_B\sigma -\omega_1}^\infty \!\! \frac{d\omega_2}{\omega_2} \  \F_{ab}(u(\sigma),\omega_1,\omega_2)\ .
}
The  integrals over $d^4x$ in~\Eq{eq:Fab} are performed by using the following results:
\eqa{
\int d^4x \,\frac{d^4 p}{(2\pi)^4} e^{i (k_\omega - p)\cdot x} \ \frac{x^\mu (\slashed p + m_s)}{(p^2-m_s^2)^2} &=&
\frac{-i  \gamma^\mu}{(k_\omega^2-m_s^2)^2}  + \frac{4 i k_\omega^\mu (\slashed k_\omega + m_s)}{(k_\omega^2-m_s^2)^3} 
\ ,\\[2mm]
\int d^4x \,\frac{d^4 p}{(2\pi)^4} e^{i (k_\omega - p)\cdot x} \ \frac{x^\mu x^\nu (\slashed p + m_s)}{(p^2-m_s^2)^2} &=&
\frac{4\big[k_\omega^\mu \gamma^\nu + k_\omega^\nu \gamma^\mu + g^{\mu\nu} (\slashed k_\omega+m_s)\big]}{(k_\omega^2-m_s^2)^3} 
- \frac{24 k_\omega^\mu k_\omega^\nu (\slashed k_\omega + m_s)}{(k_\omega^2-m_s^2)^4}  \ ,\nonumber\\
}
which follow from~\Eq{eq:d4xd4p}. After the integration over $d^4x$ and the traces in~\Eq{eq:Fab} have been performed,
one identifies the various invariant amplitudes. These depend on $k_\omega$ as in the two-particle case discussed above,
and the Borel transformation over $k^2$ is done analogously.
In particular, in analogy with~\Eq{eq:terms}, we have
\eq{
\frac{m_B}{2}
 \int _0^{m_B \sigma}  \!\!d\omega_1 \int_{m_B\sigma -\omega_1}^\infty \!\! \frac{d\omega_2}{\omega_2} \  \F_{ab}(u(\sigma),\omega_1,\omega_2)
= \sum_n  
\frac{\bar \sigma^n \hat I^{(i)}_n(\sigma)}{(k_\omega^2-m_s^2)^{n+1}}\ ,
}
which gives rise to the same structure in~\Eq{eq:FOPE} for the 3-particle contributions. 
The results for all relevant choices of the currents $\Gamma_{a,b}$ are collected in~\App{OPEexpressions}. 

\subsection{Explicit derivation of \Eq{eq:MasterDuality}}

We start considering (without justification) the integral:
\eq{
\int_{\sigma_1}^{\sigma_2} d\sigma \frac{f(\sigma)}{(k_\omega^2 - m_s^2)^\ell}
= (-1)^\ell \int_{\sigma_1}^{\sigma_2} d\sigma \frac{g(\sigma)}{(s- k^2)^\ell}\ .
}
where $g(\sigma) = f(\sigma)/\bar \sigma^\ell$. The equality of the two integrals follows from~\Eq{eq:s(sigma)}. We want to derive a formula for the integral 
\eq{
\I_\ell(g;\sigma_1,\sigma_2) \equiv \int_{\sigma_1}^{\sigma_2} d\sigma \frac{g(\sigma)}{(s- k^2)^\ell}
}
for any function $g(\sigma)$ and any integer $\ell\ge 1$, and any real numbers $0<\sigma_1<\sigma_2<1$.

We start by integrating by parts once:
\eq{
\I_\ell(g;\sigma_1,\sigma_2) =
\frac1{\ell-1} \bigg\{  \frac{\eta(\sigma_1)g(\sigma_1)}{(s_1-k^2)^{\ell-1}} - \frac{\eta(\sigma_2)g(\sigma_2)}{(s_2-k^2)^{\ell-1}} \bigg\}
+ \frac{\I_{\ell-1}(\D_\eta[g],\sigma_1,\sigma_2)}{\ell-1}
}
where $s_i=s(\sigma_i)$.
Then we infer the result of integrating by parts until the integrand has a single pole:
\eq{
\label{induction}
\I_\ell(g;\sigma_1,\sigma_2) =
S_\ell(\sigma_1) - S_\ell(\sigma_2)
+ \frac{\I_1(\D_\eta^{\ell-1}[g];\sigma_1,\sigma_2)}{(\ell-1)!}
}
with
\eq{
S_\ell(\sigma_i) \equiv \sum_{n=1}^{\ell-1}
\frac{(\ell-1-n)!}{(\ell-1)!} \ 
\frac{\eta(\sigma_i)\D_\eta^{n-1}[g](\sigma_i)}{(s_i-k^2)^{\ell-n}}
}
and where the notation is:
\eq{
\D_\eta [g](x) = \frac{d}{d\sigma} [\eta(\sigma) g(\sigma)]_{\sigma=x}\ ,\quad
\D_\eta^2 [g](x) = \frac{d}{d\sigma} \Big[\eta(\sigma) \frac{d}{d\sigma} [\eta(\sigma) g(\sigma)]\Big]_{\sigma=x}\ ,\quad \dots
}
\Eq{induction} can be proven by induction easily.

\bigskip

Let us now consider $\I_\ell(g;0,1)$ and assume duality holds for $s>s_0$.
The integral $\I_1$ in~\Eq{induction} can be written as an integral over $s\in (m_s^2,\infty)$, while $\eta(1)=0$ (which implies that $S_\ell(1)=0$),
and the remaining surface term $S_\ell(0)$ has a singularity at $k^2=m_s^2 \ll s_0$, not supposed to be contained in the OPE integral above $s_0$.
Therefore the duality subtraction should be applied only on the $\I_1$ term:
\eq{
\I_\ell(g;0,1) \to S_\ell(0) + \frac{\I_1(\D_\eta^{\ell-1}[g];0,\sigma_0)}{(\ell-1)!}\ .
}
We can rewrite the r.h.s of the previous equation as:
\eq{
S_\ell(0) + \frac{\I_1(\D_\eta^{\ell-1}[g];0,\sigma_0)}{(\ell-1)!} =
S_\ell(0) + \big[ \I_\ell(g;0,\sigma_0) - S_\ell(0) + S_\ell(\sigma_0)  \big] = 
S_\ell(\sigma_0) + \I_\ell(g;0,\sigma_0)
}
which gives the prescription for the duality approximation:
\eq{
\int_{0}^{1} d\sigma \frac{\bar\sigma^\ell\, g(\sigma)}{(k_\omega^2 - m_s^2)^\ell} \to
(-1)^\ell \Bigg\{
\sum_{n=1}^{\ell-1}
\frac{(\ell-n-1)!}{(\ell-1)!} \ 
\frac{\eta(\sigma_0)\D_\eta^{n-1}[g](\sigma_0)}{(s_0-k^2)^{\ell-n}}
+ \int_{0}^{\sigma_0} d\sigma \frac{g(\sigma)}{(s - k^2)^\ell}
\bigg\}\ .
}
Applying the Borel transformation in the variable $k^2$ leads exactly to~\Eq{eq:MasterDuality}.

\section{OPE expressions}
\label{OPEexpressions}

We present here the OPE expressions on the right-hand side of the sum rules, including two- and three-particle contributions up to twist-4,
as discussed in \App{BLCDAs}. The generic form for any form factor is written as:
\eqa{
&&\hspace{-15mm} \P_i^{(T),\text{OPE}}(q^2,\sigma_0,M^2) = \nonumber\\
&&\sum_{n\ge 0} \ \frac{f_B m_B}{(M^2)^n} \ \bigg\{ \int\limits_0^{\sigma_0} d\sigma~e^{-s(\sigma)/M^2}
I_{i,n}^{(T)}(\sigma)
+ \sum_{\ell\ge 0}
\eta(\sigma_0) \D_\eta^\ell[I_{i,n+\ell+1}^{(T)}](\sigma_0) \ e^{-s_0/M^2}
\bigg\}
\ ,
\label{eq:FOPE}
}
where the functions $I^{(T)}_n$ are a sum of two- and three-particle contributions:
\eq{
I_{i,n}^{(T)}(\sigma) = I_{i,n}^{(2)(T)}(\sigma) + 
\int _0^{m_B \sigma}  \!\!d\omega_1 \int_{m_B\sigma -\omega_1}^\infty \!\! \frac{d\omega_2}{\omega_2}\, I_{i,n}^{(3)(T)}(\sigma,\omega_1,\omega_2)\ .
}
The operator $\D_\eta$ is defined in~\Eq{eq:Deta}, and 
\eqa{
&& \eta(\sigma) = \frac{\bar\sigma^2}{\bar\sigma^2 m_B^2-(q^2-m_s^2)}\ , \quad
\hat s(\sigma) = \sigma - \frac{\sigma \hat q^2 -\hat m_s^2}{\bar \sigma}\ , \nonumber\\
&& \sigma(s) = \frac12 \bigg\{ 1+\hat s - \hat q^2 - \sqrt{(1-\hat s+\hat q^2)^2 - 4(\hat q^2-\hat m_s^2)}  \bigg\}\ ,
}
with $\hat s \equiv s/m_B^2$, $\hat q^2 \equiv q^2/m_B^2$ and $\hat m_s \equiv m_s/m_B$.

\bigskip

The full expressions for the coefficients $I_{i,n}^{(2)(T)}(\sigma)$ and
$I_{i,n}^{(3)(T)}(\sigma,\omega_1,\omega_2)$ are given in electronic format as a supplementary file called `{\tt OPEcoefficients.m}' (see below for more details). For easy reference and comparison, we reproduce here only the results for the two-particle coefficients
$I_{i,n}^{(2)(T)}(\sigma)$ for $q^2=0$ and in the limit $m_s\to 0$:
\eqa{
I_{\perp,0}^{(2)}(\sigma) &=& \frac{\phi_+}{\bar\sigma}\ , \quad
I_{\|,0}^{(2)}(\sigma) = \frac{m_B^2\,\phi_+}2 + \frac{2\,g_+}{\bar\sigma^2}\ , \quad
I_{-,0}^{(2)}(\sigma) = \frac{(\sigma-\bar\sigma)\,\phi_+}{\bar\sigma}\ ,
\nonumber\\
I_{t,0}^{(2)}(\sigma) &=& -\frac{m_b\,\sigma\, \bar\Phi_\pm}{\bar\sigma^2} -\frac{m_b m_B\sigma\,g_+}{\bar\sigma}\ , \quad
I_{\perp,0}^{(2)T}(\sigma) = m_B\,\phi_+ \ , \quad
\nonumber\\
I_{\|,0}^{(2)T}(\sigma) &=& \frac{m_B^3\bar\sigma\,\phi_+}2 - \frac{2m_B\,g_+}{\bar\sigma}
-\frac{2\, \bar G_\pm}{\bar\sigma^2} \ , \quad
I_{-,0}^{(2)T}(\sigma) = -\frac{m_B(1+\sigma)\,\phi_+}{\bar\sigma}
-\frac{2 \sigma\, \bar\Phi_\pm}{\bar\sigma^2} \ ,
\nonumber\\[4mm]
I_{\perp,1}^{(2)}(\sigma) &=& -\frac{4\,g_+}{\bar\sigma^2}\ , \ 
I_{\|,1}^{(2)}(\sigma) = -\frac{2m_B^2\, g_+}{\bar\sigma}
-\frac{4m_B\,\bar G_{\pm}}{\bar\sigma^2}\ , \ 
I_{-,1}^{(2)}(\sigma) = -\frac{2 m_B \sigma\, \bar\Phi_\pm}{\bar\sigma} -\frac{4(\sigma-\bar\sigma)\,g_+}{\bar\sigma^2}\ , \quad
\hspace{-1cm} \nonumber\\
I_{t,1}^{(2)}(\sigma) &=& \frac{m_b m_B^2\sigma\,\bar\Phi_\pm}{\bar\sigma}
+ \frac{4m_b m_B \sigma\,g_+}{\bar\sigma^2} + \frac{4m_b(1+\sigma)\,\bar G_\pm}{\bar\sigma^3}\ ,\ 
I_{\|,1}^{(2)T}(\sigma) = -2m_B^3\,g_+ - \frac{2m_B^2\, \bar G_\pm}{\bar\sigma}\ ,
\hspace{-1cm} \nonumber\\
I_{-,1}^{(2)T}(\sigma) &=& -\frac{2m_B^2\sigma\,\bar\Phi_\pm}{\bar\sigma}
+ \frac{4m_B (1+\sigma)\,g_+}{\bar\sigma^2} + \frac{4(5\sigma-1)\,\bar G_\pm}{\bar\sigma^3}\ ,\ 
I_{\perp,1}^{(2)T}(\sigma) = -\frac{4m_B\,g_+}{\bar\sigma}-\frac{4\, \bar G_\pm}{\bar\sigma^2}\ ,
\hspace{-1cm} \nonumber\\[4mm]
I_{-,2}^{(2)}(\sigma) &=& \frac{8m_B\sigma\,\bar G_{\pm}}{\bar\sigma^2}\ , \quad
I_{t,2}^{(2)}(\sigma) = -\frac{4m_b m_B^2\sigma\,\bar G_{\pm}}{\bar\sigma^2}\ , \quad
I_{-,2}^{(2)T}(\sigma) = \frac{8m_B^2\sigma\,\bar G_{\pm}}{\bar\sigma^2}\ ,
\label{eq:OPEfunctions}
}
where for brevity we have omitted the arguments of the LCDAs, $\phi_+\equiv \phi_+(m_B\sigma)$, etc.
These results can be easily extracted from the ancillary Mathematica package `{\tt OPEcoefficients.m}'. For example, the expression for $I_{\|,1}^{(2)T}(\sigma)$ given in~\Eq{eq:OPEfunctions} is obtained by typing in a Mathematica notebook:

\bigskip

{\tt IparT[2,1]/.(<<"OPEcoefficients.m")/.\{ms -> 0,q2 -> 0\}}

\bigskip

\noindent The arguments in brackets are such that $I^{(k)}_{\perp,n}$={\tt Iperp[k,n]}, for example.
For the three-particle contributions, one needs to take into account that
$\bar u \equiv 1-u$ and $u = (\sigma m_B -\omega_1)/\omega_2$.
The syntax is otherwise obvious by looking at the given expressions.


\section{Results for OPE coefficients in different models}
\label{sec:OPEnumerics}

In this appendix we expand on the results of~\Sec{sec:OPEz} and collect the results for the OPE
coefficients $\kappa_i^{(T),\rm OPE}$ and $\eta_i^{(T),\rm OPE}$  in the three different models for the $B$-meson LCDAs presented in~\App{sec:modelsLCDAs}. This is shown in Tables~\ref{tab:OPEparsV} and~\ref{tab:OPEparsT}, which complement~\Tab{tab:OPEpars} by adding the corresponding results in Models IIA and IIB. The conclusion from the results presented here is the estimate of model-dependence quoted at the end of~\Sec{sec:OPEz}, which is obtained simply by taking the maximum spread of the central values in the three models for each parameter.

\begin{table}
\centering
\setlength{\tabcolsep}{10pt}
\begin{tabular}{@{}clll@{}}
\toprule[0.7mm]
$B$-LCDA Model & $\ M^2=1.00\GeV^2$ & $\ M^2=1.25\GeV^2$ & $\ M^2=1.50\GeV^2$ \\
\midrule[0.7mm]
\multirow{2}{*}{Model MI}
& $\kappa_\perp^{\rm OPE}=+0.007(4)$ 
& $\kappa_\perp^{\rm OPE}=+0.008(5)$ 
& $\kappa_\perp^{\rm OPE}=+0.009(5)$ \\
& $\eta_\perp^{\rm OPE}=-0.010(14)$ 
& $\eta_\perp^{\rm OPE}=-0.012(17)$ 
& $\eta_\perp^{\rm OPE}=-0.013(19)$ \\\midrule
\multirow{2}{*}{Model MIIA}
& $\kappa_\perp^{\rm OPE}=+0.006(4)$ 
& $\kappa_\perp^{\rm OPE}=+0.008(4)$ 
& $\kappa_\perp^{\rm OPE}=+0.009(5)$ \\
& $\eta_\perp^{\rm OPE}=-0.024(19)$ 
& $\eta_\perp^{\rm OPE}=-0.029(23)$ 
& $\eta_\perp^{\rm OPE}=-0.034(26)$ \\\midrule
\multirow{2}{*}{Model MIIB}
& $\kappa_\perp^{\rm OPE}=+0.007(4)$ 
& $\kappa_\perp^{\rm OPE}=+0.008(5)$ 
& $\kappa_\perp^{\rm OPE}=+0.009(5)$ \\
& $\eta_\perp^{\rm OPE}=-0.020(17)$ 
& $\eta_\perp^{\rm OPE}=-0.023(20)$ 
& $\eta_\perp^{\rm OPE}=-0.027(23)$ \\\midrule[0.5mm]
\multirow{2}{*}{Model MI}
& $\kappa_\|^{\rm OPE}=+0.100(58)$ 
& $\kappa_\|^{\rm OPE}=+0.120(69)$ 
& $\kappa_\|^{\rm OPE}=+0.137(78)$ \\
& $\eta_\|^{\rm OPE}=+0.246(85)$ 
& $\eta_\|^{\rm OPE}=+0.304(108)$ 
& $\eta_\|^{\rm OPE}=+0.355(128)$ \\\midrule
\multirow{2}{*}{Model MIIA}
& $\kappa_\|^{\rm OPE}=+0.092(52)$ 
& $\kappa_\|^{\rm OPE}=+0.112(63)$ 
& $\kappa_\|^{\rm OPE}=+0.128(73)$ \\
& $\eta_\|^{\rm OPE}=+0.083(15)$ 
& $\eta_\|^{\rm OPE}=+0.102(18)$ 
& $\eta_\|^{\rm OPE}=+0.118(20)$ \\\midrule
\multirow{2}{*}{Model MIIB}
& $\kappa_\|^{\rm OPE}=+0.098(55)$ 
& $\kappa_\|^{\rm OPE}=+0.118(66)$ 
& $\kappa_\|^{\rm OPE}=+0.135(76)$ \\
& $\eta_\|^{\rm OPE}=+0.146(39)$ 
& $\eta_\|^{\rm OPE}=+0.181(51)$ 
& $\eta_\|^{\rm OPE}=+0.212(62)$ \\\midrule[0.5mm]
\multirow{2}{*}{Model MI}
& $\kappa_-^{\rm OPE}=-0.004(3)$ 
& $\kappa_-^{\rm OPE}=-0.004(4)$ 
& $\kappa_-^{\rm OPE}=-0.005(5)$ \\
& $\eta_-^{\rm OPE}=-0.020(17)$ 
& $\eta_-^{\rm OPE}=-0.025(21)$ 
& $\eta_-^{\rm OPE}=-0.029(24)$ \\\midrule
\multirow{2}{*}{Model MIIA}
& $\kappa_-^{\rm OPE}=-0.003(3)$ 
& $\kappa_-^{\rm OPE}=-0.003(4)$ 
& $\kappa_-^{\rm OPE}=-0.004(4)$ \\
& $\eta_-^{\rm OPE}=-0.018(18)$ 
& $\eta_-^{\rm OPE}=-0.021(22)$ 
& $\eta_-^{\rm OPE}=-0.024(26)$ \\\midrule
\multirow{2}{*}{Model MIIB}
& $\kappa_-^{\rm OPE}=-0.003(3)$ 
& $\kappa_-^{\rm OPE}=-0.004(4)$ 
& $\kappa_-^{\rm OPE}=-0.004(4)$ \\
& $\eta_-^{\rm OPE}=-0.019(18)$ 
& $\eta_-^{\rm OPE}=-0.023(22)$ 
& $\eta_-^{\rm OPE}=-0.026(26)$ \\\midrule[0.5mm]
\multirow{2}{*}{Model MI}
& $\kappa_t^{\rm OPE}=-0.043(9)$ 
& $\kappa_t^{\rm OPE}=-0.052(11)$ 
& $\kappa_t^{\rm OPE}=-0.060(12)$ \\
& $\eta_t^{\rm OPE}=+0.210(30)$ 
& $\eta_t^{\rm OPE}=+0.249(34)$ 
& $\eta_t^{\rm OPE}=+0.282(37)$ \\\midrule
\multirow{2}{*}{Model MIIA}
& $\kappa_t^{\rm OPE}=-0.045(10)$ 
& $\kappa_t^{\rm OPE}=-0.054(12)$ 
& $\kappa_t^{\rm OPE}=-0.062(14)$ \\
& $\eta_t^{\rm OPE}=+0.220(41)$ 
& $\eta_t^{\rm OPE}=+0.264(47)$ 
& $\eta_t^{\rm OPE}=+0.301(52)$ \\\midrule
\multirow{2}{*}{Model MIIB}
& $\kappa_t^{\rm OPE}=-0.045(10)$ 
& $\kappa_t^{\rm OPE}=-0.055(12)$ 
& $\kappa_t^{\rm OPE}=-0.063(13)$ \\
& $\eta_t^{\rm OPE}=+0.216(35)$ 
& $\eta_t^{\rm OPE}=+0.257(39)$ 
& $\eta_t^{\rm OPE}=+0.291(42)$ \\
\bottomrule[0.7mm]
\end{tabular}
\caption{\it Results for the OPE coefficients in the $z$-expansion of vector, axial-vector and timelike-helicity form factors in the three different models considered for $B$-meson LCDAs.}
\label{tab:OPEparsV}
\end{table}

\begin{table}
\centering
\setlength{\tabcolsep}{10pt}
\begin{tabular}{@{}clll@{}}
\toprule[0.7mm]
$B$-LCDA Model & $\ M^2=1.00\GeV^2$ & $\ M^2=1.25\GeV^2$ & $\ M^2=1.50\GeV^2$ \\
\midrule[0.7mm]
\multirow{2}{*}{Model MI}
& $\kappa_\perp^{T,\rm OPE}=+0.036(21)$ 
& $\kappa_\perp^{T,\rm OPE}=+0.043(25)$ 
& $\kappa_\perp^{T,\rm OPE}=+0.050(29)$ \\
& $\eta_\perp^{T,\rm OPE}=-0.056(73)$ 
& $\eta_\perp^{T,\rm OPE}=-0.065(85)$ 
& $\eta_\perp^{T,\rm OPE}=-0.071(94)$ \\\midrule
\multirow{2}{*}{Model MIIA}
& $\kappa_\perp^{T,\rm OPE}=+0.034(19)$ 
& $\kappa_\perp^{T,\rm OPE}=+0.041(23)$ 
& $\kappa_\perp^{T,\rm OPE}=+0.047(27)$ \\
& $\eta_\perp^{T,\rm OPE}=-0.122(95)$ 
& $\eta_\perp^{T,\rm OPE}=-0.148(115)$ 
& $\eta_\perp^{T,\rm OPE}=-0.171(133)$ \\\midrule
\multirow{2}{*}{Model MIIB}
& $\kappa_\perp^{T,\rm OPE}=+0.036(20)$ 
& $\kappa_\perp^{T,\rm OPE}=+0.043(24)$ 
& $\kappa_\perp^{T,\rm OPE}=+0.049(28)$ \\
& $\eta_\perp^{T,\rm OPE}=-0.103(85)$ 
& $\eta_\perp^{T,\rm OPE}=-0.122(101)$ 
& $\eta_\perp^{T,\rm OPE}=-0.139(114)$ \\\midrule[0.5mm]
\multirow{2}{*}{Model MI}
& $\kappa_\|^{T,\rm OPE}=+0.492(290)$ 
& $\kappa_\|^{T,\rm OPE}=+0.589(346)$ 
& $\kappa_\|^{T,\rm OPE}=+0.671(393)$ \\
& $\eta_\|^{T,\rm OPE}=+1.347(458)$ 
& $\eta_\|^{T,\rm OPE}=+1.662(582)$ 
& $\eta_\|^{T,\rm OPE}=+1.939(693)$ \\\midrule
\multirow{2}{*}{Model MIIA}
& $\kappa_\|^{T,\rm OPE}=+0.461(261)$ 
& $\kappa_\|^{T,\rm OPE}=+0.556(316)$ 
& $\kappa_\|^{T,\rm OPE}=+0.638(364)$ \\
& $\eta_\|^{T,\rm OPE}=+0.499(78)$ 
& $\eta_\|^{T,\rm OPE}=+0.614(96)$ 
& $\eta_\|^{T,\rm OPE}=+0.713(112)$ \\\midrule
\multirow{2}{*}{Model MIIB}
& $\kappa_\|^{T,\rm OPE}=+0.488(276)$ 
& $\kappa_\|^{T,\rm OPE}=+0.586(331)$ 
& $\kappa_\|^{T,\rm OPE}=+0.671(379)$ \\
& $\eta_\|^{T,\rm OPE}=+0.828(230)$ 
& $\eta_\|^{T,\rm OPE}=+1.025(298)$ 
& $\eta_\|^{T,\rm OPE}=+1.200(360)$ \\\midrule[0.5mm]
\multirow{2}{*}{Model MI}
& $\kappa_-^{T,\rm OPE}=-0.021(19)$ 
& $\kappa_-^{T,\rm OPE}=-0.025(23)$ 
& $\kappa_-^{T,\rm OPE}=-0.028(26)$ \\
& $\eta_-^{T,\rm OPE}=-0.098(103)$ 
& $\eta_-^{T,\rm OPE}=-0.121(126)$ 
& $\eta_-^{T,\rm OPE}=-0.141(146)$ \\\midrule
\multirow{2}{*}{Model MIIA}
& $\kappa_-^{T,\rm OPE}=-0.015(16)$ 
& $\kappa_-^{T,\rm OPE}=-0.018(20)$ 
& $\kappa_-^{T,\rm OPE}=-0.021(23)$ \\
& $\eta_-^{T,\rm OPE}=-0.087(108)$ 
& $\eta_-^{T,\rm OPE}=-0.104(135)$ 
& $\eta_-^{T,\rm OPE}=-0.118(160)$ \\\midrule
\multirow{2}{*}{Model MIIB}
& $\kappa_-^{T,\rm OPE}=-0.018(18)$ 
& $\kappa_-^{T,\rm OPE}=-0.021(22)$ 
& $\kappa_-^{T,\rm OPE}=-0.025(25)$ \\
& $\eta_-^{T,\rm OPE}=-0.091(109)$ 
& $\eta_-^{T,\rm OPE}=-0.110(134)$ 
& $\eta_-^{T,\rm OPE}=-0.126(157)$ \\
\bottomrule[0.7mm]
\end{tabular}
\caption{\it Results for the OPE coefficients in the $z$-expansion of tensor form factors in the three different models considered for $B$-meson LCDAs.}
\label{tab:OPEparsT}
\end{table}

As mentioned in~\Sec{sec:OPEz}, the quoted parametric uncertainties in the OPE coefficients shown in~Tables~\ref{tab:OPEpars},~\ref{tab:OPEparsV} and~\ref{tab:OPEparsT} are strongly correlated.
These correlations are described by one $42\times 42$ correlation matrix for each of the three models.
We do not find it illustrative to display in the text these correlation matrices. However, we wish to make available the correlation matrix for Model~I in electronic format, 
since it is convenient for reproducing the results in~\Sec{sec:BtoK*NWL} without the need to evaluate all the OPE functions.  
The corresponding file is a Mathematica package called `{\tt CorrMatrixOPEparameters.m}', which is available from the authors upon request. The order of the correlation coefficients in this matrix is given by:
\eq{
\{
\underbrace{\kappa_\perp^{\rm OPE}, \eta_\perp^{\rm OPE}, \kappa_\|^{\rm OPE}, \eta_\|^{\rm OPE},\dots,
\kappa_-^{T,\rm OPE}, \eta_-^{T,\rm OPE}}_{M^2=1}\,,\,
\underbrace{\kappa_\perp^{\rm OPE}, \eta_\perp^{\rm OPE},\dots}_{M^2=1.25}\,,\,
\underbrace{\kappa_\perp^{\rm OPE}, \eta_\perp^{\rm OPE},\dots}_{M^2=1.5}
\}\ ,
}
where the order among the parameters gathered under each horizontal brace is given by $\{\xi_\perp,\xi_\|,\xi_-,\xi_t,\xi_\perp^T,\xi_\|^T,\xi_-^T\}$.
For example, the element $(7,24)$ in this matrix contains the correlation coefficient:
\eq{
\text{corr}\Big(\kappa_t^{\rm OPE}[M^2=1\GeV], \eta_\perp^{T,\rm OPE}[M^2=1.25\GeV]\Big) = 0.96\ .
}
When using this correlation matrix one must take into account the degeneracies that exist among several of the form factor parameters.

\section{Beyond the narrow-width approximation}
\label{app:beyondNWL}

\subsection{Breit-Wigner model with fixed widths}

We consider the convolution of a Breit-Wigner function $D(s,\Gamma)$ with a smooth function $f(s)$
\begin{equation}
J=\int_{s_{\rm min}}^{s_{\rm max}} \frac{ds}{2\pi} D(s,\Gamma) f(s)\ ,
\qquad D(s,\Gamma)=\frac{2m\Gamma}{(s-m^2)^2+\Gamma^2 m^2}\ ,
\label{eq:J}
\end{equation}
which in the narrow-width limit $\Gamma\to 0$  becomes
\begin{equation}
J \to\ J_0=\int_{s_{\rm min}}^{s_{\rm max}} ds\ \delta(s-m^2) f(s)=f(m^2)\ .
\end{equation}
We want to determine the $\op(\Gamma)$ correction to this limit, corresponding to $J-J_0$ at first order in $\Gamma$. To this end, we can use a similar approach as in \Ref{Uhlemann:2008pm}, expressing the difference between the integral and its narrow-width limit as
\begin{equation}
J-J_0=\int_{s_{\rm min}}^{s_{\rm max}} \frac{ds}{2\pi} D(s,\Gamma) [f(s)-f(m^2)]-\alpha J_0\ ,
\end{equation}
with 
\begin{eqnarray}
\alpha&=&1-\int_{s_{\rm min}}^{s_{\rm max}} \frac{ds}{2\pi} D(s,\Gamma)
  =1+\frac{1}{\pi}\arctan\frac{m^2-s_{\rm max}}{m\Gamma}-\frac{1}{\pi}\arctan\frac{m^2-s_{\rm min}}{m\Gamma}
\nonumber\\
 & =&\frac{m\Gamma}{\pi}\frac{s_{\rm max}-s_{\rm min}}{(s_{\rm max}-m^2)(m^2-s_{\rm min})} + \op(\Gamma^2)\ .
 \label{eq:alpha}
\end{eqnarray}
We then have
\begin{eqnarray}
\frac{J-J_0}{J_0}=\int_{s_{\rm min}}^{s_{\rm max}} \frac{ds}{2\pi} D(s,\Gamma) \left[\frac{f(s)}{f(m^2)}-1\right]-\alpha
 =\int_{s_{\rm min}}^{s_{\rm max}} \frac{ds}{2\pi} D(s,\Gamma) [g(s)-g(m^2)]-\alpha\ ,\quad
\end{eqnarray}
where
\begin{equation}
g(s)=\frac{f(s)}{f(m^2)}\ .
\end{equation}
Performing a Taylor expansion of the smooth function $g(s)$, we write
\begin{eqnarray}
\frac{J-J_0}{J_0}=-\alpha+\int_{s_{\rm min}}^{s_{\rm max}} \frac{ds}{2\pi} D(s,\Gamma) 
 \left[(s-m^2)g'(m^2)+\sum_{n\geq 2} \frac{(s-m^2)^n}{n!} g^{(n)}(m^2)\right]\ .
 \end{eqnarray}
Using the expression for $\alpha$ in~\Eq{eq:alpha}, and integrating each of the remaining terms, we find
\eqa{
\frac{J-J_0}{J_0}&=&
  -\frac{m\Gamma}{\pi}\frac{s_{\rm max}-s_{\rm min}}{(s_{\rm max}-m^2)(m^2-s_{\rm min})}
  +\frac{2m\Gamma g'(m^2)}{4\pi}\log\frac{(s_{\rm max}-m^2)^2+m^2\Gamma^2}{(s_{\rm min}-m^2)^2+m^2\Gamma^2}
\nonumber\\
&& + \sum_{n\geq 2} \frac{1}{n!} g^{(n)}(m^2) K_n\ ,
}
with $K_n$ defined as
\begin{equation}
K_n=\int_{s_{\rm min}}^{s_{\rm max}} \frac{ds}{2\pi} D(s,\Gamma) (s-m^2)^n\ .
\end{equation}
We should take the expansion of the above expressions up to $\op(\Gamma^2)$. In particular,
\begin{equation}
K_n=\frac{1}{(2\pi)}\frac{2m\Gamma}{n-1}[(s_{\rm max}-m^2)^{n-1}-(s_{\rm min}-m^2)^{n-1}]+\op(\Gamma)\ .
\end{equation}
We now define the intermediate function
\begin{equation}
G(s)=\sum_{n\geq 2} g^{(n)}(m^2) \frac{1}{n!}  \frac{1}{n-1} (s-m^2)^{n-1}
\end{equation}
such that
\begin{eqnarray}
G'(s)&=&\sum_{n\geq 2} g^{(n)}(m^2) \frac{1}{n!}   (s-m^2)^{n-2}
 =\frac{1}{(s-m^2)^2}[g(s)-g(m^2)-g'(m^2)(s-m^2)]\ ,\quad
 \\
G(s)&=&\int_{m^2}^{s} d\tau   \frac{1}{(\tau-m^2)^2}[g(\tau)-g(m^2)-g'(m^2)(\tau-m^2)]\ .
\end{eqnarray}
This leads to
\begin{eqnarray}
\frac{J-J_0}{J_0}&=&
  -\frac{m\Gamma}{\pi}\frac{s_{\rm max}-s_{\rm min}}{(s_{\rm max}-m^2)(m^2-s_{\rm min})}
  +\frac{2m\Gamma g'(m^2)}{2\pi}\log\frac{s_{\rm max}-m^2}{m^2-s_{\rm min}}
\nonumber\\
&&  \qquad+\frac{2m\Gamma}{2\pi} \sum_{n\geq 2} \frac{1}{(n-1)n!} g^{(n)}(m^2) [(s_{\rm max}-m^2)^{n-1}-(s_{\rm min}-m^2)^{n-1}]+\op(\Gamma^2)\nonumber\\
&=& -\frac{m\Gamma}{\pi}\frac{s_{\rm max}-s_{\rm min}}{(s_{\rm max}-m^2)(m^2-s_{\rm min})}
  +\frac{2m\Gamma g'(m^2)}{2\pi}\log\frac{s_{\rm max}-m^2}{m^2-s_{\rm min}}
  \nonumber\\
&&\qquad  +\frac{2m\Gamma}{2\pi} [G(s_{\rm max})-G(s_{\rm min})] +\op(\Gamma^2)\ ,
\end{eqnarray}
and therefore
\begin{eqnarray}
\frac{J-J_0}{J_0}
&=& \frac{m\Gamma}{\pi f(m^2)}
 \Bigg[-f(m^2)\frac{s_{\rm max}-s_{\rm min}}{(s_{\rm max}-m^2)(m^2-s_{\rm min})}
 \nonumber\\
&&\qquad\quad
    +f'(m^2)\log\frac{s_{\rm max}-m^2}{m^2-s_{\rm min}}+F(s_{\rm max},m)-F(s_{\rm min},m)\Bigg]+\op(\Gamma^2)\ ,
\end{eqnarray}
with the following definition of $F(s,m)$
\begin{equation}
F(s,m)=\int_{m^2}^{s} d\tau   \frac{1}{(\tau-m^2)^2}[f(\tau)-f(m^2)-f'(m^2)(\tau-m^2)]\ .
\end{equation}
The validity of this formula can be checked for polynomial expressions without difficulty, and it provides the relative error due to the narrow-width approximation.

\subsection{The case of $s$-dependent widths}

We can adapt the formalism to a Breit-Wigner resonance with  a $s$-dependent width
\begin{equation}
J=\int_{s_{\rm min}}^{s_{\rm max}} ds\ \tilde{D}(s,\Gamma) \Phi(s)\ ,
\qquad \tilde{D}(s,\Gamma)=\frac{1}{\pi}\frac{\sqrt{s}\Gamma(s)}{(s-m^2)^2+\Gamma(s)^2 s}\ ,
\end{equation}
where $\Gamma(s)$ is a function of the form
\begin{equation}
\Gamma(s)=\Gamma\times \gamma(s)\ , \qquad \Gamma=\Gamma(m^2)\ , \qquad \gamma(m^2)=1\ .
\end{equation}
We can recover the structure in~\Eq{eq:J} by defining
\begin{equation}
f(s)=\Phi(m^2)\phi(s)\rho(s,\Gamma)\ ,
\  \phi(s)=\frac{\Phi(s)}{\Phi(m^2)}\ ,
\  \rho(s,\Gamma)=\gamma(s)\frac{\sqrt{s}}{m}\frac{(s-m^2)^2+\Gamma^2 m^2}{(s-m^2)^2+\Gamma^2\gamma(s)^2 s}\ .
\end{equation}
Now $f(s)$ depends also on $\Gamma$, but this dependence starts only at $\op(\Gamma^2)$:
\begin{equation}
\rho(s,\Gamma)=\rho(s) + \op(\Gamma^2/m^2)\ ,\qquad
\rho(s)=\gamma(s)\frac{\sqrt{s}}{m}\ .
\end{equation}
These functions fulfill the relations
\begin{equation}
\rho(m^2)=1\ ,
\quad \rho'(m^2)=\frac{1}{2m^2}+\gamma'(m^2)\ ,
\quad \phi(m^2)=1\ ,
\quad f(m^2)=\Phi(m^2)\ .
\end{equation}
We can now apply the previous analysis, with $J_0=\Phi(m^2)$ and
\begin{eqnarray}
&&\frac{J-J_0}{J_0}
= \frac{\Gamma}{m}\frac{1}{\pi}
\Bigg[-\frac{m^2(s_{\rm max}-s_{\rm min})}{(s_{\rm max}-m^2)(m^2-s_{\rm min})}
+m^2[\phi'(m^2)+\rho'(m^2,\Gamma)] \log\frac{s_{\rm max}-m^2}{m^2-s_{\rm min}} 
\nonumber\\
&&\hspace{4cm}
+\tilde{F}(s_{\rm max},m,0)-\tilde{F}(s_{\rm min},m,0)\Bigg]+\op(\Gamma^2)\ ,
\end{eqnarray}
with the definition of $\tilde{F}(s,m,\Gamma)$
\begin{equation}
\tilde{F}(s,m,\Gamma)=\int_{1}^{s/m^2} d\tau  \frac{1}{(\tau-1)^2}\left[\phi(m^2\tau)\rho(m^2\tau)-1-(\tau-1)m^2[\phi'(m^2)+\rho'(m^2)]\right]\ .
\end{equation}
These are the final formulas used in \Sec{sec:NWL} to calculate the $\op(\Gamma)$
corrections to the narrow-width limit.

\section{Kinematics for $B\to K\pi\ell\ell$}\label{app:kinematics}

We consider the decay $\bar B^0 \to K^-(k_1) \pi^+(k_2) \ell^-(q_1) \ell^+(q_2)$. We follow 
the ``theory'' conventions for $B\to K^*(\to K\pi)\ell\ell$ as defined in \Ref{Gratrex:2015hna} (in agreement with \Ref{Altmannshofer:2008dz}) to define the three angles $\theta_K$, $\theta_\ell$ and $\phi$. In particular, the definition of $\theta_K$ agrees with that in \Sec{sec:FormFactors}. In the $B$-meson rest frame
we obtain, for the 4-vectors of interest for the leptonic side:
\eq{
q^\mu=\left(\begin{array}{c}
\frac{m_B^2-k^2+q^2}{2 m_B}\\
0\\
0\\
\frac{\sqrt\lambda}{2m_B}
\end{array}
\right)\ ,
\qquad
\bar{q}^\mu=\left(\begin{array}{c}
-\frac{\sqrt{\lambda_q \lambda}}{2m_B q^2}\cos\theta_\ell\\
\sqrt\frac{\lambda_q}{q^2}\sin\theta_\ell\cos\phi\\
\sqrt\frac{\lambda_q}{q^2}\sin\theta_\ell\sin\phi\\
-\frac{\sqrt{\lambda_q}}{2m_B q^2}(m_B^2-k^2+q^2)\cos\theta_\ell
\end{array}
\right) \ ,
}
and on the hadronic side:
\eq{
k^\mu=\left(\begin{array}{c}
\frac{m_B^2+k^2-q^2}{2 m_B}\\
0\\
0\\
-\frac{\sqrt\lambda}{2m_B}
\end{array}
\right)\ ,
\qquad
\bar{k}^\mu=\left(\begin{array}{c}
\frac{\sqrt{\lambda_{K\pi} \lambda}}{2m_B k^2}\cos\theta_K\\
-\sqrt\frac{\lambda_{K\pi}}{k^2}\sin\theta_K\\
0\\
-\frac{\sqrt{\lambda_{K\pi}}}{2m_B k^2}(m_B^2+k^2-q^2)\cos\theta_K
\end{array}
\right)\ , 
}
which leads to the following basis structures in the Lorentz decomposition of form factors and transversity amplitudes:
\eq{
k_t^\mu=\left(\begin{array}{c}
\frac{m_B^2-k^2+q^2}{2m_B\sqrt{q^2}}\\
0\\
0\\
\frac{\sqrt{\lambda}}{2m_B\sqrt{q^2}}
\end{array}
\right)\,,
\ 
k_0^\mu=\left(\begin{array}{c}
-\frac{\sqrt{\lambda}}{2m_B\sqrt{q^2}}\\
0\\
0\\
-\frac{m_B^2-k^2+q^2}{2m_B\sqrt{q^2}}
\end{array}
\right)\,,
\ 
k_\perp^\mu=\left(\begin{array}{c}
0\\
0\\
i\frac{\lambda_{K\pi}}{k^2}\sin\theta_K\\
0
\end{array}
\right)\,,
\ 
k_\|^\mu=\left(\begin{array}{c}
0\\
-\frac{\lambda_{K\pi}}{k^2}\sin\theta_K\\
0\\
0
\end{array}
\right)\,.
}


\end{document}